# FROM PQCD TO NEUTRON STARS: MATCHING

# EQUATIONS OF STATE TO CONSTRAIN GLOBAL

# STAR PROPERTIES

by

**TYLER GORDA**

B.S., Rutgers, the State University of New Jersey, 2011

M.S., University of Colorado Boulder, 2014

A thesis submitted to the

Faculty of the Graduate School of the

University of Colorado in partial fulfillment

of the requirements for the degree of

Doctor of Philosophy

Department of Physics

2016

This thesis entitled:

From pQCD to neutron stars: matching equations of state to constrain global star properties

written by Tyler Gorda

has been approved for the Department of Physics

———————————————————————

Prof. Paul Romatschke

———————————————————————

Prof. Anna Hasenfratz

Date ————————————————

The final copy of this thesis has been examined by the signatories, and we find that both the content and the form meet acceptable presentation standards of scholarly work in the above-mentioned discipline.

Gorda, Tyler (Ph.D., Physics)

From pQCD to neutron stars: matching equations of state to constrain global star properties

Thesis directed by Prof. Paul Romatschke

The equation of state (EoS) of quantum chromodynamics (QCD) at zero temperature can be calculated in two different perturbative regimes: for small values of the baryon chemical potential $\mu$, one may use chiral perturbation theory (ChEFT); and for large values of $\mu$, one may use perturbative QCD (pQCD). Each of these theories is controlled, predictive, and has much theoretical development. There is, however, a gap for $\mu \in (0.97\ \text{GeV}, 2.6\ \text{GeV})$, where these theories becomes non-perturbative, and where there is currently no known microscopic description of QCD matter. Unfortunately, this interval obscures the values of $\mu$ found within the cores of neutron stars (NSs).

In this thesis, we argue that thermodynamic matching of the ChEFT and pQCD EoSs is a legitimate way to obtain quantitative constraints on the non-pertubative QCD EoS in this intermediate region. Within this framework, one pieces together the EoSs coming from ChEFT (or another low-energy description) and pQCD in a thermodynamically consistent manner to obtain a band of allowed EoSs. This method trades qualitative modeling for quantitative constraints: one attempts no microscopic characterization of the underlying matter.

In this thesis, we argue that this method is an effective, verifiable, and systematically improvable way to explore and characterize the interior of NSs. First, we carry out a simplified matching procedure in QCD-like theories that can be simulated on the lattice without a sign problem. Our calculated pressure band serves as a prediction for lattice-QCD practitioners and will allow them to verify or refute the simplified procedure. Second, we apply the state-of-the-art matched EoS of Ref. [1] to rotating NSs. This allows us to obtain bounds on observable NS properties, as well as point towards future observations that would more tightly constrain the current state-of-the-art EoS band. Finally, as evidence of the ability to improve the procedure, we carry out calculations in pQCD to improve the zero-temperature pressure. We calculate the full $\mathcal{O}(g^6 \ln^2 g)$ contribution to the pQCD pressure for $n_f$ massless quarks, as well as a significant portion of the $\mathcal{O}(g^6 \ln g)$ piece and even some of the $\mathcal{O}(g^6)$ piece.

# DEDICATION

This thesis is dedicated with warmest appreciation to my past teachers and educators of all subjects.

# ACKNOWLEDGEMENTS

I would like to thank Paul Romatschke for his help and guidance over the last five years: especially for his encouragements to participate in a variety of conferences and workshops and the personal freedom he has given me to explore whatever physics I find interesting. I would also like to thank Oscar Henriksson, Andrew Koller, and Paige Warmker for discussing many an interesting physics topic over the years, and for helping me clarify many deep concepts. Many thanks as well to Hans Bantilan, with whom I have also had many exciting and fruitful discussions, and who constantly reminds me that mathematical clarity of thought not only has a place in physics, but is vital to it. Lastly, I thank Aleksi Vuorinen, Ioan Ghişoiu, and Aleksi Kurkela for many stimulating conversations about physics, which I look forward to continuing.

Finally, this thesis draws from two papers on which I have been an author. For those papers, I wish to acknowledge Gert Aarts, Tom DeGrand, Simon Hands, Yuzhi Liu, Marco Panero, Paul Romatschke, Andreas Schmitt, and Aleksi Vuorinen for many helpful discussions and suggestions.

# CONTENTS

**CHAPTER**

**APPENDIX**

# TABLES

**TABLE**

# FIGURES

**FIGURE**

# CHAPTER 1

# GENERAL INTRODUCTION

Quantum chromodynamics (**QCD**) is the microscopic theory of the strong nuclear force. Fundamentally, it is a theory of interacting quarks and gluons, but by extension, it describes the physics underpinning any objects composed of these elemental fields. Thus, in principle, the theory of QCD describes a vast range of objects and phenomena. This ranges from phenomena impacting our low-energy, everyday world (such as the structure and scattering of hadrons, the binding of baryons to one another in nuclei, and the structure and properties of nuclei in atoms) to much more exotic and energetic objects and processes (such as nuclear processes in the core of typical stars, the structure and properties of compact neutron stars (NSs), and supernovae). Unfortunately, in practice, most of these properties are not directly computable from the QCD path integral itself (and even those that are require significant effort). The reason for this is that QCD is **non-perturbative** at low energies. This means that only at high energies are the fundamental quark and gluon fields useful descriptions of QCD: at the low energies relevant to our everyday world, the weakly interacting degrees of freedom (i.e., the **quasiparticles**) are instead the baryons and mesons. This change of degrees of freedom as one proceeds down in energy from colored quarks and gluons to color-neutral hadrons is called **confinement**.

As a rule of thumb, perturbative (pQCD) calculations from the QCD path integral itself are typically valid for energies much higher than the so-called $\Lambda_{\text{QCD}}$ **scale**, which is around 250 MeV.

That is, roughly speaking, for energy scales E such that $E \gg \Lambda_{QCD}$, perturbation theory is valid. This is, of course a very rough estimate, for what constitutes "much greater than" in the problem is sometimes complicated or unclear. One can look at the running coupling constant $\alpha_s(\mu)$ and see when this becomes of order one, but this just shifts the problem to defining what "of order one" means. In any case, according to the Particle Data Group [2], $\alpha_s(m_Z) = 0.1148 \pm 0.0007$, and this increases to $\alpha_s(\sim 5\,\text{GeV}) = 0.2$ and $\alpha_s(\sim 1.5\,\text{GeV}) = 0.3$. Thus, $E \gg \Lambda_{QCD}$ more or less translates to $E > $ a few GeV. In what follows, we shall refer to this energy scale of a few GeV as the **perturbative scale: $\Lambda_{pQCD}$**.

On the low-energy side of the spectrum, the quasiparticles are hadrons. Again, precisely what determines "low-energy" requires some care, but it is approximately 1.2 GeV. This scale corresponds roughly to the mass of the lightest non-pionic hadrons. This distinction between pions and the rest of the hadrons is relevant because the pions are anomalously light for a reason: there is a symmetry of the QCD Lagrangian that is not manifest in the ground state. This symmetry is called **chiral symmetry**, and the aforementioned energy scale is referred to as the **chiral symmetry breaking scale $\Lambda_{CSSB}$**. At low energies, another perturbative theory exists called **chiral effective theory** (ChEFT) which allows one to compute low-energy properties and processes within a controlled framework [3, 4].

There is however, a pronounced gap in energies between $\Lambda_{CSSB}$ and $\Lambda_{pQCD}$, which is out of reach of any perturbative framework. To explore this region, one can resort to model Lagrangians that incorporate some relevant phenomena, such as confinement or chiral symmetry breaking, but in these models there is no controlled perturbative framework connected to the fundamental physics. Within the context of the condensed matter of QCD, or thermal field theory (TQFT), this non-perturbative region unfortunately obscures some very interesting physics. For example, this region obscures the chiral and confinement transitions, meaning that researchers are unable to study these transitions in a first-principled way throughout the entire phase plane. (There is a non-perturbative numerical technique called **lattice QCD**, but its techniques are not currently applicable outside of a narrow slice of the phase plane. This will be discussed in more detail in Chap. 2.) In addition, at small temperatures (T $\ll \Lambda_{QCD}$) and large densities n or baryon chemical ($\mu \gg \Lambda_{QCD}$), this non-perturbative regime obscures the region of the phase diagram applicable to the interiors of NSs.

On the one hand, this is a blessing: measurements of the properties of NSs can shed light on a region of the QCD phase diagram that is inaccessible to current theoretical techniques. On the other hand, NSs are extremely complex physical systems, combining fundamental QCD microphysics with thermodynamics and general relativity (and often extreme electromagnetism as well), and thus extracting the properties of NSs from observations is very challenging. Often, much theoretical modeling is necessary to perform this extraction, and since there is little theoretical control over the models, there are large uncertainties. Some properties determined primarily by the crust are understood in some detail, for the crust is of a low density and thus ChEFT or other techniques can describe it well. We do have limited knowledge of possible exotic phases of nuclear matter in the crust, and what effects crustal microphysics has on transport properties (see Ref. [5, 6, 7] for reviews).

As one moves into the cores of NSs, however, direct theoretical knowledge is lost. There are theoretical speculations of phase transitions to deconfined quark matter or even to the hypothesized ground state of QCD, strange quark matter [8, 9], within the cores, but the models and ideas in this region are, for the most part, phenomenological and not derived from QCD itself in a controlled way. That is, most of the models used to predict the structure of NSs use microphysics that is not connected rigorously to QCD.

However, there is a technique to reach the NS-region of the QCD phase diagram *if one is not interested in the microphysics*: thermodynamic matching [10, 1, 11]. One can hope to constrain bulk thermodynamic properties, such as the equation of state (EoS), along the entire $\mu$-axis using the two controlled perturbative regimes at low and high $\mu$. The EoS in the non-perturbative middle region will have to match the perturbative EoSs at the edges (meaning that the pressures of the two phases must be equal at the matching points), and throughout, there are restrictions coming from thermodynamic consistency and stability. For example, the energy density $\varepsilon$ (and the pressure P) must be monotonically increasing functions of $\mu$, and on either side of a matching point, the stable phase must be the one with the higher $P$. This simple prescription allows one to obtain a band of permitted EoSs in the non-perturbative regime. With this in hand, one may investigate bulk, global properties of NSs, including relations between the total mass $M$; circumferential, equatorial radius $R_e$; frequency of rotation $f$; and the moment of inertia $I$. We stress that these predictions are made essentially from the underlying QCD theory, and are thus constrained by first-principles, controlled physics.

It is this remarkable story that we wish to tell in this thesis: one can constrain properties of NSs governed by non-perturbative regimes of QCD by thermodynamically matching EoSs determined by *controlled, perturbative*-QCD physics. There are three main goals that we wish to accomplish in this thesis, in addition to identifying thermodynamic matching as an approach to real, physically interesting problems. First, we will show that this remarkably simplistic idea of matching EoSs can be checked by lattice "QCD" in some QCD-like theories. This is an important result, for it allows one to verify whether or not this simplistic, but seemingly powerful approach can really achieve what it claims to achieve. Second, we will extend applications of the state-of-the-art matched QCD EoS results of Kurkela *et al*. [1] and Fraga *et al*. [11] to rotating NSs. This will give us insight into which astronomical observations would provide additional constraints on the QCD EoS in the intermediate, non-perturbative regime (not to mention that this extension will provide important bounds on global NS properties). Third, we will calculate higher-order corrections to the zero-temperature pQCD EoS, which may be used in the future to conduct more refined matching studies of the QCD EoS in the non-perturbative regime.

The structure of this thesis is as follows. In Chap. 2 we begin with the QCD phase diagram, detailing which regions of the phase diagram permit direct theoretical investigation. We shall also outline the various phases and phase transitions that have been theoretically proposed to exist, as well as those which have been observed. In this chapter, we will be especially interested in the $T = 0$ region of the phase diagram, for $T = 0$ is an excellent approximation for NSs after their violent formation period [12, 13]. In Chap. 3, we discuss the perturbative approaches to understanding QCD, and in particular, for calculating thermodynamic quantities. We provide a comprehensive outline of thermal pQCD and explain how one may compute $P$ as a perturbative series in the strong coupling constant $g = \sqrt{4\pi\alpha_s}$. In the process, we discuss a class of diagrams known as the "plasmon" or "ring-sum" diagrams, and compute the $g^4 \ln g$ contribution to the zero-temperature pQCD EoS, as an exercise. We also provide a brief introduction to low-energy ChEFT in this chapter. Next, we take up the topic of EoS matching in Chap. 4. This chapter investigates questions about the matching procedure itself. Specifically, we investigate if there are ways to test if such a simple prescription actually works, using exotic "QCD-like" theories. In addition, we describe the current state-of-the-art matching procedure carried out in Refs. [1, 11]. Chap. 5 contains an brief overview of NSs, including a general discussion of NS structure within the framework of general relativity. This chapter also includes a survey of connections between the EoS band of Kurkela *et al.* [1] and global properties of rotating NSs. We examine the bounds on observable NS properties that follow from the EoS band, and we discuss future observations that can further constrain the band. Finally, in Chap. 6, we present improvements to the pQCD EoS at $T = 0$. We present a derivation of the $\mathcal{O}(g^6 \ln^2 g)$ term in the pressure for an arbitrary number $n_f$ of massless quarks from the plasmon sum terms, as well as parts of the $\mathcal{O}(g^6 \ln g)$ and $\mathcal{O}(g^6)$ terms. Finally, Chap. 7 contains our conclusions.

Let us now begin our approach towards the aforementioned goals with a survey of the QCD phase diagram.

## Notation

A brief word on notation in this thesis: we will work in the mostly minus convention, with the Minkowski metric $\eta_{\mu\nu} = \text{diag}(+1, -1, -1, -1)$. Four-vectors shall be denoted $P^\mu$, with components $P^\mu = (p^0, \vec{p})$. We shall denote the number of flavors of quarks as $n_f$, and when we generalize to QCD-like theories, $N$ shall denote the number of possible colors for a quark in the fundamental representation (e.g. SU(N)). Gluonic color indices shall be drawn from the beginning of the Latin alphabet: $a$, $b$, $c$, ...; flavor indices from the middle of the Latin alphabet: $f$, $g$, $h$, ...; and quark color indices from the beginning of the Greek alphabet: $\alpha$, $\beta$, $\gamma$, .... Space-time indices shall be drawn from the middle of the Greek alphabet: $\mu$, $\nu$, $\rho$, .... In this thesis, Boltzmann's constant $k$, Planck's reduced constant $\hbar$, and the speed of light $c$ will be set equal to one throughout.

# CHAPTER 2

# THE PHASE DIAGRAM OF QCD

In this chapter, we present a survey of the phase diagram of QCD, with an emphasis on the μ-axis. We will start with a brief overview of QCD itself (more details will be given in Chap. 3), and mention which parts of the phase diagram are accessible by direct calculations. This will consist of three different regions: first, regions on and near the T-axis, which are accessible to lattice-QCD techniques; second, the $T \gg \Lambda_{pQCD}$ region on and near the T-axis and the $\mu \gg \Lambda_{pQCD}$ region on and near the μ-axis, which are accessible by pQCD; and third, the $T \ll \Lambda_{CSSB}$, $\mu \ll \Lambda_{CSSB}$ region, accessible by ChEFT. In these regions, we will review what is known, with emphasis on the thermodynamic phases. We will also detail the main phase-transition regions, including those that are not well-understood (e.g., the confinement-deconfinement and chiral symmetry breaking transitions). We will finish with a focus on the non-perturbative region on and near the μ-axis, which is relevant to NSs.

## 2.1    The general structure of QCD

QCD is a theory of $n_f = 6$ flavors (f = u, d, s, c, b, t) of massive quarks with color indices $\alpha = 1, 2, 3$ interacting via an SU(3) gauge boson (the gluon) with color indices $a = 1, 2, \ldots, 8$. We shall depict these fields as $\psi_f^\alpha$ and $\mathcal{A}_\mu^a$ respectively. The **Lagrangian** is

$$\mathcal{L}_{QCD} = \sum_f \overline{\psi}_f^\alpha \left( \delta_{\alpha\beta} \left( i \gamma^\mu \partial_\mu - m_f \right) + g \gamma^\mu \mathcal{A}_\mu^a T_{\alpha\beta}^a \right) \psi_f^\beta - \frac{1}{4} F_{\mu\nu}^a F^{a\mu\nu}, \tag{2.1}$$

where here and in what follows, repeated color indices are always summed over. Here, $\gamma^\mu$ are the Dirac gamma matrices, $g$ is the gluon coupling constant, $m_f$ is the mass of the $f^{th}$ quark, $\overline{\psi} = \psi^\dagger \gamma^0$, $F_{\mu\nu}$ is the gluonic **field strength tensor**

$$F_{\mu\nu}^a = \partial_\mu \mathcal{A}_\nu^a - \partial_\nu \mathcal{A}_\mu^a - g f^{abc} \mathcal{A}_\mu^b \mathcal{A}_\nu^c, \tag{2.2}$$

with $f^{abc}$ to be defined below, and $T_{\bullet\bullet}^a$ are the **generators** of SU(3) in the same (fundamental) representation as the quarks (as can be seen from the lower indices in Eq. (2.1)). In what follows, we shall denote a general representation as R. Since in Chap. 4, we shall be considering different fermionic representations, we will take a moment here to remind the reader of the properties of these $T_R^a$. Regardless of the representation (i.e., range of the lower indices) these matrices satisfy commutation relations

$$[T_R^a, T_R^b] = i f^{abc} T_R^c, \tag{2.3}$$

where the $f^{abc}$ are the **structure constants** of SU(3). The $f^{abc}$ are completely antisymmetric and are defined by

$$f^{111} = 1, \quad f^{147} = f^{165} = f^{246} = f^{257} = f^{345} = f^{376} = \frac{1}{2}, \quad f^{458} = f^{678} = \frac{\sqrt{3}}{2}. \tag{2.4}$$

For the fundamental representation of SU(3), the generators are given by $T_f^a = \lambda^a / 2$ where the $\lambda^a$ are the **Gell-Mann matrices**:

$$\lambda^1 = \begin{pmatrix} 0 & 1 & 0 \\ 1 & 0 & 0 \\ 0 & 0 & 0 \end{pmatrix}, \quad \lambda^2 = \begin{pmatrix} 0 & -i & 0 \\ i & 0 & 0 \\ 0 & 0 & 0 \end{pmatrix}, \quad \lambda^3 = \begin{pmatrix} 1 & 0 & 0 \\ 0 & -1 & 0 \\ 0 & 0 & 0 \end{pmatrix}, \tag{2.5}$$

$$\lambda^4 = \begin{pmatrix} 0 & 0 & 1 \\ 0 & 0 & 0 \\ 1 & 0 & 0 \end{pmatrix}, \quad \lambda^5 = \begin{pmatrix} 0 & 0 & -i \\ 0 & 0 & 0 \\ i & 0 & 0 \end{pmatrix}, \tag{2.6}$$

$$\lambda^6 = \begin{pmatrix} 0 & 0 & 0 \\ 0 & 0 & 1 \\ 0 & 1 & 0 \end{pmatrix}, \quad \lambda^7 = \begin{pmatrix} 0 & 0 & 0 \\ 0 & 0 & -i \\ 0 & i & 0 \end{pmatrix}, \quad \lambda^8 = \frac{1}{\sqrt{3}} \begin{pmatrix} 1 & 0 & 0 \\ 0 & 1 & 0 \\ 0 & 0 & -2 \end{pmatrix}. \tag{2.7}$$

In the fundamental representation, the $T_f^a$ are normalized such that

$$\text{Tr}\left(T_f^a T_f^b\right) = T_f \delta^{ab}, \tag{2.8}$$

$$T_{\alpha\beta}^a T_{\beta\gamma}^a = C_f \delta_{\alpha\gamma}, \tag{2.9}$$

$$f^{acd} f^{bcd} = C_A \delta^{ab}, \tag{2.10}$$

with $T_f = \frac{1}{2}$, $C_f = \frac{3^2-1}{2\cdot3} = \frac{4}{3}$, and $C_A = 3$. Eq. (2.8) and Eq. (2.9) can be extendend to an arbitrary representation R; in that case, these equations define new constants $T_R$ and $C_R$.

A few words about these factors are in order here. First, they are frequently written in terms of N for a general SU(N) gauge theory, it which case one has $T_f = \frac{1}{2}$, $C_f = \frac{N^2-1}{2N}$, and $C_A = N$. Secondly, the quark fields $\psi_f^\alpha$ are commonly combined into a single quark flavor vector $\vec{\psi}^\alpha$. In this case, the flavor index f disappears from the fields and appears in the $T_f$: $T_f = \frac{n_f}{2}$ instead in this case, since the trace in Eq. (2.8) will also trace over the flavor index as well. Lastly, the "A" in Eq. (2.10) stands for "adjoint". The generators of the adjoint representation are defined by the structure constants:

$$(T_A^a)_{bc} = -i f^a_{bc}. \tag{2.11}$$

(Note that for the structure constants, whether an index is up or down makes no difference.) This definition means that Eq. (2.10) is a special case of the generalized Eq. (2.9).

One final note must be made about the general structure of QCD. Although there are indeed $n_f = 6$ flavors of quarks in nature, only $n_f = 3$ are ever probed at zero temperature in nature. As will be discussed in the following section, in NSs, the quark chemical potential $\mu_q$ is always much below the mass of the c quark. This means that only the three lightest quarks (u, d, s) are active within dense nuclear matter.

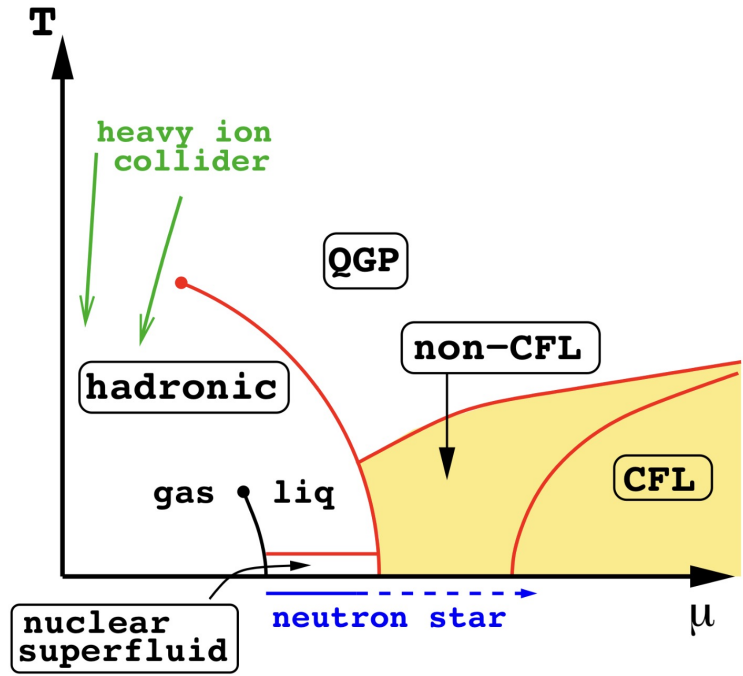

FIGURE 2.1: A cartoon of the QCD phase diagram. Taken from Ref. [14].

## 2.2  The phase diagram of nuclear matter

In Fig. 2.1 we present a cartoon of the phase diagram of QCD from Ref. [14]. There are many phases labelled, but most of these are less well-established than the plot suggests: all the lines drawn in red are theoretical predictions. The main phase transitions in nuclear matter are located on the curving red line, separating the regions labelled "hadronic" and "QGP" (quark-gluon plasma). This is where the confinement-deconfinement transition and chiral transitions are located. The line represents a suspected first order phase transition [5], which ends at a proposed critical point, represented by the red dot. This confinement-deconfinement transition continues in the form of a crossover all the way to the $T$-axis, where it has been studied using lattice QCD [15]. This transition effectively separates the two perturbative regimes mentioned in the introduction: pQCD and ChEFT. Across this transition, the degrees of freedom of QCD change dramatically: At high $\mu$ and $T$, the degrees of freedom are colored quarks and gluons (though they do not necessarily interact weakly), and at low $\mu$ and $T$, the degrees of freedom are the colorless hadrons of everyday experience.

### 2.2.1 Lattice QCD and the T-axis

In some senses, the most directly-accessible region of the QCD phase diagram is the entire T-axis, and the adjacent $\mu/T \ll 1$ region at large values of T. This is the region where large-scale numerical simulations on discretized space-time, known as **lattice QCD** are applicable. It is beyond the scope of this thesis to describe lattice QCD in great detail; the interested reader may refer to Montvay and Munster [16] or Kogut and Stephanov [17], the latter being an overview of the whole QCD phase plane from a lattice-QCD perspective. The essential idea of lattice QCD is to use a Monte-Carlo algorithm to generate many possible gauge field configurations on the discretized space-time lattice. Using these generated configurations, one can then calculate the expectation value of an observable $\mathcal{O}$ using a numerical approximation of the Euclidean path-integral expression

$$\langle \hat{\mathcal{O}} \rangle = \int \mathcal{D}U \, \mathcal{D}\overline{\psi} \, \mathcal{D}\psi \, \mathcal{O} \, e^{-S^E(U, \overline{\psi}, \psi)}$$
$$= \int \mathcal{D}U \, \mathcal{O} \left[ e^{-S^E_{YM}(U)} \det M(\mu) \right]. \qquad (2.12)$$

In the above expression, U is an element of the gauge group SU(3), and $S^E(U, \overline{\psi}, \psi)$ is the Euclidean action of QCD, which has been broken up as

$$S^E = S^E_{YM} + \int_x \overline{\psi} M \psi. \qquad (2.13)$$

Here $S^E_{YM}$ represents the pure Yang-Mills (gluon) action, and M represents a matrix operator which couples the fermions to the gauge field. In going from the first to the second line of Eq. (2.12), the fermionic integrals were performed, meaning that one may indeed use only the gauge fields to evaluate $\langle \hat{\mathcal{O}} \rangle$. [For the full expression of the path integral in thermal QCD, see Eq. (3.65) of this thesis, with the Euclidean Lagrangians of thermal QCD given in Eqs. (3.62), (3.63), and (3.64).] This method of calculating $\langle \hat{\mathcal{O}} \rangle$ works very well when the bracketed term in Eq. (2.12) is a real, positive number, for then one can view Eq. (2.12) as a weighted sum over gauge field configurations. More importantly, when this bracketed term is real and positive, one can **importance sample**. That is, when generating configurations within the simulation, one can weight the probability of accepting or rejecting a configuration by this $e^{-S_{YM}^E(U)} \det M(\mu)$ factor. This avoids generating too many gauge configurations with very small probabilities. However, when $\mu > 0$, $\det M(\mu)$ becomes complex, and importance sampling breaks down. This is referred to as a **sign problem**. It is possible to extend lattice-QCD predictions to small values of $\mu/T$ using a Taylor expansion on the T-axis [18, 19], and so the lattice can probe both the T-axis and slightly off of it at high T, but the rest of the phase plane remains inaccessible to direct simulation because of this sign problem. There are many proposals to extend direct simulations to non-zero density, including Lefschetz thimbles [20], complex Langevin [21, 22, 23, 24], and strong coupling expansions [25]. For an introductory survey of non-zero density lattice-QCD ideas, see Ref. [26].

### 2.2.2 The μ-axis and phases in pQCD

The μ-axis in the QCD phase plane is theoretically accessible only at low and high values of μ. We begin from low μ first in our survey. At low values of μ (and for non-zero T), matter can be viewed (and has been observed!) as a gas of weakly-interacting, color-neutral hadrons. Exactly at $T = 0$, as μ increases, the density and pressure are zero up to an **onset transition**. This can be understood by remembering that at $T = 0$, there can be no excitations of particles of mass m until $\mu = m$. Past the onset transition, there is a predicted liquid-gas phase transition [27], for which there is experimental evidence (see Ref. [5] and the references therein). The general theoretical reason to expect a liquid-gas phase transition in nuclear matter is simply that any self-attracting system of particles with a hard core repulsion will exhibit such a phase transition at low enough temperatures. The reason is as follows. Consider a gas of particles with an intermediate-range attractive force and a short-range repulsive force. If the temperature of the system is high enough, the attractive, intermediate-range force will not substantially affect the dynamics of the particles, as they will have too much kinetic energy to be much affected. As the temperature is lowered, however, the attractive force will have a larger and larger impact on the denser areas of the gas until at some point, the pressure of the gas will have two minima at different densities, one corresponding to the less-dense gas, and the other the more-dense liquid phase. (This is exactly what happens in a van der Waals gas.) Hadrons do have an attractive interaction (as evidenced by the fact that they bind into nuclei) and a hard core repulsion, and thus a liquid-gas transition is expected. A critical endpoint of the liquid-gas transition line is expected, corresponding to the point where the densities of the liquid and gaseous phases are no longer different.

Beyond the onset transition and the liquid-gas transition, a nuclear superfluid is expected [6, 7], as well as various inhomogeneous "pasta phases" where the nuclei stretch and rearrange into different structures to minimize energy, with extra (superfluid) neutrons flowing in the empty spaces between (see Refs. [5, 6], and Ref. [28] and references therein). These structures and phases very likely exist in the crusts of neutron stars, and perhaps even deeper, but beyond a certain depth, these descriptions should break down, and it is unknown what microphysics takes over.

Starting from asymptotically high values of $\mu$ also begins with understanding and ends in non-perturbative physics. At high values of $\mu$, the weakly interacting degrees of freedom are quarks and gluons, and pQCD is applicable. At high enough densities, there is another general theoretical prediction for the phase of nuclear matter: a color superconductor. As is nicely described in Ref. [14], this state is expected on quite general theoretical grounds. Since the quasi-particles in pQCD are quarks and gluons, the relevant dynamics are confined to the quark Fermi surface. However, the Fermi surface has a Cooper instability. Since there are some gluon-exchange channels that are attractive (as is evidenced by baryon formation at lower chemical potentials), quarks form BCS pairs, and the ground state will be a superconductor. As noted in Ref. [14], in this case, the Cooper pairs are bound by the fundamental interactions between the fermions themselves, unlike in an electrical superconductor, and so the binding is much more robust. At the highest densities relevant to NSs (meaning that the quark degrees of freedom consist of the $u$, $d$, and $s$ quarks only; see below), a **color-flavor-locked phase (CFL)** can be shown to have the lowest energy. In this phase, the normally independent SU(3) symmetries of color (c) and left and right flavor rotations (L and R respectively) become locked together: only the vector subgroup of $SU(3)_c \otimes SU(3)_L \otimes SU(3)_R$ is a symmetry of the ground state. In fact, the full hidden symmetry or symmetry breaking pattern is [14, 29]

$$SU(3)_c \otimes SU(3)_L \otimes SU(3)_R \otimes U(1)_B \rightarrow SU(3)_{c+L+R} \otimes \mathbb{Z}_2, \qquad (2.14)$$

where $U(1)_B$ is the $U(1)$ symmetry associated baryon conservation in QCD. This symmetry breaking proceeds via the formation of a condensate

$$\langle \psi_f^\alpha C \gamma^5 \psi_g^\beta \rangle \neq 0. \qquad (2.15)$$

Here, the Latin subscripts are flavor indices, $C$ is the charge-conjugation matrix, and $\gamma^5 = i \gamma^0 \gamma^1 \gamma^2 \gamma^3$ is the usual operator associated with projections onto left and right chiral fields. This CFL phase, like the hadronic phase *breaks the chiral flavor symmetry*, though in a different way. (The details of the hadronic chiral symmetry breaking, as well as the chiral flavor symmetry of QCD in general, will be discussed in Sec. 3.2.2 below.) In the CFL phase, there are condensates of L quarks paired with L quarks and of R quarks paired with R quarks. The former locks $SU(3)_L$ to $SU(3)_c$, and the latter locks $SU(3)_R$ to $SU(3)_c$. Since both of the flavor rotations are locked to the same color rotation, axial flavor rotations (i.e., the ones that act oppositely on L and R quarks) are not a symmetry of the CFL ground state: chiral symmetry is indeed hidden or spontaneously broken.

It can be shown [14] that this CFL phase is the ground state of QCD matter in nature at sufficiently high densities. Values of $\mu$ much beyond 500 MeV do not exist in nature, and so the heavier three quarks (c, b, t) are never active in dense matter [14, 30]. As $\mu$ is lowered from the region of the CFL ground state, the large value of $m_s$ compared to $m_u$ and $m_d$ begins to stress the CFL pairing. CFL pairing eventually becomes energetically disfavored, and possibly one or more different pairings set in [30]. This occurs in the region marked "non-CFL" in Fig. 2.1. The precise microphysical details of this region are at present theoretically unknown. There are proposals of other types of color superconductivity that may be the ground state of nuclear matter at some points in the "non-CFL" region, but it is unknown if these are the ground state near the confinement-deconfinement transition line [30].

Between the nuclear superfluid phase and the CFL phase lies a region of unknown behavior: towards larger $\mu$ there are proposals for valid microscopic descriptions, but towards smaller $\mu$, the microscopic description is unknown. It is even unknown whether or not there is deconfined quark matter within NS cores. It is this central, non-perturbative region on the $\mu$-axis that describes the physics of NS cores, and it is this region of the QCD EoS that we wish to investigate in this work. In Chap. 4 below, we shall discuss the EoS-matching approach that Kurkela *et al*. [1] have used to constrain the intermediate, non-perturbative QCD EoS (as well as our original work on EoS matching in QCD-like theories that can be simulated on the lattice without a sign problem). Before approaching the matching details, however, we will review the perturbative regimes at high and low $\mu$ (at $T = 0$) where controlled calculations are possible. We proceed to this topic in Chap. 3.

# CHAPTER 3

# PERTURBATIVE EXPLORATIONS OF QCD

In this chapter, we describe the perturbative regimes at low and high $\mu$ (at $T = 0$) that allow for controlled calculations within QCD. First, in Sec. 3.1, we present the details of thermal pQCD starting from quantum statistical mechanics. We discuss the derivation of the path-integral representation of the partition function for both bosons and fermions, and we explain how to compute the pressure as a sum of bubble Feynman diagrams. We also discuss a specific class of diagrams, called "ring-sum" or "plasmon" diagrams, and compute the lowest-order contribution to the QCD pressure coming from these diagrams, as an exercise. A calculation of a higher-order contribution coming from these diagrams will be detailed in Chap. 6. In Sec. 3.2, we provide a brief overview of two low-energy, effective descriptions of QCD: ChEFT and the hadron resonance gas. Following this chapter, we shall proceed to the details of how one may match these perturbative regimes thermodynamically in Chap. 4.

## 3.1   PQCD at nonzero temperature and density

### 3.1.1   The partition function of a quantum field

In the vacuum, the fundamental quantity governing the dynamics of a quantum field $\phi$ with a Lagrangian $\mathcal{L}$ is the **generating functional**, which can be written as a path integral over

field configurations

$$Z = \int \mathcal{D}\phi\, e^{-i\int d^4x\, \mathcal{L}(\phi, \partial_\mu \phi)}. \tag{3.1}$$

In matter (i.e. in a thermodynamic ensemble), this is no longer the quantity of interest. For a thermodynamic system, the **partition function** of a system contains all the possible information about the system:

$$Z \equiv \text{Tr}\left[e^{-\beta(\hat{H}-\mu\cdot\hat{N})}\right] = \int d\phi \langle\phi|e^{-\beta(\hat{H}-\mu\cdot\hat{N})}|\phi\rangle, \tag{3.2}$$

where here, $\beta = 1/T$, $\hat{H}$ is the **Hamiltonian** of the statistical system, $\hat{N}$ is a conserved number operator, $\mu$ is the chemical potential associated with $\hat{N}$, and the $|\phi\rangle$ are a complete set of states of the system satisfying

$$\hat{\phi}|\phi\rangle = \phi|\phi\rangle. \tag{3.3}$$

(N.b. that the second equality in Eq. (3.2) is only true for bosons, as are the state definitions (3.3). This will be discussed more below). Incredibly, for a quantum field theory, the partition function (3.2) can be written in a form that is almost exactly identical to the generating functional (3.1). This remarkable fact is connected to the similar forms of the time evolution operator $\exp(-i\hat{H}t)$ and the Boltzmann factor $\exp(-\beta(\hat{H}-\mu\cdot\hat{N}))$. (Note that without $\mu$ there is a suggestive correspondence

$$i\, t \leftrightarrow \beta, \tag{3.4}$$

which can indeed be made more precise.)

### 3.1.2    A derivation of the partition function for bosonic and fermionic fields

To derive the form of the partition function for a bosonic quantum field, we follow Kapusta and Gale [31]. (We will deal with the fermionic result afterwards.) Because of the suggestive analogy between time in the time evolution operator and inverse temperature in the Boltzmann factor, we divide the interval $[0, \beta]$ into N pieces (with the intent of letting $N \to \infty$), and we insert both a complete set of momentum-conjugate states $|\pi\rangle$ and a complete set of states $|\phi\rangle$ at each division such that each Boltzmann factor has a $\langle\pi|$ on its left and a $|\phi\rangle$ on its right. We also define $\Delta\tau = \beta/N$, and number the insertions $1, 2, \ldots, N$, from right to left. This gives, for a Boltzmann factor inserted between two different states $\phi_a$ and $\phi_b$:

$$\langle\phi_b|e^{-\beta(\hat{H}-\mu\cdot\hat{N})}|\phi_a\rangle = \lim_{N\to\infty}\int\prod_{i=1}^{N}\left(\frac{d\pi_i\,d\phi_i}{2\pi}\right)\langle\phi_1|\pi_N\rangle\langle\pi_N|e^{-\Delta\tau(\hat{H}-\mu\cdot\hat{N})}|\phi_N\rangle\times$$
$$\langle\phi_N|\pi_{N-1}\rangle\langle\pi_{N-1}|e^{-\Delta\tau(\hat{H}-\mu\cdot\hat{N})}|\phi_{N-1}\rangle\langle\phi_{N-1}|\pi_{N-2}\rangle\times$$
$$\cdots\times\langle\phi_2|\pi_1\rangle\langle\pi_1|e^{-\Delta\tau(\hat{H}-\mu\cdot\hat{N})}|\phi_1\rangle\langle\phi_1|\phi_a\rangle. \tag{3.5}$$

Now,

$$\langle\phi_i|\phi_j\rangle = \delta(\phi_i - \phi_j), \tag{3.6}$$

$$\langle\phi_{i+1}|\pi_i\rangle = \exp\left(i\int d^3x\,\pi_i(\vec{x})\phi_{i+1}(\vec{x})\right), \tag{3.7}$$

and if $\Delta\tau \ll 1$,

$$\langle\pi_j|e^{-\Delta\tau(\hat{H}-\mu\cdot\hat{N})}|\phi_j\rangle \approx \langle\pi_j|1 - \Delta\tau(\hat{H}-\mu\cdot\hat{N})|\phi_j\rangle \tag{3.8}$$

$$= \langle\pi_j|\phi_j\rangle\left[1 - \Delta\tau(H - \mu\cdot N)\right] \tag{3.9}$$

$$= \exp\left(\int d^3x\left\{-i\,\pi_j(\vec{x})\phi_j(\vec{x}) - \Delta\tau\left[\mathcal{H}(\pi_j,\phi_j) - \mu\cdot\mathcal{N}(\pi_j,\phi_j)\right]\right\}\right), \tag{3.10}$$

where here $\mathcal{H}(\pi,\phi)$ and $\mathcal{N}(\pi,\phi)$ are the Hamiltonian and number densities

$$\hat{H} = \int d^3x\,\mathcal{H}(\hat{\pi},\hat{\phi}), \tag{3.11}$$

$$\hat{N} = \int d^3x\,\mathcal{N}(\hat{\pi},\hat{\phi}), \tag{3.12}$$

respectively, evaluated at the eigenvalues $\pi$ and $\phi$. Plugging Eq. (3.10) back into the expanded amplitude (3.5), we find

$$\langle\phi_b|e^{-\beta(\hat{H}-\mu\cdot\hat{N})}|\phi_a\rangle = \lim_{N\to\infty}\int\prod_{i=1}^{N}\left(\frac{d\pi_i\,d\phi_i}{2\pi}\right)\delta(\phi_1 - \phi_a)\cdot$$
$$\exp\left\{-\Delta\tau\sum_{j=1}^{N}\int d^3x\left[\mathcal{H}(\pi_j,\phi_j) - i\,\pi_j\frac{(\phi_{j+1}-\phi_j)}{\Delta\tau} - \mu\cdot\mathcal{N}(\pi_j,\phi_j)\right]\right\},$$
$$\tag{3.13}$$

where we have defined $\phi_{N+1} \equiv \phi_b$. Taking the limit $N \to \infty$, we obtain a path integral

$$\langle \phi_b | e^{-\beta(\hat{H} - \mu \cdot \hat{N})} | \phi_a \rangle = \int \mathcal{D}\pi \int_{\phi(\vec{x},0)=\phi_a(\vec{x})}^{\phi(\vec{x},\beta)=\phi_b(\vec{x})} \mathcal{D}\phi \cdot$$
$$\exp\left\{ -\int_0^\beta d\tau \int d^3x \left[ \mathcal{H}(\pi,\phi) - i\,\pi \frac{\partial\phi}{\partial\tau} - \mu \cdot \mathcal{N}(\pi,\phi) \right] \right\}. \tag{3.14}$$

This expression is quite general, and will work for any bosonic field component. Let us now for a moment assume that there is no conserved number operator or chemical potential. Eq. (3.14) then becomes

$$\langle \phi_b | e^{-\beta\hat{H}} | \phi_a \rangle = \int \mathcal{D}\pi \int_{\phi(\vec{x},0)=\phi_a(\vec{x})}^{\phi(\vec{x},\beta)=\phi_b(\vec{x})} \mathcal{D}\phi \exp\left\{ -\int_0^\beta d\tau \int d^3x \left[ \mathcal{H}(\pi,\phi) - i\,\pi \frac{\partial\phi}{\partial\tau} \right] \right\}. \tag{3.15}$$

Since we are assuming a bosonic field, we know that $\mathcal{H}$ will be of the form

$$\mathcal{H} = \frac{1}{2}\pi^2 + \frac{1}{2}(\vec{\nabla}\phi)^2 + \cdots, \tag{3.16}$$

and there will be no other $\pi$ dependence. Thus, the $\pi$ integral is a Gaussian, and the result is (dropping an infinite constant coefficient)

$$\langle \phi_b | e^{-\beta\hat{H}} | \phi_a \rangle = \int_{\phi(\vec{x},0)=\phi_a(\vec{x})}^{\phi(\vec{x},\beta)=\phi_b(\vec{x})} \mathcal{D}\phi \exp\left\{ -\int_0^\beta d\tau \int d^3x\, \mathcal{L}^E \right\}, \tag{3.17}$$

where here, $\mathcal{L}^E = -\mathcal{L}(\tau = it)$ is simply the Minkowski Lagrangian with $\eta_{\mu\nu}$ replaced by $\delta_{\mu\nu}$ (and t replaced by $\tau$). This makes rigorous the correspondence (3.4), at least in the case of bosons. The full partition function then follows from Eq. (3.17) as

$$Z = \int d\phi \langle \phi | e^{-\beta\hat{H}} | \phi \rangle = \int_{\phi(\vec{x},0)=\phi(\vec{x},\beta)} \mathcal{D}\phi \exp\left\{ -\int_0^\beta d\tau \int d^3x\, \mathcal{L}^E \right\}, \tag{3.18}$$

where the integral imposes periodic boundary conditions on the $\phi$ field in the $\tau$-direction. Bosonic fields are periodic in imaginary time.

The derivation of the path integral for fermions is very similar, though there are some additional complications due to the anticommuting nature of the fermionic variables. We shall not go into all the details there, but we will partially derive the results below. First of all, the states that one uses in the path integral are different. Defining the **coherent states**

$$|\psi\rangle = e^{-\psi\hat{a}^\dagger}|0\rangle \tag{3.19}$$

and

$$\langle\psi| = \langle 0|e^{-\hat{a}\psi^\dagger},\tag{3.20}$$

where here $\hat{a}^\dagger$ and $\hat{a}$ are the fermionic creation and annihilation operators respectively, allows one to perform integrals over Grassmann (anticommuting) fields. With these definitions, one finds that the identity and trace operations are different than the usual ones for a bosonic field. These relations are (see the texts by Laine and Vuorinen [32] or Le Bellac [33] for details)

$$\mathrm{Id} = \int\int d\psi^\dagger\, d\psi\, e^{-\psi^\dagger\psi}|\psi\rangle\langle\psi|,\tag{3.21}$$

$$\mathrm{Tr}\,(\hat{\mathcal{O}}) = \int\int d\psi^\dagger\, d\psi\, e^{-\psi^\dagger\psi}\langle-\psi|\hat{\mathcal{O}}|\psi\rangle.\tag{3.22}$$

Retracing the same steps as the bosonic relations above, but inserting Id as in Eq. (3.21) instead of alternating between complete sets of $|\phi\rangle$ and $|\pi\rangle$, one finds

$$\langle\psi_b|e^{-\beta(\hat{H}-\mu\cdot\hat{N})}|\psi_a\rangle = \int_{\overline{\psi}(\vec{x},0)=\overline{\psi}_a(\vec{x})}^{\overline{\psi}(\vec{x},\beta)=\overline{\psi}_b(\vec{x})}\mathcal{D}\overline{\psi}\int_{\psi(\vec{x},0)=\psi_a(\vec{x})}^{\psi(\vec{x},\beta)=\psi_b(\vec{x})}\mathcal{D}\psi\,\cdot$$
$$\cdot\exp\left\{-\int_0^\beta d\tau\int d^3x\left[\mathcal{H}(\overline{\psi},\psi)+\overline{\psi}\gamma^0\frac{\partial\psi}{\partial\tau}-\mu\cdot\mathcal{N}(\overline{\psi},\psi)\right]\right\}.\tag{3.23}$$

Here, we have changed variables $\psi^\dagger\to\overline{\psi}$, which produces no new factors into the integrand because $\left|\det(\gamma^0)\right|=1$. This equation is the fermionic equivalent of Eq. (3.14). To proceed further, let us first assume that there is no conserved current and that the fermionic field $\psi$ is a Dirac field. Then we will have

$$\mathcal{L} = \overline{\psi}(i\,\gamma^\mu\partial_\mu - m)\psi + \cdots,\tag{3.24}$$

where there is no $\partial_0\psi$ dependence in the $\cdots$. Then

$$\pi = \frac{\partial\mathcal{L}_{\mathrm{Dirac}}}{\partial(\partial_0\psi)} = \overline{\psi}\,i\,\gamma^0 = i\,\psi^\dagger,\tag{3.25}$$

giving

$$\mathcal{H} = \pi\partial_0\psi - \mathcal{L} = \overline{\psi}(-i\,\vec{\gamma}\cdot\vec{\nabla}+m)\psi + \cdots,\tag{3.26}$$

since the time component cancels out. Plugging this into Eq. (3.23) then yields the analogue to Eq. (3.17)

$$\langle\psi_b|e^{-\beta\hat{H}}|\psi_a\rangle = \int_{\overline{\psi}(\vec{x},0)=\overline{\psi}_a(\vec{x})}^{\overline{\psi}(\vec{x},\beta)=\overline{\psi}_b(\vec{x})}\mathcal{D}\overline{\psi}\int_{\psi(\vec{x},0)=\psi_a(\vec{x})}^{\psi(\vec{x},\beta)=\psi_b(\vec{x})}\mathcal{D}\psi\,\exp\left\{-\int_0^\beta d\tau\int d^3x\,\mathcal{L}^{\mathrm{E}}\right\},\tag{3.27}$$

where, in this case we define

$$\mathcal{L}^{\mathrm{E}} = \overline{\psi}(\widetilde{\gamma}_\mu \partial_\mu + \mathfrak{m})\psi - \cdots \tag{3.28}$$

to be the Euclidean Lagrangian. In this case, in addition to changing the metric $\eta_{\mu\nu}$ to $\delta_{\mu\nu}$ (and t to $\tau$), we have to also change the Dirac matrices to **Euclidean Dirac matrices**

$$\widetilde{\gamma}_0 \equiv \gamma^0, \qquad \widetilde{\gamma}_i \equiv -i\,\gamma^i, \tag{3.29}$$

which are appropriately named, for they satisfy the same algebra as the Minkowskian Dirac matrices, but for the *Euclidean* metric:

$$\{\widetilde{\gamma}_\mu, \widetilde{\gamma}_\nu\} = \delta_{\mu\nu}. \tag{3.30}$$

From this it follows that all of the Euclidean Dirac matrices are Hermitian (unlike in the Minkowskian case, in which only $\gamma^0$ is). We thus find that for fermions the path integral is over a Euclidean version of the Minkowski Lagrangian as well.

Using the different fermionic trace identity (3.22) gives the partition function:

$$Z = \int\!\!\int d\psi^\dagger \, d\psi \, e^{-\psi^\dagger \psi} \langle -\psi | e^{-\beta \hat{H}} | \psi \rangle = \int\!\!\int_{\substack{\psi(\vec{x},0)=-\psi(\vec{x},\beta) \\ \overline{\psi}(\vec{x},0)=-\overline{\psi}(\vec{x},\beta)}} \mathcal{D}\overline{\psi}\mathcal{D}\psi \exp\left\{ -\int_0^\beta d\tau \int d^3x \, \mathcal{L}^{\mathrm{E}} \right\}, \tag{3.31}$$

where the integral now imposes **antiperiodic** boundary conditions on the $\psi$ field in the $\tau$-direction. Fermions are antiperiodic in imaginary time.

### 3.1.3 Changes to the Euclidean Lagrangians with a chemical potential

To derive the forms of Eqs. (3.18) and (3.31) with a conserved current (and conjugate chemical potential $\mu$), we need to use the specific form of the bosonic and fermionic conserved currents. For a complex field $\chi$ whose Lagrangian is invariant under the relation ($\alpha \in \mathbb{R}$)

$$\chi \to e^{-i\,\alpha}\chi, \qquad \chi^\dagger \to e^{i\,\alpha}\chi^\dagger, \tag{3.32}$$

the general conserved current is given by [34]

$$\mathcal{N} = \frac{\partial \mathcal{L}}{\partial(\partial^0 \chi)}\frac{\delta \chi}{\delta \alpha} + \frac{\partial \mathcal{L}}{\partial(\partial^0 \chi^*)}\frac{\delta \chi^*}{\delta \alpha}. \tag{3.33}$$

For a bosonic field $\phi$ this becomes

$$\mathcal{N}_b = -i\,\phi\,\partial_0\phi + i\,\phi^*\partial_0\phi^*, \tag{3.34}$$

which after the Legendre transformation becomes

$$\mathcal{N}_b = -i\,\phi\,\pi + i\,\phi^*\pi^*; \tag{3.35}$$

and for fermions, the current is

$$\mathcal{N}_f = -i\,\overline{\psi}\,i\,\gamma_0\psi = \overline{\psi}\gamma_0\psi. \tag{3.36}$$

In the bosonic case, this means that the integrand in the argument of the exponential in Eq. (3.14) becomes (working in terms of the normalized real and imaginary parts $\phi = (\phi_1 + i\,\phi_2)/\sqrt{2}$

$$\begin{aligned}
\mathcal{H} - i\sum_i \pi_i\partial_\tau\phi_i - \mu\cdot\mathcal{N}_b =\; & \frac{\pi_1^2}{2} + \frac{\pi_2^2}{2} - i\,\pi_1\partial_\tau\phi_1 - i\,\pi_2\partial_\tau\phi_2 + \mu(\pi_2\phi_1 - \pi_1\phi_2) + \cdots \tag{3.37}\\
=\; & \frac{1}{2}\big[\pi_1 - 2(i\,\partial_\tau\phi_1 + \mu\phi_2)\big]^2 + \frac{1}{2}\big[\pi_2 - 2(i\,\partial_\tau\phi_2 - \mu\phi_1)\big]^2 \\
& + \frac{1}{2}(\partial_\tau\phi_1 - i\,\mu\phi_2)^2 + \frac{1}{2}(\partial_\tau\phi_2 + i\,\mu\phi_1)^2 + \cdots \tag{3.38}\\
=\; & \frac{\widetilde{\pi}_1^2}{2} + \frac{\widetilde{\pi}_2^2}{2} + \frac{1}{2}(\partial_\tau\phi_1 - i\,\mu\phi_2)^2 + \frac{1}{2}(\partial_\tau\phi_2 + i\,\mu\phi_1)^2 + \cdots \tag{3.39}
\end{aligned}$$

where here we have suppressed even the spacial part of the kinetic term, and in the last step we have redefined the $\pi_1$ and $\pi_2$ integrals to remove the constant shift factors. Performing the $\pi_i$ integrals then yields the changes

$$\partial_\tau\phi_1 \to \partial_\tau\phi_1 - i\,\mu\phi_2, \qquad \partial_\tau\phi_2 \to \partial_\tau\phi_2 + i\,\mu\phi_1, \tag{3.40}$$

in $\mathcal{L}^{\text{E}}$. Or, going back to the complex notation,

$$\partial_\tau\phi \to (\partial_\tau - \mu)\phi, \qquad \partial_\tau\phi^* \to (\partial_\tau + \mu)\phi^*, \tag{3.41}$$

in $\mathcal{L}^{\text{E}}$. For fermions, we find a similar result almost immediately from Eq. (3.23):

$$\mathcal{H} + \psi^\dagger\partial_\tau\psi - \mu\cdot\mathcal{N}_f = 0 + \psi^\dagger\partial_\tau\psi - \mu\overline{\psi}\gamma_0\psi + \cdots, \tag{3.42}$$

$$= \overline{\psi}\,\widetilde{\gamma}_0(\partial_\tau - \mu)\psi + \cdots, \tag{3.43}$$

where once again we have suppressed the spacial parts of the kinetic term here and have also used the fact that $\left(\gamma^0\right)^2 = 1$ and $\widetilde{\gamma}_0 = \gamma^0 = \gamma_0$. Thus, for fermions, adding a chemical potential amounts to the change

$$\partial_\tau \psi \to (\partial_\tau - \mu)\psi \tag{3.44}$$

in $\mathcal{L}^E$, just as in Eq. (3.41) above for bosons.

### 3.1.4 The propagators an nonzero temperature and density

To obtain the propagators for bosons and fermions at nonzero temperature and density, we simply look at the Fourier components of the exponentials in Eqs. (3.14) and (3.23). Observe that for $T > 0$, the $\tau$ integral is over a compact interval. Therefore, the Fourier transform of this component is a Fourier series. Moreover, because the boundary conditions for bosons and fermions on this interval are different (bosons are periodic while fermions are antiperiodic), the allowed values of frequency components of these fields are different. For bosons, the allowed frequencies are

$$\omega_n^b = 2\pi T n, \tag{3.45}$$

and for fermions, the allowed frequencies are

$$\omega_n^f = 2\pi T \left(n + \frac{1}{2}\right). \tag{3.46}$$

In either case, these frequencies are referred to as **Matsubara frequencies**. For a free charged scalar field, the integral in the exponential reads

$$\int_0^\beta d\tau \int d^3x \left\{ \left[(\partial_\tau + \mu)\phi^*\right]\left[(\partial_\tau - \mu)\phi\right] + (\vec{\nabla}\phi^*) \cdot (\vec{\nabla}\phi) + m^2\phi^*\phi \right\}$$
$$= \int_0^\beta d\tau \int d^3x \left\{ \phi^*\left[-(\partial_\tau - \mu)^2 - \nabla^2 + m^2\right]\phi \right\}, \tag{3.47}$$

where we have integrated by parts. This means that the propagator for a charged scalar field is

$$D_b(\omega, \vec{k}) = \frac{1}{(\omega + i\mu)^2 + \vec{k}^2 + m^2}. \tag{3.48}$$

For free fermions, the derivation is again more straightforward, for it can be read directly from the integral in the exponential

$$\int_0^\beta d\tau \int d^3x \left\{ \overline{\psi} \left[ \widetilde{\gamma}_0 (\partial_\tau - \mu) + \vec{\widetilde{\gamma}} \cdot \vec{\nabla} + m \right] \psi \right\}. \tag{3.49}$$

The propagator is thus

$$D_f(\omega, \vec{k}) = \frac{1}{i\widetilde{\gamma}_0(\omega + i\mu) + i\vec{\widetilde{\gamma}} \cdot \vec{k} + m} = \frac{-i\widetilde{\gamma}_0(\omega + i\mu) - i\vec{\widetilde{\gamma}} \cdot \vec{k} + m}{(\omega + i\mu)^2 + \vec{k}^2 + m^2}. \tag{3.50}$$

We can write these in a more compact form and think about them in a different way by looking at the case with no conserved current (i.e., the $\mu = 0$ case), and by defining the Euclidean four momentum $P = (\omega, \vec{k})$. Then, if we define the propagators as

$$D_b(\omega, \vec{k}) = \frac{1}{P^2 + m^2}, \qquad D_f(\omega, \vec{k}) = \frac{-i\slashed{P} + m}{P^2 + m^2}, \tag{3.51}$$

where $\slashed{P} = \widetilde{\gamma}_\mu P^\mu$ is the obvious generalization of the Feynman slash, we see that Eqs. (3.48) and (3.50) imply that for $\mu \neq 0$, $\omega$ *acquires an imaginary part*

$$D_b(\omega, \vec{k}) \to D_b(\omega + i\mu, \vec{k}), \tag{3.52}$$

$$D_f(\omega, \vec{k}) \to D_f(\omega + i\mu, \vec{k}). \tag{3.53}$$

This is often a more useful way to think of the effect of chemical potential: **$\mu$ adds an imaginary part $i\mu$ to the Matsubara frequencies**. To simplify future equations, we adopt the notation $\widetilde{P}$ to mean that $P_0$ is shifted by $i\mu$ (and the rest of the components remain unchanged). Therefore, we can write the above general propagators as

$$D_b(\widetilde{P}) = \frac{1}{\widetilde{P}^2 + m^2}, \qquad D_f(\widetilde{Q}) = \frac{-i\slashed{\widetilde{Q}} + m}{\widetilde{Q}^2 + m^2}. \tag{3.54}$$

Note that the change $P \to \widetilde{P}$ only occurs for the particles that have a conserved current/chemical potential. If there are bosons and fermions in a theory with only one of them having a chemical potential (such as in QCD), then only that one type of particle will have shifted time components.

### 3.1.5 The pressure of interacting quantum fields at nonzero temperature and density

As with quantum fields in vacuum, interacting quantum fields can be described using perturbation theory, Wick's theorem, and the Feynman diagram formalism. This can be seen as follows. The (grand canonical) pressure in statistical mechanics is given by

$$P \equiv -\Omega/V = T/V \ln Z \tag{3.55}$$

where $\Omega$ is the grand potential, $V$ is the volume, and $Z$ is the partition function of the system. Now, if we have an interacting quantum field $\chi$, with the (Euclidean) Action

$$S_{\text{free}} + S_{\text{I}}, \tag{3.56}$$

where $S_{\text{free}}$ is the free-field action, and $S_{\text{I}}$ contains the interaction terms, the partition function will be of the form

$$Z = \int \mathcal{D}\chi^* \int \mathcal{D}\chi \, e^{-\left(S_{\text{free}} + S_{\text{I}}\right)}, \tag{3.57}$$

which can be manipulated up as

$$
\begin{aligned}
Z &= \int \mathcal{D}\chi^* \int \mathcal{D}\chi \, e^{-S_{\text{free}}} e^{-S_{\text{I}}} \\
&= \int \mathcal{D}\chi^* \int \mathcal{D}\chi \, e^{-S_{\text{free}}} \sum_{n=0}^{\infty} \frac{(-1)^n}{n!} S_{\text{I}}^n,
\end{aligned}
\tag{3.58}
$$

so that

$$\ln Z = \ln Z_0 + \ln\left\{ 1 + \sum_{n=1}^{\infty} \frac{(-1)^n}{n!} \left[ \frac{\int \mathcal{D}\chi^* \int \mathcal{D}\chi \, e^{-S_{\text{free}}} S_{\text{I}}^n}{\int \mathcal{D}\chi^* \int \mathcal{D}\chi \, e^{-S_{\text{free}}}} \right] \right\}, \tag{3.59}$$

where here $Z_0$ is the partition function of the free part only (i.e., the denominator of the fraction in Eq. (3.59)). Thus we see that as in the vacuum case, what is necessary is to compute structures of the form

$$\frac{\int \mathcal{D}\chi^* \int \mathcal{D}\chi \, e^{-S_{\text{free}}} S_{\text{I}}^n}{\int \mathcal{D}\chi^* \int \mathcal{D}\chi \, e^{-S_{\text{free}}}}. \tag{3.60}$$

Since all that is different between this and the vacuum case is the change $i \to -1$ in the exponent, the shift in the energies $\omega \to \omega + i\mu$ in the case of $\mu \neq 0$, and the change from the Minkowski to Euclidean Lagrangian, the whole machinery of Feynman diagrams still goes through, just with slightly different Feynman rules. In the next section, we specialize to thermal QCD and write down the thermal Feynman rules.

### 3.1.6 Feynman diagrams in thermal QCD

QCD at nonzero temperature is defined by the path-integral partition function

$$Z_{QCD} = \int\int_{\text{anti-periodic}} \prod_f \mathcal{D}\overline{\psi}_f \mathcal{D}\psi_f \int_{\text{periodic}} \mathcal{D}\mathcal{A}_\mu^a \, e^{-\int_0^\beta d\tau \int d^3x \, \mathcal{L}_{QCD}^E}, \tag{3.61}$$

where

$$\mathcal{L}_{QCD}^E = \sum_f \overline{\psi}_f^\alpha \left( \delta_{\alpha\beta} \left( \widetilde{\gamma}_\mu \partial_\mu + m_f \right) - i \, g \widetilde{\gamma}_\mu \mathcal{A}_\mu^a T_{\alpha\beta}^a \right) \psi_f^\beta + \frac{1}{4} F_{\mu\nu}^a F^{a\mu\nu}, \tag{3.62}$$

and where $F_{\mu\nu}^a$ has the usual form of Eq. (2.2), and again, $\eta_{\mu\nu}$ has been replaced by $\delta_{\mu\nu}$ (and t by $\tau$). For non-zero density, the $\partial_\tau \to \partial_\tau - \mu$ prescription of Eq. (3.49) holds here as well, of course. This expression (3.61) is what one may expect from the forms of bosonic and fermionic path integrals in Eqs. (3.18) and (3.31), but in reality, there are substantial subtleties in the derivation brought about by gauge invariance. We refer the interested reader to Ref. [32] for a complete derivation.

As is the case in vacuum, in order to do perturbative calculations, one must restrict the gauge freedom further by introducing an explicit gauge-fixing term and so-called **Faddeev-Popov ghosts**. We skip the details of this (see Refs. [32, 31, 33] for the derivations), for as we shall discuss below, we will not need to do calculations with ghosts in this work. In a general covariant gauge, one must use not just $\mathcal{L}_{QCD}^E$ as the Lagrangian, but

$$\mathcal{L}_{pQCD}^E = \mathcal{L}_{QCD}^E + \mathcal{L}_{\text{gauge-fixed}}^E, \tag{3.63}$$

with the gauge-fixing Lagrangian given by

$$\mathcal{L}_{\text{gauge-fixed}}^E = \frac{1}{2\xi} \partial_\mu \mathcal{A}_\mu^a \partial_\nu \mathcal{A}_\nu^a + \partial_\mu \overline{c}^a \partial_\mu c^a + g \, f^{abc} \partial_\mu \overline{c}^a \mathcal{A}_\mu^b c^c, \tag{3.64}$$

where the $c^a$ are the Grassmann, scalar ghost fields, and $\xi$ is a free parameter that must drop out of any gauge-invariant observable. The full pQCD partition function is then given by

$$Z_{pQCD} = \int\int_{\text{anti-periodic}} \prod_f \mathcal{D}\overline{\psi}_f \mathcal{D}\psi_f \int_{\text{periodic}} \mathcal{D}\mathcal{A}_\mu^a \int\int_{\text{periodic}} \mathcal{D}\overline{c}^a \, \mathcal{D}c^a \, e^{-\int_0^\beta d\tau \int d^3x \, \mathcal{L}_{pQCD}^E}. \tag{3.65}$$

Note that the ghost fields $c^a$, $\bar{c}^a$, though fermionic, are *periodic* in imaginary time. This is in keeping with the nature of the ghost fields: they are Grassmann variables with bosonic properties otherwise.

Because of the similarities between the gauge-fixed Minkowskian Lagrangian and the Euclidean Lagrangian of Eqs. (3.63), (3.62), and (3.64), one can practically modify the Minkowskian Feynman rules by eye to arrive at the Euclidean Feynman rules. They are [1]

$$\psi_\beta \xrightarrow{\quad Q \quad} \overline{\psi}_\alpha \;=\; \delta_{\alpha\beta} \frac{-i\,\widetilde{\not{Q}}+m}{\widetilde{Q}^2+m^2}, \tag{3.66}$$

$$\mathcal{A}_\mu^a \underset{P}{\overset{\text{\textbf{0000000}}}{\sim}} \mathcal{A}_\nu^b \;=\; \frac{\delta_{ab}}{P^2}\left[\delta_{\mu\nu}-(1-\xi)\frac{P_\mu P_\nu}{P^2}\right], \tag{3.67}$$

$$c^a \xrightarrow[\;\;K\;\;]{\cdots\cdots} \overline{c}^b \;=\; \frac{\delta^{ab}}{K^2}, \tag{3.68}$$

$$\mathcal{A}_\mu^a \;=\; g\,\widetilde{\gamma}_\mu T_{\alpha\beta}^a, \tag{3.69}$$

$$\mathcal{A}_\mu^b \;=\; -i\,g\,f^{abc}K_\mu, \tag{3.70}$$

---

[1] Feynman diagrams created with the `TikZ-Feynman` package [35].

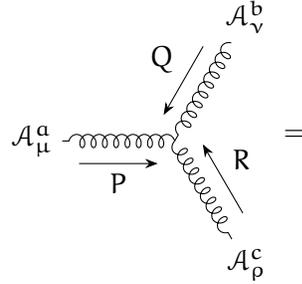

$$
= \quad i\, g\, f^{abc} \Big[ \delta_{\mu\rho}(P-R)_\nu
$$
$$
+ \delta_{\nu\mu}(Q-P)_\rho
$$
$$
+ \delta_{\rho\nu}(R-Q)_\mu \Big], \tag{3.71}
$$

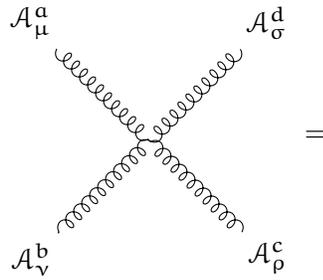

$$
= \quad -g^2 \Big[ f^{ade} f^{ebc}(\delta_{\mu\nu}\delta_{\sigma\rho} - \delta_{\mu\rho}\delta_{\sigma\nu})
$$
$$
+ f^{abe} f^{edc}(\delta_{\mu\sigma}\delta_{\nu\rho} - \delta_{\mu\rho}\delta_{\nu\sigma})
$$
$$
+ f^{ace} f^{edb}(\delta_{\mu\sigma}\delta_{\rho\nu} - \delta_{\mu\nu}\delta_{\rho\sigma}) \Big]. \tag{3.72}
$$

In addition to these Feynman rules, one must also satisfy momentum conservation at the vertices, and one must also integrate over every internal propagator momentum. For bosons (and ghosts) this is achieved by the sum-integral

$$
T \sum_{n=-\infty}^{\infty} \int \frac{d^3\vec{p}}{(2\pi)^3} \Big[ \cdots \Big]_{P_0 = \omega_n^b}, \tag{3.73}
$$

and for fermions

$$
T \sum_{n=-\infty}^{\infty} \int \frac{d^3\vec{q}}{(2\pi)^3} \Big[ \cdots \Big]_{Q_0 = \omega_n^f + i\,\mu}. \tag{3.74}
$$

We thus see that because of the compact interval in the $\tau$ (i.e., imaginary time) direction, one must perform a sum-integral rather than just a sum.

Another important point that must be mentioned here is that in evaluating the pressure (3.55), since one is dealing directly with the logarithm of the partition function, **one need only compute connected bubble diagrams**. One need only compute the connected diagrams because $W = \ln(Z)$ is the generator of the connected diagrams, as in the vacuum, and one need only compute the bubble diagrams (i.e., those with no external legs) because there is no operator insertion inside $Z$. (Note that another common name for these bubble diagrams is "vacuum diagrams", but we choose to reserve that name for any diagram evaluated at $T = 0$ and $\mu = 0$.) We now have all the machinery necessary to compute the pressure in pQCD.

### 3.1.7 Restriction to zero temperature

Since in this thesis we are primarily interested in high-density, low-temperature regime, we shall now restrict ourselves to $T = 0$ and discuss what simplifications follow from this. Sending $T \to 0$ means that $\beta \to \infty$, and therefore the integral over $\tau$ is no longer over a compact interval. This means that the aforementioned sum-integrals turn into sums again:

$$T \sum_{n=-\infty}^{\infty} \int \frac{d^3\vec{p}}{(2\pi)^3} \left[ \cdots \right]_{P_0 = \omega_n^b} \to \int \frac{d^4 P}{(2\pi)^4} \left[ \cdots \right] \equiv \int_P, \tag{3.75}$$

and for fermions

$$T \sum_{n=-\infty}^{\infty} \int \frac{d^3\vec{p}}{(2\pi)^3} \left[ \cdots \right]_{Q_0 = \omega_n^f + i\mu} \to \int \frac{d^4 Q}{(2\pi)^4} \left[ \cdots \right]_{\tilde{Q}} \equiv \int_{\tilde{Q}}. \tag{3.76}$$

Here we have defined shortened definitions for the integrals over fields with and without chemical potentials. This change from sum-integral to integral is simplifying because sum-integrals are (usually) most easily evaluated by integrating over a function with poles in the complex plane at the values $\{\omega_n^b : n \in \mathbb{Z}\}$ or $\{\omega_n^f + i\mu : n \in \mathbb{Z}\}$. However, at zero temperature, one need only evaluate a single integral without inserting the function with the poles.

Another important simplification that happens at zero temperature is that many Feynman diagrams no longer contribute. This can be seen by a heuristic explanation. At $T > 0$, all the fields in the theory can be excited by thermal fluctuations: fermions, gauge bosons, and ghosts. Therefore, all possible bubble diagrams contribute at $T > 0$. (By this we mean, all these diagrams have a dependence on $T$ and thus differ from their vacuum values.) At $T = 0$ (and $\mu > 0$), only fermions can be directly populated by "thermal" fluctuations, since they are the only particles with a chemical potential. Thus, only bubble diagrams with fermions contribute at $\mu > 0$ (meaning that only diagrams with fermions have a dependence on $\mu$ and differ from their vacuum values). This vastly simplifies calculations. We also need not concern ourselves with ghost fields in this work, for at $T = 0$ they enter into the pQCD pressure through the gluon polarization tensor at $\mathcal{O}(g^2)$, and this polarization tensor has already been calculated in the literature for massless [36] and massive [37] quarks. In our calculation in Chap. 6, the $\mathcal{O}(g^2)$ gluon polarization tensor is all that is needed to extract the $g^6 \ln^2 g$ contribution to the pQCD pressure, and so we do not need to calculate any further diagrams with ghost fields.

### 3.1.8 Ring/plasmon diagrams and their contribution to the pressure at zero temperature

Let us now compute a piece of the pQCD pressure, one that comes from a class of diagrams instead of a single diagram: the $\mathcal{O}(g^4 \ln g)$ piece. This will serve not only to illustrate the Feynman rules and machinery described above, but it will also serve as a prelude to the calculation of the $\mathcal{O}(g^6 \ln^2 g)$ piece of the pQCD pressure in Chap. 6.

First, recall that in QCD, one defines the **polarization tensor** or **self-energy $\Pi^{\mu\nu}$** of the gluon field as

$$\Pi^{\mu\nu} = \text{\includegraphics{}} + \text{\includegraphics{}} + \text{\includegraphics{}} + \text{\includegraphics{}} + \cdots \qquad (3.77)$$

where all the terms shown are $\mathcal{O}(g^2)$, and the $\cdots$ represents higher-order terms. At $T = 0$, only the first term has a matter contribution, since the other diagrams have no dependence on $\mu$. That is, at $T = 0$, one may split

$$\Pi^{\mu\nu} = \left( \text{\includegraphics{}} + \text{\includegraphics{}} + \text{\includegraphics{}} + \text{\includegraphics{}} + \cdots \right)_{\text{vac}}$$
$$+ \left( \text{\includegraphics{}} + \cdots \right)_{\text{mat}}. \qquad (3.78)$$

We will define these two terms as $\Pi^{\mu\nu}_{\text{vac}}$ and $\Pi^{\mu\nu}_{\text{mat}}$, respectively, and from now on, we shall only consider $\Pi^{\mu\nu}_{\text{mat}}$.

Now consider a diagram of the form

$$(3.79)$$

where there are a total of $n$ insertions of the gluon polarization tensor $\Pi_{mat}^{\mu\nu}$. This diagram corresponds to the expression (with $\xi = 1$ for simplicity)

$$(3.79) = \int_P \frac{\text{Tr}\,(\Pi_{mat})^n}{(P^2)^n}. \tag{3.80}$$

Here, the trace is over the space-time indices; i.e., we have used the simplified notation $\text{Tr}\,(\Pi_{mat})^n = \Pi_{mat}^{\mu\nu}\Pi_{mat}^{\nu\rho}\cdots\Pi_{mat}^{\sigma\mu}$, for the contraction of the $n$ polarization tensors. In medium, and, more specifically, for $\mu > 0$,

$$\lim_{P\to 0} \Pi_{mat}^{\mu\nu} \neq 0, \tag{3.81}$$

that is, **the gluon acquires an in-medium mass**. This is a standard result in thermal field theory [32, 31, 33], which is actually quite technically involved for QCD. We refer the interested reader to the aforementioned references for details. Because of this result, we see that the diagram in Eq. (3.79) will be infrared divergent for $2n - 4 \geqslant 1$, or $n > 2$, and moreover, the diagrams will become more strongly divergent for larger $n$.

This diagram (3.79) enters the partition function with a coefficient

$$\frac{(n-1)!}{2} \times \frac{(-1)^n}{n!} = \frac{1}{2} \times \frac{(-1)^n}{n}, \tag{3.82}$$

where here the $(n-1)!/2$ comes from the ways of arranging the $n$ $\Pi_{mat}$ blobs in a circle (the factor of 2 is because the handedness of the circle does not matter), and the $(-1)^n/n!$ is from the original $(-1)^n/n!$ in Eq. (3.59). Because of the form of Eqs (3.79) and (3.82), it is possible to combine the infrared divergent diagrams for every $n > 2$. We define the **plasmon** or **ring-sum** contribution to

be

$$\frac{\Omega_{\text{plas}}}{V} = \frac{1}{2}\left(\frac{1}{2}\ \overset{\text{(diagram)}}{\Pi_{\text{mat}}\ \Pi_{\text{mat}}}\ -\ \frac{1}{3}\ \overset{\text{(diagram)}}{\Pi_{\text{mat}}}\ +\ \frac{1}{4}\ \overset{\text{(diagram)}}{\Pi_{\text{mat}}}\ -\ \cdots\right) \tag{3.83}$$

$$= \frac{1}{2}\int_P\left[\sum_{n=2}^{\infty}\frac{(-1)^n}{n}\text{Tr}\left(\Pi_{\text{mat}}\mathcal{D}_0\right)^n\right] = \frac{1}{2}\int_P\text{Tr}\left[\sum_{n=2}^{\infty}\frac{(-1)^n}{n}\left(\Pi_{\text{mat}}\mathcal{D}_0\right)^n\right]$$

$$= \frac{1}{2}\int_P\text{Tr}\left[\ln\left(1+\Pi_{\text{mat}}\mathcal{D}_0\right) - \Pi_{\text{mat}}\mathcal{D}_0\right], \tag{3.84}$$

where we have written $\mathcal{D}_0$ for the gluon propagator, and we have again used the Tr notation. Using the standard definitions [31]

$$F_{\text{mat}}(K) = \Pi_{\text{mat}}^{\mu\nu}(K)P_{L\mu\nu} = \frac{K^2}{\vec{k}^2}\Pi_{\text{mat}}^{00}(k), \tag{3.85}$$

and

$$G_{\text{mat}}(K) = \frac{1}{2}\Pi_{\text{mat}}^{\mu\nu}(K)P_{T\mu\nu} = \frac{1}{2}\left(\Pi_{\text{mat}\,\mu}^{\mu}(K) - \frac{K^2}{\vec{k}^2}\Pi_{\text{mat}}^{00}(K)\right), \tag{3.86}$$

where $P_{L\mu\nu}$ and $P_{T\mu\nu}$ are the longitudinal and transverse projectors, respectively, and $\Pi_{\mu}^{\mu} = \Pi_0^0 - \Pi_i^i$, we can write the plasmon sum (3.84) as

$$\frac{\Omega_{\text{plas}}}{V} = \frac{d_A}{2}\int_K\left[\ln\left(1 - \frac{F_{\text{mat}}(K)}{K^2}\right) + \frac{F_{\text{mat}}(K)}{K^2} + 2\ln\left(1 - \frac{G_{\text{mat}}(K)}{K^2}\right) + 2\frac{G_{\text{mat}}(K)}{K^2}\right], \tag{3.87}$$

Here, $d_A = 8$ is the dimension of the adjoint representation of SU(3) and appears because of the contraction over the gluon color indices. In most of what follows, we will assume that $\Pi^{\mu\nu}$ is known only to order $g^2$. Moreover, to conserve space in the following derivation, we will omit the $G$ terms, which have precisely the same form as the $F$ terms but with a factor of 2 multiplying them. To evaluate the four-dimensional integral (3.87), we combine the four Euclidean dimensions $(K_0, K_1, K_2, K_3)$ into radial and angular coordinates

$$\int_K \equiv \int\frac{d^4K}{(2\pi)^4} = \int_0^{\infty}dK\frac{K^3}{(2\pi)^4}\int d\Omega_3, \tag{3.88}$$

and use the $SO(3) \times \mathbb{Z}_2$ symmetry of the integral to write the measure as

$$\int_K = \int_0^\infty dK \frac{K^3}{(2\pi)^4} \int d\Omega_2 \int_0^\pi d\Phi \sin^2(\Phi) = \frac{2}{(2\pi)^3} \int_0^\infty dK\, K^2 \int_0^{\pi/2} d\Phi \sin^2(\Phi), \tag{3.89}$$

where here $\Phi$ is the four-dimensional polar angle.

To extract the $g^4 \ln g$ piece of $\Omega_{\text{plas}}$, it is enough to know the behavior of the functions $F_{\text{mat}}(K, \Phi)$ and $G_{\text{mat}}(K, \Phi)$ at $K = 0$ [31, 10]. To isolate this behavior, we can set $K = 0$ wherever possible and subtract off the divergent behavior at large $K$ to arrive at

$$\frac{\Omega_{\text{plas}}}{V} = \frac{d_A}{(2\pi)^3} \int_0^\infty dK\, K^2 \int_0^{\pi/2} d\Phi \sin^2(\Phi) \left[ \ln\left(1 - \frac{F_{\text{mat}}(K = 0, \Phi)}{K^2}\right) + \frac{F_{\text{mat}}(K = 0, \Phi)}{K^2} \right.$$
$$\left. + \frac{F_{\text{mat}}(K = 0, \Phi)}{2K^2(K^2 + \chi^2)} \right] + \mathcal{O}(g^4). \tag{3.90}$$

Here $\chi$ is a fictitious mass scale introduced in order to preserve the IR behavior of the integrand. Since it is fictitious, physical observables must be independent of it. The reason for writing things in this manner is that (3.90) is analytically integrable in $K$, giving

$$\frac{\Omega_{\text{plas}}}{V} = \frac{d_A}{2(2\pi)^3} \int_0^{\pi/2} d\Phi \sin^2(\Phi) \left[ F_{\text{mat}}^2(K = 0, \Phi) \left(-\frac{1}{2} + \ln\left(\frac{-F_{\text{mat}}(K = 0, \Phi)}{\chi^2}\right)\right) \right] + \mathcal{O}(g^4). \tag{3.91}$$

Since $F = g^2(\cdots)$, the $g^4 \ln g$ piece can be isolated as

$$\frac{\Omega_{\text{plas}}}{V} = \frac{d_A\, g^4 \ln g}{(2\pi)^3} \int_0^{\pi/2} d\Phi \sin^2(\Phi) \left[ 2\frac{G_{\text{mat}}^2(K = 0, \Phi)}{g^4} + \frac{F_{\text{mat}}^2(K = 0, \Phi)}{g^4} \right]. \tag{3.92}$$

This has been evaluated further in the massless quark case by Freedman and McLerran [38] and Vuorinen [39], and in the massive case by Kurkela *et al.* [10].

Manipulations of this kind will lead us to the $\mathcal{O}(g^6 \ln^2 g)$ piece of the pressure in Chap. 6. There, we shall isolate the subleading behavior near $K = 0$, which will allow us to extract the higher-order terms. In the meantime, let us move to small $\mu$ and give a brief overview of the perturbative description of nuclear matter in that regime.

## 3.2    Low-energy effective theories of QCD

As has been already discussed, at small $\mu$, the effective degrees of freedom of nuclear matter are hadrons. At very low $\mu$ and T, one can correctly [40, 41, 42, 43] describe these degrees of freedom as a **hadron resonance gas** (HRG): a non-interacting collection of hadrons. One can even describe some aspects of the hadrons using the quark model. In Chap. 4 we shall make use of this group-theoretic description to construct the hadrons in QCD-like theories: SU(N) gauge theories with N = 2, 3, 4; $n_f = 2$, 3; and with quarks in the fundamental or two-index, antisymmetric representation. For the moment, we shall keep the discussion of the HRG as general as possible. For a more controlled, robust theoretical description of low-energy QCD, one may use ChEFT instead. This effective theory constrains the interactions between hadrons by using the chiral symmetry of massless QCD as a guiding principle. In this section, we will provide a brief introduction to each of these topics so that the reader may have some understanding of the low-energy theories later used in the thermodynamic matching of Chap. 4.

### 3.2.1    The HRG equation of state

In this section, we shall keep the discussion of the HRG as general as possible: we shall consider both T = 0 and $\mu = 0$, and we shall consider color-neutral hadrons in a general SU(N) gauge theory with quarks in an arbitrary representation. In doing so, results in this section will be applicable to our discussions of QCD-like theories in Sec. 4.1.

The low-T pressure (at $\mu = 0$) in a general QCD-like theory is given by considering the system to be a free gas of hadrons. Moreover, the statistics of the hadrons may be ignored, so that the distribution functions may all be assumed to be Boltzmann factors. In that case, we have

$$P_{\mathrm{HRG}}(T) = T \sum_{i \in H} g_i \int \frac{d^3 \vec{p}}{(2\pi)^3} e^{-\sqrt{\vec{p}^2 + m_i^2}/T} = T^4 \sum_{i \in H} \frac{g_i}{2\pi^2} \left(\frac{m_i}{T}\right)^2 K_2\left(\frac{m_i}{T}\right), \qquad (3.93)$$

where here the sum is over the hadron spectrum of the theory; $g_i$ and $m_i$ are the degeneracy and the mass of the $i^{\mathrm{th}}$ hadron, respectively; and $K_2$ is a modified Bessel function of the second kind.

The low-$\mu$ pressure (at $T = 0$) can be calculated in a similar way, but in this case the statistics of the particles cannot be ignored. For $T = 0$ and $\mu > 0$, the only particles that contribute to the partition function are particles containing no antiquarks, which we denote by B. In theories with quarks in the fundamental representation, these are simply the baryons, whereas for theories with quarks in exotic representations, there are more particles fitting this description (see Sec. 4.1.2 for a detailed discussion of hadrons in specific QCD-like theories). As such, in this section we shall refer to all the particles in B as baryons. Taking the $T \to 0$ limit in the fermionic-baryon ($\eta = 1$) or bosonic-baryon ($\eta = -1$) case yields

$$
\begin{aligned}
P_{HRG}(\mu) &= \lim_{T \to 0} T \sum_{i \in B} g_i \eta \int \frac{d^3\vec{p}}{(2\pi)^3} \ln \left( 1 + \eta \, e^{\left( \mu r_i - \sqrt{\vec{p}^2 + m_i^2} \right)/T} \right) \\
&= \sum_{i \in B} g_i \eta \int \frac{d^3\vec{p}}{(2\pi)^3} \ln \left[ \lim_{T \to 0} \left( 1 + \eta \, e^{\left( \mu r_i - \sqrt{\vec{p}^2 + m_i^2} \right)/T} \right)^T \right] \\
&= \sum_{i \in B} g_i \eta \int_0^{\sqrt{(\mu r_i)^2 - m_i^2}} \frac{\vec{p}^2 \, d|\vec{p}|}{2\pi^2} \left( \mu r_i - \sqrt{\vec{p}^2 + m_i^2} \right) \theta(\mu r_i - m_i) \\
&= \eta \sum_{i \in B} \frac{g_i}{48\pi^2} \left[ \mu r_i \sqrt{(\mu r_i)^2 - m_i^2} \left( 2(\mu r_i)^2 - 5m_i^2 \right) \right. \\
&\qquad\qquad \left. + 3m_i^4 \cosh^{-1} \left( \frac{m_i}{\sqrt{(\mu r_i)^2 - m_i^2}} \right) \right] \theta(\mu r_i - m_i), \qquad (3.94)
\end{aligned}
$$

where here $\theta$ is the Heaviside step function and $r_i = N_i/N_b$, with $N_i$ being the number of quarks in the $i$th particle, and $N_b$ being maximum number of quarks contained in a particle in B. (Particles satisfying $N_i = N_b$ are what we would normally call baryons). This result is correct for the fermionic-baryon case, but this formula gives negative P in the bosonic-baryon case when $\mu > \min_{i \in B} m_i/r_i$. This is because, in the bosonic case, a condensate of the $i^{th}$ baryon forms when $\mu = m_i/r_i$. (This has been numerically investigated in the two-color case by Hands $et\ al.$ [44] and analytically by Kogut $et\ al.$ [45] in all QCD-like theories with pseudoreal fermions.) In fact, in the completely non-interacting case it is nonsensical for $\mu$ to exceed $\min_{i \in B} m_i/r_i$. Since the hadrons in these theories are composite particles, they are not truly non-interacting, and we can have $\mu > \min_{i \in B} m_i/r_i$.

To make sense of this case, we consider the bosons as a (complex) quantum field $\Phi$ with a $|\Phi|^4$ repulsive interaction. For simplicity, we consider each baryon to be an independent field, and we examine the case of a scalar field (degeneracies may easily be incorporated at the end). A single baryon then has the Lagrangian density

$$\mathcal{L} = (\partial_\mu \Phi^\dagger)(\partial^\mu \Phi) - m^2 \Phi^\dagger \Phi - \lambda(\Phi^\dagger \Phi)(\Phi^\dagger \Phi), \tag{3.95}$$

with $\lambda > 0$. Following Kapusta and Gale [31], we introduce a baryon chemical potential $\mu_r$ and explicitly factor out the zero momentum mode

$$\Phi = \xi + \chi, \tag{3.96}$$

where $\xi \in \mathbb{R}$ is a constant and the constant Fourier component of $\chi$ satisfies $\chi_{n=0}(\vec{p} = 0) = 0$. One may think of $\xi$ as the condensate field and $\chi$ as the fluctuations about the vacuum state. We also write the fluctuations in terms of the normalized real and imaginary parts

$$\chi = \frac{1}{\sqrt{2}}(\chi_1 + i\chi_2). \tag{3.97}$$

In terms of these new variables, the Euclidean Lagrangian density becomes

$$\mathcal{L} = -\frac{1}{2}\left(\frac{\partial \chi_1}{\partial \tau} - i\mu_r \chi_2\right)^2 - \frac{1}{2}\left(\frac{\partial \chi_2}{\partial \tau} + i\mu_r \chi_1\right)^2 - \frac{1}{2}\nabla^2 \chi_1 - \frac{1}{2}\nabla^2 \chi_2$$
$$-\frac{1}{2}(6\lambda\xi^2 + m^2)\chi_1^2 - \frac{1}{2}(2\lambda\xi^2 + m^2)\chi_2^2 - U(\xi) + \mathcal{L}_I, \tag{3.98}$$

where $\tau$ is the Euclidean time, $\mathcal{L}_I$ contains interacting terms in $\chi$ (which we henceforth ignore), and

$$U(\xi) = (m^2 - (\mu_r)^2)\xi^2 + \lambda\xi^4. \tag{3.99}$$

We thus see from (3.99) that for $\mu_r < m$ the state $\xi_0 = 0$ is the stable vacuum and (3.98) describes a system of particles and antiparticles of equal masses. However, for $\mu_r > m$, the stable vacuum becomes

$$\xi_0^2 = \frac{(\mu_r)^2 - m^2}{2\lambda}, \tag{3.100}$$

and (3.98) describes a collection of two particles with differing masses: $\overline{m}_1^2 = 3(\mu r)^2 - 2m^2$ and $\overline{m}_2^2 = (\mu r)^2$ respectively. Because of the chemical potential, the dispersion relation of the latter is gapless, and Goldstone's theorem is satisfied. At zero temperature, the pressure is simply

$$P_{HRG}(\mu) = U(\xi)\big|_{\xi = \xi_0} = \frac{1}{4\lambda}\left((\mu r)^2 - m^2\right)^2 \theta(\mu r - m). \tag{3.101}$$

This gives us the dependence of the pressure on $\mu$, but we still have not set the coupling constant $\lambda$. We set it as follows. According to Ref. [39] (the relevant generalized equation is reproduced later in this thesis as Eq. (4.14)), we see that the Fermi–Dirac pressure for a quark in the theory $(N, n_f)$ with representation R becomes

$$P_{fd} = \frac{d_R}{12\pi^2}\left(\frac{\mu r}{N_b}\right)^4, \tag{3.102}$$

with $d_R$ being the dimension of the fermionic representation. Thus, for a single degree of freedom (recalling that a fermionic quark has two degrees of freedom) one has

$$P_{fd} = \frac{1}{24\pi^2 N_b^4}(\mu r)^4. \tag{3.103}$$

Thus, in order for $P_{HRG} \to P_{fd}$ when $\mu \to \infty$, we must have for a single scalar baryon

$$P_{HRG}(\mu) = \frac{1}{24\pi^2 N_b^4}\left((\mu r)^2 - m^2\right)^2 \theta(\mu r - m), \tag{3.104}$$

and so for a theory with bosonic baryons we have

$$P_{HRG}(\mu) = \sum_{i \in B} \frac{g_i}{24\pi^2 N_b^4}\left((\mu r_i)^2 - m_i^2\right)^2 \theta(\mu r_i - m_i). \tag{3.105}$$

### 3.2.2  Chiral effective theory

ChEFT is a perturbative effective field theory of low-energy QCD that builds an effective Lagrangian using **chiral symmetry** as a constraint [34, 46]. Chiral symmetry is a symmetry of the QCD Lagrangian (3.62) or (2.1) *with massless quarks* under transformations of the internal flavor indices. In this section, we will detail chiral symmetry and its breaking in real-world QCD (i.e., as an SU(3) gauge theory). Later, in Sec. 4.1.2.2, we will briefly discuss chiral symmetry breaking in QCD-like theories: SU(N) gauge theories in which the fermions are in different representations.

The Euclidean Lagrangian (3.62) in the massless-quark limit can be written as

$$\mathcal{L}_{QCD}^E = \sum_f \overline{\psi}_{fL}^\alpha \slashed{D}_{\alpha\beta} \psi_{fL}^\beta + \overline{\psi}_{fR}^\alpha \slashed{D}_{\alpha\beta} \psi_{fR}^\beta + \frac{1}{4} F_{\mu\nu}^a F^{a\mu\nu}, \tag{3.106}$$

where we have defined the covariant derivative $D_{(\alpha\beta)\mu}$ as

$$D_{(\alpha\beta)\mu} \equiv \delta_{\alpha\beta} \partial_\mu - i g \mathcal{A}_\mu^a T_{\alpha\beta}^a, \tag{3.107}$$

and we have broken up the Dirac spinors $\psi_f$ into the L- and R-chirality parts. In the form of Eq. (3.106), it is clear that the L and R pieces of the quark flavors transform independently in the Lagrangian, that is, the Lagrangian is invariant under the global, chiral flavor-symmetry group

$$U(n_f)_L \otimes U(n_f)_R = SU(n_f)_L \otimes SU(n_f)_R \otimes U(1)_L \otimes U(1)_R. \tag{3.108}$$

These symmetries can be recast into vector and axial forms, where here "vector" means that the symmetry acts the same on the L and R fields and where "axial" means that the symmetry acts oppositely on the L and R fields:

$$SU(n_f)_V \otimes SU(n_f)_A \otimes U(1)_V \otimes U(1)_A, \tag{3.109}$$

Though Eq. (3.109) is the chiral flavor symmetry of the Lagrangian, it is not the full flavor symmetry of the path integral: the path-integral measure for the quark fields does not respect the $U(1)_A$ symmetry. (One says that the $U(1)_A$ symmetry is *anomalous*.) Thus, the full chiral flavor symmetry of QCD is

$$SU(n_f)_V \otimes SU(n_f)_A \otimes U(1)_V. \tag{3.110}$$

For completeness, we point out here that the $SU(n_f)_V$ symmetry is the symmetry of isospin (for $n_f = 2$) and the $U(1)_V$ is frequently written $U(1)_B$ (e.g., as was done in the discussion of chiral symmetry breaking in the CFL phase in Sec. 2.2.2 above), as it is the symmetry leading to baryon number conservation.

Before discussing which of the chiral flavor symmetries of massless QCD become hidden in the ground state, let us discuss the addition of quark masses. The addition of a mass term

$$\sum_f m_f^2 \overline{\psi}_f \psi_f \equiv \sum_f m_f^2 (\overline{\psi}_{fL} \psi_{fR} + \overline{\psi}_{fR} \psi_{fL}) \tag{3.111}$$

breaks the $SU(n_f)_A$ flavor symmetry explicitly, and if the masses of the quarks are different, then even the $SU(n_f)_V$ symmetry is explicitly broken. However, the symmetries can still be approximately valid if the quark masses are not "too large". The relevant energy scale with which to compare the quark masses is $\Lambda_{QCD}$. Thus, for real QCD, a massless Lagrangian is a good approximation for $n_f = 2$, and a questionable, but sometimes still passable description for $n_f = 3$. For the moment, let us continue to keep $n_f$ general, but with the understanding that $n_f = 2$ or $3$ depending on the desired accuracy. After discussing spontaneous symmetry breaking in the massless (often called "chiral") limit, we will return to the quark masses.

The low-energy (i.e., hadronic) ground state of massless QCD respects only a subgroup of the full symmetry group (3.110). The $SU(n_f)_A$ symmetry is hidden or spontaneously broken. For each generator of the hidden global symmetry, one finds a massless Nambu–Goldstone boson. Since the hidden $SU(n_f)_A$ symmetry has $n_f^2 - 1$ generators, one will have $n_f^2 - 1$ massless particles in low-energy massless QCD. The axial chiral flavor symmetry is broken by the formation of a **chiral condensate** of each quark flavor

$$\langle \overline{\psi}_f \psi_f \rangle \equiv \langle \overline{\psi}_{f\,L} \psi_{f\,R} + \overline{\psi}_{f\,R} \psi_{f\,L} \rangle = v^3 \sim \Lambda_{QCD}^3 \qquad \text{(no sum on f)}. \tag{3.112}$$

Note that this couples L quarks to R antiquarks and vice versa, which is different than the condensate in the CFL phase, discussed in Sec. 2.2.2 above. The appearance of $\Lambda_{QCD}$ here also explains why $\Lambda_{QCD}$ is the relevant energy scale with which to compare the quark masses.

A perturbative **effective theory** for the low-energy degrees of freedom can be systematically constructed from knowledge of the full chiral flavor symmetry group (3.110) and the hidden symmetry group $SU(n_f)_A$: to do this, one constructs the most general Lagrangian consistent with the chiral symmetry breaking pattern. One begins by defining a unitary operator that acts of the flavor space

$$U_{fg}(x) = \exp\left( 2\,i\,\frac{\pi^a(x)\tau^a_{fg}}{F_\pi} \right), \tag{3.113}$$

where the $\pi^a$ are the dynamical fields (the three pions in the case of $n_f = 2$; and the three pions, four kaons, and one eta in the case of $n_f = 3$), the $\tau^a_{fg}$ are the generators of $SU(n_f)_V$, and $F_\pi$ is a constant which can later be determined by experiment. Note that this means that the $\pi^a$ transform in the adjoint representation of the preserved $SU(n_f)_V$ symmetry. Here, $U$ transforms as

$$U \to g_L U g_R^\dagger, \tag{3.114}$$

with $g_L \in SU(n_f)_L$ and $g_R \in SU(n_f)_R$. From this object $U$ one can construct the most general Lagrangian that exhibits the chiral flavor symmetry breaking pattern of QCD [34, 46]

$$\begin{aligned}
\mathcal{L}_\chi = &\frac{F_\pi^2}{4} \text{Tr} \left[ (\partial_\mu U)(\partial_\mu U)^\dagger \right] + L_1 \text{Tr} \left[ (\partial_\mu U)(\partial_\mu U)^\dagger \right]^2 \\
&+ L_2 \text{Tr} \left[ (\partial_\mu U)(\partial_\nu U)^\dagger \right] \text{Tr} \left[ (\partial_\nu U)^\dagger (\partial_\mu U) \right] \\
&+ L_3 \text{Tr} \left[ (\partial_\mu U)(\partial_\mu U)^\dagger (\partial_\nu U)(\partial_\nu U)^\dagger \right] + \cdots .
\end{aligned} \tag{3.115}$$

Here, $L_1$, $L_2$, and $L_3$ are additional constants that must be fit to experiment. In this expression, the first term (with only two derivatives) is of lower order than the next three terms (with four derivatives): these derivatives turn into powers of $P^\mu/F_\pi \propto P^\mu/\Lambda_{CSSB}$ in amplitudes and thus provide a natural way to order the terms by relevance. This natural ordering procedure is what allows ChEFT to be predictive and effective.

As an aside, we note that by Goldstone's theorem, the light Nambu–Goldstone bosons $\pi^a$ should be labelled by the *broken* generators of $SU(n_f)_V \otimes SU(n_f)_A$, not the generators of the *preserved* symmetry as occurs in Eq. (3.113). We refer the interested reader to Weinberg [47] for the discussion and resolution of this apparent discrepancy. The essence of the discussion there is as follows. For a general group $G$ breaking down to a subgroup $H \subset G$, it is indeed the case that the Nambu–Goldstone bosons are labelled by the cosets of $G/H$. However, for chiral symmetries, one can simplify the derivation of the effective Lagrangian to the procedure described above in which the Nambu–Goldstone bosons are instead labelled by the unbroken symmetry. Ref. [47] provides all the details of both prescriptions and clearly illustrates the connection between them.

Finally, let us discuss the addition of the quark mass terms back into the QCD Lagrangian. As mentioned above, these terms explicitly break the chiral flavor symmetry of massless QCD. Nevertheless, these terms can be incorporated into the effective Lagrangian (3.115) by viewing the

mass term

$$\mathcal{L}_{\text{mass}} = \sum_{f,g} \overline{\psi}_f M_{fg} \psi_g, \quad M = \text{diag}(m_1, m_2, \dots, m_{n_f}) \tag{3.116}$$

as defining another (dynamical-field) operator $M$ that acts on flavor space. If one assumes that $M$ transforms in the same way as $U$ above under $SU(n_f)_L \otimes SU(n_f)_R$, namely

$$M \to g_L M g_R^\dagger, \tag{3.117}$$

then one can write additional terms with the $M$ matrix that respect the $SU(n_f)_L \otimes SU(n_f)_R$ symmetry [34, 46, 47]:

$$\mathcal{L}_{\chi,\text{mass}} = \frac{v^3}{2} \text{Tr} \left( MU + M^\dagger U^\dagger \right) + \cdots, \tag{3.118}$$

with the $\cdots$ containing higher-order terms in both $M$ and $U$ (see Refs. [47, 46]), and $v$ is the same constant that appears in the chiral condensate (3.112).

The Lagrangians (3.115) and (3.118) describe the anomalously light mesons in low-energy QCD, but do not include baryons. One may incorporate nucleons into ChEFT by including additional nucleon fields and describing the coupling between nucleons and mesons. Defining $\xi = \sqrt{U}$, and letting $\Psi$ denote the nucleon field, then the meson-nucleon Lagrangian is [4]

$$\mathcal{L}_{\pi N} = \overline{\Psi} \Big[ i \gamma^\mu (\partial_\mu + \Gamma_\mu) - m_N + \frac{g_A}{2} \gamma^\mu \gamma_5 u_\mu \Big] \Psi + \cdots, \tag{3.119}$$

where the $\cdots$ again represents higher-order terms,

$$\Gamma_\mu = \frac{1}{2} \big[ \xi^\dagger, \partial_\mu \xi \big], \tag{3.120}$$

and

$$u_\mu = i \big\{ \xi^\dagger, \partial_\mu \xi \big\}. \tag{3.121}$$

Once again, this Lagrangian can be systematically improved by powers of $P^\mu / \Lambda_{\text{CSSB}}$, and is thus a useful tool for controlled, low-energy calculations.

We will end our sketch of ChEFT here. Since ChEFT is also a quantum field theory, the machinery of Sec. 3.1.5 and Sec. 3.1.6 allows one to calculate the EoS for low-energy QCD using the above effective Lagrangians. The reader should now at least appreciate that this effective theory indeed provides a controlled, perturbative description of low-energy QCD. If the reader is interested in further details, we direct them to the books of Donoghue *et al.* [46] or Weinberg [47], or the detailed ChEFT reviews in Refs. [3, 4].

We now turn to the problem of finding the EoS in the intermediate regime between the HRG or ChEFT descriptions and pQCD using thermodynamic matching.

# CHAPTER 4

# MATCHING EQUATIONS OF STATE

Since the region of the QCD phase diagram applicable to NS interiors is out of reach of the perturbative approaches discussed in the previous chapter, a different approach must be used. The approach advocated in this thesis is one of **matching EoSs**: Since the $T = 0$ axis has controlled, first-principled theories that are effective for $\mu \ll \Lambda_{CSSB}$ and for $\Lambda_{pQCD} \ll \mu$, one straightforward approach is to simply try to fit together the EoSs of these two theories to shed light on the non-perturbative, middle region. This approach was taken in Ref. [10] using the simpler HRG at low $\mu$ in place of ChEFT, and in Refs. [1, 11] using the more comprehensive ChEFT at low values of $\mu$.

This sophistication of this matching procedure is variable. At the simplest level, one may match the low- and high-energy EoSs simply by using each EoS until the point where they intersect. In this method, it is often preferable to include a bag constant in the pQCD pressure as an extra degree of freedom when matching. In addition, one may choose to specify the order of the phase transition, or the latent heat of the phase transition if first order. In a more sophisticated matching procedure, one may cut the low- and high-energy EoSs before they intersect (perhaps at the point where they each have some fixed uncertainty), and attempt to extrapolate between them with a simple EoS (e.g., a polytropic equation of state; see Sec. 4.2 below). This method anticipates the breakdown of each EoS in the central, non-perturbative region and allows for a wider range of behavior in that region. In this chapter, we will highlight both of these approaches.

In addition to highlighting both of these approaches, the current chapter serves two additional purposes. The first additional purpose of this chapter is to critically examine this "EoS-matching" paradigm. In Sec. 4.1, we take up this task by examining both how accurately the simple Hadron Resonance Gas plus pQCD matching (HRG+pQCD) reproduces the lattice $\mu = 0$ result, and by providing *predictions* of the $T = 0$ EoS in QCD-like theories with different numbers of colors and/or different representations of quarks. Crucially, the theories for which we calculate the $T = 0$ EoS do not suffer from a sign problem in lattice QCD, and thus these theories can be simulated for any value of $\mu$. It is one of the goals of this chapter to provide the lattice with definitive predictions that will be able to validate or rule out the simple HRG+pQCD matching prescription. The work from this first section has been published previously in Ref. [48]. The second additional purpose of this chapter, which we address in Sec. 4.2, is to describe the more sophisticated matching procedure of Kurkela *et al.* [1] in some detail and to highlight the main ideas. This section will serve as a basis of our applications in Chap. 5, in which we use the matched EoS of Kurkela *et al.* [1] to constrain properties of rotating NSs.

## 4.1    Matching in QCD-like theories accessible to lattice QCD

Knowledge about condensed-matter properties of QCD in thermodynamic equilibrium is required for the interpretation of experimental and observational data in cosmology, high-energy nuclear physics, and the physics of neutron stars. While tremendous progress has been made for the case of high temperature and small baryon densities using direct simulations in lattice QCD [40, 41, 42], much less is known for the case of small temperature and large densities. The reason for this shortcoming is that the so-called sign problem prohibits the direct simulation of QCD at large density using established importance sampling techniques. While established techniques fail, several recent techniques have been studied that at least in principle could permit one to cal-culate thermodynamic properties from first principles in QCD at large density. These techniques include Lefschetz thimbles [20], complex Langevin [21, 22, 23, 24], strong coupling expansion [25], and hadron resonance gas plus perturbative QCD ("HRG+pQCD" in the following) [49, 50, 10]. In this section, we propose a series of 'control studies' in QCD-like theories (in particular two-color QCD with two fundamental flavors and four-color QCD with two flavors in the two-index, antisymmetric representation), which—despite not corresponding to the actual theory of strong interactions realized in nature—have the advantage of not suffering from a sign problem, and are thus amenable to direct simulations using established lattice-QCD techniques. We then proceed to calculate thermodynamic properties in these QCD-like theories in one of the above non-traditional approaches (HRG+pQCD, Refs. [49, 50, 10]), which effectively makes predictions for possible future lattice-QCD studies that can be used to validate or falsify this HRG+pQCD approach. Since two- and four-color QCD are qualitatively similar to three-color QCD, we furthermore expect the level of agreement between lattice QCD and HRG+pQCD in the two- or four-color cases to be roughly comparable to the three-color QCD case, thus offering an indirect validation of non-traditional methods for QCD at large densities.

We note here that, as has been true throughout this thesis, we are only interested in bulk, thermodynamic properties in the following. Moreover, the HRG+pQCD approach followed in this section will be unable to describe the details of the phase-transition region, in particular, its order. This is not our goal. This section will serve as an illustration of the simpler matching approach described above.

This topic is organized as follows. In Sec. 4.1.1 we give the EoS in pQCD by stating the pressure P as a function of temperature T at baryon chemical potential $\mu = 0$ and P as a function of $\mu$ at T = 0. This section is essentially a compilation of what has been derived in the literature. Sec. 4.1.2 contains an explanation of how the hadrons in the theories listed above are computed. (For a general overview of the HRG EoS, see Sec. 3.2.1 in the previous chapter.) Sec. 4.1.3 contains a description of how we perform the matching between these two asymptotic EoSs, and in Sec. 4.1.4 we discuss the results of this matching.

### 4.1.1 pQCD equation of state

We are interested in calculating the pressure P along the T- and μ-axes in a general SU(N) gauge theory with $n_f$ massless fermions. In particular, we are interested in the theories $(N, n_f) = (2, 2), (3, 3)$, and $(4, 2)$ with quarks in the fundamental representation (fundamental) and $(4, 2)$ with quarks in the two-index, antisymmetric representation (antisymmetric). In order to constrain the pressure of these theories, we derive the asymptotic behavior for both low and high T or μ and then match these behaviors using basic thermodynamics. At high T or high μ, the EoS can be calculated using (resummed) pQCD and at low T or μ, the EoS of the theory is to good approximation [40, 41, 42, 43] that of a HRG (a non-interacting collection of the hadrons of that theory). In the intermediate regime, the EoSs can be constructed by matching the high-/low-energy asymptotic behavior using the criterion that the pressure P must increase as a function of T or as a function of μ (see Ref. [10]). More details of the matching procedure will be discussed below.

The high-T, pQCD EoS can be calculated by following the equations and procedure of Kajantie *et al.* [51, 52] and Vuorinen [39] with the resummation modifications described by Blaizot *et al.* [53] (cf. Ref. [54] for a different approach to the resummed pQCD EoS). We first define the following group-theory terms to be used in all future pQCD expressions:

$$C_A = N, \tag{4.1}$$

$$d_A = N^2 - 1, \tag{4.2}$$

and

$$C_{\text{fundamental}} = \frac{N^2 - 1}{2N}, \qquad C_{\text{antisymmetric}} = \frac{(N-2)(N+1)}{N}, \tag{4.3}$$

$$T_{\text{fundamental}} = \frac{n_f}{2}, \qquad T_{\text{antisymmetric}} = \frac{(N-2)}{2} n_f, \tag{4.4}$$

$$d_{\text{fundamental}} = N n_f, \qquad d_{\text{antisymmetric}} = \frac{N(N-1)}{2} n_f. \tag{4.5}$$

(Note that we are using the vector flavor notation $\vec{\psi}^\alpha$ here.) In all of the expressions that follow, we let group-theory terms with a subscript R denote the fermionic group-theory invari-ants, which must be replaced by the corresponding fundamental or antisymmetric representation group-theory invariants above as needed. In terms of these group-theory terms, the pQCD pressure at $\mu = 0$ in these theories can be written

$$P_{pQCD}(T) = P_{sb}(T) + P_{hard}(T) + P_{EQCD}(T). \qquad (4.6)$$

Here, the $P_{sb}$ the Stefan-Boltzmann pressure given by

$$P_{sb}(T) = \frac{\pi^2 T^4}{45} \left( d_A + \frac{7}{4} d_R \right). \qquad (4.7)$$

To 3-loop order, $P_{hard}$ is given by Braaten and Nieto [55] as

$$
\begin{aligned}
P_{hard}(T) = \frac{\pi^2 d_A}{9} T^4 \Bigg\{ & - \left( C_A + \frac{5}{2} T_R \right) \frac{\alpha_s}{4\pi} \\
& + \Bigg( C_A^2 \left[ 48 \ln \frac{\Lambda_E}{4\pi T} - \frac{22}{3} \ln \frac{\overline{\Lambda}}{4\pi T} + \frac{116}{5} + 4\gamma + \frac{148}{3} \frac{\zeta'(-1)}{\zeta(-1)} - \frac{38}{3} \frac{\zeta'(-3)}{\zeta(-3)} \right] \\
& + C_A T_R \left[ 48 \ln \frac{\Lambda_E}{4\pi T} - \frac{47}{3} \ln \frac{\overline{\Lambda}}{4\pi T} + \frac{401}{60} - \frac{37}{5} \ln 2 + 8\gamma + \frac{74}{3} \frac{\zeta'(-1)}{\zeta(-1)} - \frac{1}{3} \frac{\zeta'(-3)}{\zeta(-3)} \right] \\
& + T_R^2 \left[ \frac{20}{3} \ln \frac{\overline{\Lambda}}{4\pi T} + \frac{1}{3} - \frac{88}{5} \ln 2 + 4\gamma + \frac{16}{3} \frac{\zeta'(-1)}{\zeta(-1)} - \frac{8}{3} \frac{\zeta'(-3)}{\zeta(-3)} \right] \\
& + C_R T_R \left[ \frac{105}{4} - 24 \ln 2 \right] \Bigg) \left( \frac{\alpha_s}{4\pi} \right)^2 \Bigg\},
\end{aligned}
\qquad (4.8)
$$

where $\Lambda_E$ is the factorization scale between the hard and soft modes, and $\alpha_s$ is the strong coupling constant squared over $4\pi$ in the $\overline{MS}$ renormalization scheme at the scale $\overline{\Lambda} = \sqrt{(2\pi T)^2 + (\mu)^2}$. This is given by [56, 10]

$$\alpha_s(\overline{\Lambda}) = \frac{4\pi}{\beta_0 L} \left( 1 - \frac{\beta_1}{\beta_0^2} \frac{\ln L}{L} \right), \qquad L = \ln \left( \overline{\Lambda}^2 / \Lambda_{\overline{MS}}^2 \right), \qquad (4.9)$$

with

$$\beta_0 = \frac{11}{3} C_A - \frac{4}{3} T_R, \quad \beta_1 = \frac{34}{3} C_A^2 - 4 C_R T_R - \frac{20}{3} C_A T_R, \qquad (4.10)$$

where $\Lambda_{\overline{MS}}$ is the $\overline{MS}$ renormalization point (to be set later). In all the results, we set $\Lambda_E = \overline{\Lambda}$ and vary $\overline{\Lambda}$ about the aforementioned value by a factor of two (cf. the end of Sec. 4.1.3). Finally, $P_{EQCD}$ is given by

$$P_{EQCD}(T) = \frac{d_A}{4\pi} T \left( \frac{1}{3} m_E^3 - \frac{C_A}{4\pi} \left( \ln \frac{\Lambda_E}{2 m_E} + \frac{3}{4} \right) g_E^2 m_E^2 - \left( \frac{C_A}{4\pi} \right)^2 \left( \frac{89}{24} - \frac{11}{6} \ln 2 + \frac{1}{6} \pi^2 \right) g_E^4 m_E \right),$$

$$(4.11)$$

where

$$
\begin{aligned}
m_E^2 = \frac{4\pi}{3}\alpha_s T^2 \bigg\{ & C_A + T_R \\
& + \Big[ C_A^2 \left( \frac{5}{3} + \frac{22}{3}\gamma + \frac{22}{3}\ln\frac{\overline{\Lambda}}{4\pi T} \right) + C_A T_R \left( 3 - \frac{16}{3}\ln 2 + \frac{14}{3}\gamma + \frac{14}{3}\ln\frac{\overline{\Lambda}}{4\pi T} \right) \\
& + T_R^2 \left( \frac{4}{3} - \frac{16}{3}\ln 2 - \frac{8}{3}\gamma - \frac{8}{3}\ln\frac{\overline{\Lambda}}{4\pi T} \right) - 6 C_R T_R \Big] \left( \frac{\alpha_s}{4\pi} \right) \bigg\},
\end{aligned}
\tag{4.12}
$$

and

$$
g_E^2 = 4\pi\alpha_s T. \tag{4.13}
$$

The zero-temperature pQCD EoS is more straightforward in the sense that resummation of the strict perturbative series is not required. The result is given in Ref. [39] by

$$
\begin{aligned}
P_{\text{pQCD}}(\mu) = \frac{1}{4\pi^2} \bigg( & \sum_f \mu_f^4 \bigg\{ \frac{d_R}{3 n_f} - d_A \left( \frac{2 T_R}{n_f} \right)\left( \frac{\alpha_s}{4\pi} \right) - d_A \left( \frac{2 T_R}{n_f} \right)\left( \frac{\alpha_s}{4\pi} \right)^2 \Big[ \frac{2}{3}(11 C_A - 4 T_R)\ln\frac{\overline{\Lambda}}{\mu_f} \\
& + \frac{16}{3}\ln 2 + \frac{17}{4}\left( \frac{C_A}{2} - C_R \right) + \frac{1}{36}(415 - 264\ln 2)C_A - \frac{8}{3}\left( \frac{11}{6} - \ln 2 \right) T_R \Big] \bigg\} \\
& - d_A \left( \frac{2 T_R}{n_f} \right)\left( \frac{\alpha_s}{4\pi} \right)^2 \left\{ \left( 2\ln\frac{\alpha_s}{4\pi} - \frac{22}{3} + \frac{16}{3}\ln 2\,(1 - \ln 2) + \delta + \frac{2\pi^2}{3} \right)(\mu^2)^2 + F(\mu) \right\} \bigg) \\
& + \mathcal{O}(\alpha_s^3 \ln\alpha_s),
\end{aligned}
\tag{4.14}
$$

where the sum is over all the quark flavors in the theory, $\mu_f$ is the f-quark chemical potential, $\mu^2 = \sum_f \mu_f^2$, and

$$
\begin{aligned}
F(\mu) = -2\mu^2 \left( \frac{2 T_R}{n_f} \right) \sum_f \mu_f^2 \ln\frac{\mu_f^2}{\mu^2} + \frac{2}{3}\left( \frac{2 T_R}{n_f} \right)^2 \sum_{f>g} \bigg\{ & (\mu_f - \mu_g)^2 \ln\frac{|\mu_f^2 - \mu_g^2|}{\mu_f \mu_g} \\
& + 4\mu_f \mu_g(\mu_f^2 + \mu_g^2)\ln\frac{(\mu_f + \mu_g)^2}{\mu_f \mu_g} - (\mu_f^4 - \mu_g^4)\ln\frac{\mu_f}{\mu_g} \bigg\},
\end{aligned}
\tag{4.15}
$$

with the constant $\delta$ having the value $\delta = -0.85638320933$. In what follows we always set all of the quark chemical potentials equal to each other, so that $\mu_f = \mu/N_b$ for each flavor f, where $N_b$ is the number of quarks in a baryon. Note that this means that some of the terms in (4.15) do not contribute.

Let us pause here to mention that we are not including a color superconductivity (CSC)

phase in the EoS at $T = 0$. Including a CSC phase amounts to adding a term of the form

$$P_{CSC} = \frac{\Delta^2 \mu^2}{3\pi^2} \qquad (4.16)$$

to P [14, 57, 58]. Here, $\Delta$ is the superconducting energy gap. In the three-color case, this contribution to the pressure adds a correction of at most ten percent.

### 4.1.2    Hadron resonance gas spectra in the QCD-like theories

The EoS of a HRG was discussed previously in Sec. 3.2.1; see that section for the details of the construction. Below, we shall discuss the hadron spectrum in each of the exotic QCD-like theories that we listed above: $(N, n_f) = (2, 2), (3, 3)$, and $(4, 2)$ with fundamental quarks and $(N, n_f) = (4, 2)$ with antisymmetric quarks.

#### 4.1.2.1    Determining the hadron spectrum

For the three-color $(N, n_f) = (3, 3)$ fundamental case, we use the real world spectrum of hadrons up to 2.25 GeV [59]. For the two- and four-color theories with two fundamental quarks and the four-color theory with two antisymmetric quarks, we determine the hadrons using group-theoretic arguments and Fermi statistics (in the case of objects composed of quarks only). We explicitly ignore the glueballs in these theories because they tend to be more massive than the lightest hadrons [60]. For the two- and four-color theories, we set the scale using the string tension $\sqrt{\sigma}$ and the relation between the string tension and the $\overline{MS}$ renormalization scale $\Lambda_{\overline{MS}}$ given in Ref. [61]. However, the ratios $\Lambda_{\overline{MS}}/\sqrt{\sigma}$ given in the aforementioned reference are for the pure-gauge theories. To remedy this, we scale these ratios by $\Lambda_{\overline{MS}}^{N=3}(n_f=2)/\Lambda_{\overline{MS}}^{N=3}(n_f=0)$, determined from Ref. [62]. These lead to the values

$$\Lambda_{\overline{MS}}^{N=2}(n_f=2)/\sqrt{\sigma} = 1.032 \quad \text{and} \quad \Lambda_{\overline{MS}}^{N=4}(n_f=2)/\sqrt{\sigma} = 0.723 \qquad (4.17)$$

for the fundamental theories. For the three-color theory, we use $\Lambda_{\overline{MS}} = 0.378$ GeV, as in [10].

For the four-color antisymmetric theory, we were unable to locate a result for $\Lambda_{\overline{MS}}^{N=4}(n_f = 2)/\sqrt{\sigma}$ from the lattice in the literature. Since some of the group-theory terms for the antisym-metric theory scale more strongly with the number of colors than the corresponding terms in the fundamental theory, it seems reasonable to expect that $\Lambda_{\overline{\text{MS}}}$ will scale differently with the number of quark flavors in the antisymmetric theory than in the fundamental theory. Moreover, it would be most accurate to view $\Lambda_{\overline{\text{MS}}}/\sqrt{\sigma}$ as a free parameter in the HRG+pQCD scheme that must be determined independently from the lattice. In light of these considerations, we have decided to use both the pure-glue value [61]

$$\Lambda_{\overline{\text{MS}}}^{\text{N}=4}/\sqrt{\sigma} = 0.527, \tag{4.18}$$

and the previously-given value of $\Lambda_{\overline{\text{MS}}}^{\text{N}=4}(n_f = 2)/\sqrt{\sigma}$ that we use for the four-color fundamental theory for the four-color antisymmetric theory, with the expectation that the true value will lie somewhere near this range.

For both the two- and four-color cases, the mesons are taken to be the analogues of the flavorless mesons that exist in the real world (up to a mass of about 2 GeV) whose masses are written in multiples of the string tension $\sigma_{\text{SU}(3)} = (420\,\text{MeV})^2$. In the two-color case, we mainly use the analogues of the real-world mesons, substituting the two-color masses calculated by Bali *et al.* [63] when available. (We also note here that the $\mu$-dependence of the two-color spectrum has been studied numerically in Ref. [64] and analytically in Ref. [45], though we do not need this $\mu$-dependence for our HRG+pQCD scheme.)

We now discuss in some detail how the non-meson objects in these three cases are determined. For convenience and as a summary of these sections, we list tables for all of the particles that we have included in the SU(2) and SU(4) cases in Appendix A.

**Two-color case**

In two-color QCD, the baryons are composed of two quarks with the added simplicity that the masses are degenerate with the corresponding mesons made from the same quarks [65]. Thus, the mass spectrum of the baryons is the same as the mass spectrum of mesons. However, there are fewer baryons than mesons, for there is an additional constraint imposed by Fermi statistics in the case of the baryons. Since we may view the two massless quarks as part of an isospin doublet, one sees that exchanging the two internal quarks in a baryon causes the wave function to become multiplied by

$$(-1)^{1+L+S+I}. \tag{4.19}$$

In this equation, $L$ is the angular momentum quantum number, $S$ is the spin, and $I$ is the isospin, with the additional 1 due to the fact that the quarks are in an antisymmetric color singlet. We thus see that for even $L$ the spin and isospin must be equal ($S = 0$ implies $I = 0$ and $S = 1$ implies $I = 1$), and for odd $L$ they must be the opposite in order to have a totally antisymmetric wave function. (Even though the composite baryon is itself a boson in two-color QCD, it is still a multi-particle state of fundamental fermions.) This information is enough to determine the set of hadrons in the HRG pressure (3.93).

### Four-color fundamental case

Baryons in four-color QCD with fundamental fermions consist of four quarks. In this case, to determine the masses $M$ we use the large-N expansion

$$M(J) = NA + \frac{J(J+1)}{N}B, \tag{4.20}$$

where $J$ is the total angular momentum of the baryon, and $A$, $B$ are constants independent of N [66, 67]. As pointed out by DeGrand [68] and demonstrated by Appelquist *et al.* [69], a term independent of N could be used for better agreement. However, we have no way to set the value of such a term and thus do not include it.

We find the possible values of $J$ beginning with the ground-state baryons of zero orbital angular momentum. Since we still have a isospin doublet of massless, spin-one-half quarks, we only need the group-theory expression

$$2 \otimes 2 \otimes 2 \otimes 2 = 5_S \oplus 3_M \oplus 3_M \oplus 3_M \oplus 1_A \oplus 1_A, \tag{4.21}$$

where the $5_S$ state is fully symmetric, the $3_M$ states are are symmetric in three of the four quarks and antisymmetric in the other, and the $1_A$ states are pairwise antisymmetric. Since, again, the quarks are in an antisymmetric color singlet, it must be the case that they are in a *symmetric* com-bination of spin and flavor. This means that there is a spin-2 quintet, a spin-1 triplet, and a spin-0 singlet of ground-state baryons.

We may also determine the first excited states in this simple manner by realizing that for this four-body problem there are three relevant orbital-angular-momentum quantum numbers and the first excited state corresponds to when exactly one of them is one. In order to still be in a completely antisymmetric state, either the spin or the flavor state must now be in one of the $3_M$ states while the other must be in a $5_S$ state. This means that there are a quintet of particles with $S = 1$ and a triplet of particles with $S = 2$. Combining these with an orbital angular momentum $L = 1$ yields three baryonic quintets with $J = 0, 1, 2$ and three baryonic triplets with $J = 1, 2, 3$. We did not determine the baryons for any higher excited states.

**Four-color antisymmetric case**

The hadron spectrum in the four-color theory with two antisymmetric quarks consists of two-quark objects: mesons and diquarks; four-quark objects: tetraquarks, di-mesons, and diquark-mesons; and six-quark baryons [70, 71, 72]. Since the antisymmetric representation is real, the arguments of Ref. [65] carry through here and one may conclude that all two-quark objects with the same quark content have degenerate masses and that the same holds for the four-quark objects. In addition, the four-quark objects have a mass equal to the sum of their constituent two-quark-object masses [70]. Because of this mass degeneracy, we need not determine how all of the four-quark-object degrees of freedom break up into spin and isospin multiplets; rather, we may simply combine the two-quark-object degrees of freedom in every possible way. One major difference from the two-color case, however, is that in the four-color theory with antisymmetric quarks the color-singlet state for diquarks is *symmetric*. This means that the spin-isospin locking in this theory is the opposite of the locking in the two-color theory. That is, for odd L the spin and isospin must be equal ($S = 0$ implies $I = 0$ and $S = 1$ implies $I = 1$), and for even L they must be opposite. As for the six-quark baryons, we again use the large-N expression (4.20), but with N replaced by $N_b = 6$, the number of quarks in the baryon. We include only the ground-state baryons, where isospin and spin are locked as $I = J = 3, 2, 1$, and $0$ [70].

### 4.1.2.2 Chiral symmetry breaking and the Nambu–Goldstone bosons

The lowest-mass particles in all of the aforementioned theories are precisely zero at zero quark mass. This can be understood in terms of the pattern of chiral symmetry breaking in these theories [73]. Consider an SU(N) gauge theory with $n_f$ massless fermions. For fermions in a complex representation (such as in the cases $N \geqslant 3$ with fundamental fermions—discussed above in Sec. 3.2.2), the Lagrangian density possesses the symmetry $U(n_f)_L \otimes U(n_f)_R$, corresponding to the separate left- and right-handed flavor symmetries; and for real representations (such as any N with adjoint fermions or $N = 4$ with antisymmetric fermions) or pseudoreal representations (such as in $N = 2$ with fundamental fermions), the Lagrangian density possesses the larger symmetry $U(2n_f)$. In all of these cases, the axial $U(1)_A$ symmetry is broken by an anomaly, and the remaining symmetries are spontaneously broken in the following ways. For fermions in a complex representation:

$$SU(n_f)_L \otimes SU(n_f)_R \to SU(n_f)_V; \qquad (4.22)$$

for fermions in a real representation:

$$SU(2n_f) \to O(2n_f); \qquad (4.23)$$

and for fermions in a pseudoreal representation:

$$SU(2n_f) \to Sp(2n_f). \qquad (4.24)$$

(See Refs. [73, 45] for more details.) The generators of the broken symmetries become massless Nambu–Goldstone bosons. Since $SU(n_f)$ has $n_f^2 - 1$ generators, $O(n_f)$ has $n_f(n_f - 1)/2$ generators, and $Sp(n_f)$ has $n_f(n_f + 1)/2$ generators, we see that in the three-color, three fundamental-quark case there will be 8 Nambu–Goldstone bosons (a meson octet); in the four-color, two fundamental-quark case there will be 3 Nambu–Goldstone bosons (a meson triplet); in the four-color, two antisymmetric-quark case there will be 9 Nambu–Goldstone bosons (a triplet each of mesons, diquarks, and antidiquarks); and in the two-color, two fundamental-quark case there will be 5 Nambu–Goldstone bosons (a triplet of mesons, a diquark, and an antidiquark).

In addition, recall that if the quarks in these theories are not precisely massless, then the massless Nambu–Goldstone bosons will become instead small-mass, pseudo-Nambu–Goldstone bosons. In the spectra, we are free to vary the mass of these lightest particles to see what effects this will have on the EoS of the theories. This is especially interesting for lattice practitioners. We discuss this further in Sec. 4.1.4.

### 4.1.3  Matching the pQCD and HRG equations of state

To match the two asymptotic EoSs, we employ the same technique on the T-axis as on the µ-axis. As such, let us introduce the symbol F to stand for either T or µ so that we may discuss the matching in full generality.

To perform the matching, we take the simpler approach discussed at the beginning of this chapter: we use each EoS until they intersect, and we assume that at the phase-transition point the pressures of the two phases are equal. We use the thermodynamic constraints that the pressure of a system must increase with F

$$P(F + \Delta F) \geqslant P(F), \tag{4.25}$$

and that above a phase-transition point, the physical phase is the one with the higher pressure. We also add a bag constant B to the pQCD pressure so that

$$P_{\mathrm{pQCD}}(F) = P^0_{\mathrm{pQCD}}(F) + B, \tag{4.26}$$

where $P^0_{\mathrm{pQCD}}$ is given by either (4.6)-(4.11) or (4.14). In the plots that follow, we solve the following set of two equations with two unknowns (for a given $\overline{\Lambda}$):

$$P_{\mathrm{HRG}}(F_0) = P_{\mathrm{pQCD}}(F_0, B_0), \tag{4.27}$$

$$\left.\frac{\mathrm{d}P_{\mathrm{HRG}}(F)}{\mathrm{d}F}\right|_{F=F_0} = \left.\frac{\partial P_{\mathrm{pQCD}}(F, B_0)}{\partial F}\right|_{F=F_0}. \tag{4.28}$$

The second of these equations amounts to assuming that the phase transition is of second order. By varying $\overline{\Lambda}$ between $\pi T$ and $4\pi T$ for the case $F = T$ and between $\mu/2$ and $2\mu$ in the case $F = \mu$, (4.27)-(4.28) allows us to obtain a region of possible EoSs in the $(F, P)$ plane for each theory.

### 4.1.4    Results: HRG+pQCD matching

In Fig. 4.1, we overlay the bands for the pressure and trace anomaly $\epsilon - 3P$ (with $\epsilon$ the energy density) at $\mu = 0$ that we calculate in the three-color, three-massless-quark case with lattice data from the Budapest–Marseille–Wuppertal Collaboration [41] and the HotQCD Collaboration [42] in their respective regions of validity. We observe that the lattice data agree reasonably well with the band resulting from the HRG+pQCD calculation, both for the pressure as well as for the trace anomaly.

In Fig. 4.2, we show the HRG+pQCD pressure and trace-anomaly bands at $\mu = 0$ for all four theories with the T-axis scaled by the critical temperature $T_c$, which we define to be the average of the matching temperatures for the upper and lower edge of the pQCD band to the HRG EoS. $T_c$ should thus be regarded as an estimate of the confinement-deconfinement critical temperature. We list the explicit values obtained in HRG+pQCD in Table 4.1. We see that once the temperature axis has been scaled by $T_c$, all the theories show similar behavior both for the pressure and trace

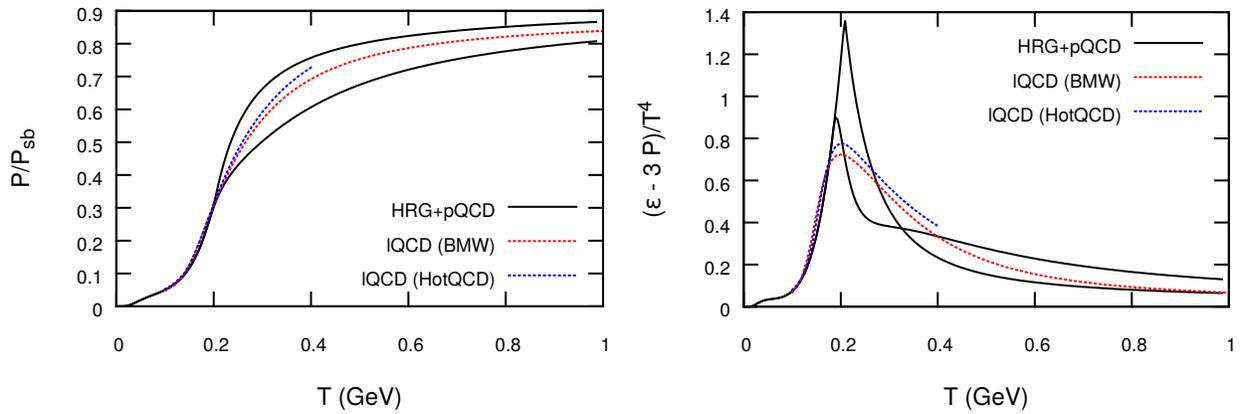

FIGURE 4.1: Normalized pressure (left) and trace anomaly (right) at $\mu = 0$ for the three-color, three-massless-quark case from HRG+pQCD in comparison to lattice-QCD data from the Budapest–Marseille–Wuppertal Collaboration [41] and the HotQCD Collaboration [42].

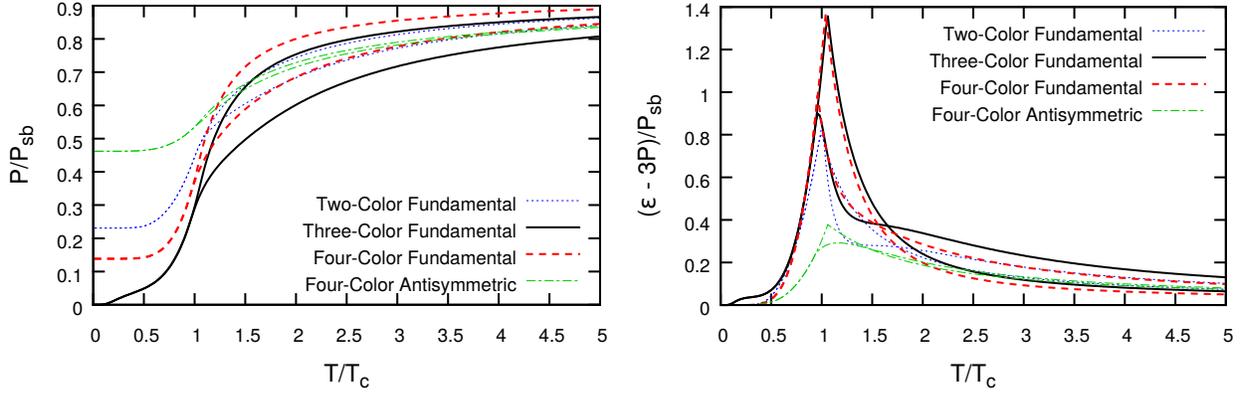

FIGURE 4.2: Normalized pressure (left) and trace anomaly (right) at $\mu = 0$ for the two-color, three-color, four-color fundamental, and four-color antisymmetric theories in HRG+pQCD. Note that the T-axis has been scaled by the critical temperature (see main text).

anomaly, a phenomenon that is well-known from pure-gauge theories [74].

The differences at low temperatures are due to the different numbers of Nambu–Goldstone bosons in the two-color and four-color theories with zero quark mass (see Sec. 4.1.2.2 or Appendix A) and the fact that in the real world there are only pseudo-Nambu–Goldstone bosons. We verified this by increasing the mass of the lightest (now pseudo-) Nambu–Goldstone bosons, which qualitatively changed the shape of the pressure curves until they matched that of the real-world, three-color theory. In Fig. 4.3, we show the pressure and trace-anomaly bands at $T = 0$ for all four theories with the $\mu$-axis scaled by the critical chemical potential $\mu_c$, again, defined to be the average of the matching chemical potential of the upper and lower edge of the pQCD band to the HRG EoS. The value of $\mu_c$ should be regarded as an estimate for the confinement-deconfinement transition, whereas the critical chemical potential for the onset transition would be given by the smallest value of $m_i/r_i$, to use the notation of Sec. 4.1.2. In the fundamental theories, this value of $m_i/r_i$ corresponds to the lightest baryon mass. Similar to the $\mu = 0$ case, the $\mu \neq 0, T = 0$ results show similar trends when scaled appropriately. Again, the different behaviors at low $\mu/\mu_c$ are due to the fact that there are Nambu–Goldstone bosons composed solely of quarks in the two-color fundamental and four-color antisymmetric theories. Again, this was tested by increasing the masses of the lightest particles.

The values of $T_c/\sqrt{\sigma}$ and $\mu_c/\sqrt{\sigma}$ for the HRG+pQCD calculations are given in Tab. 4.1.

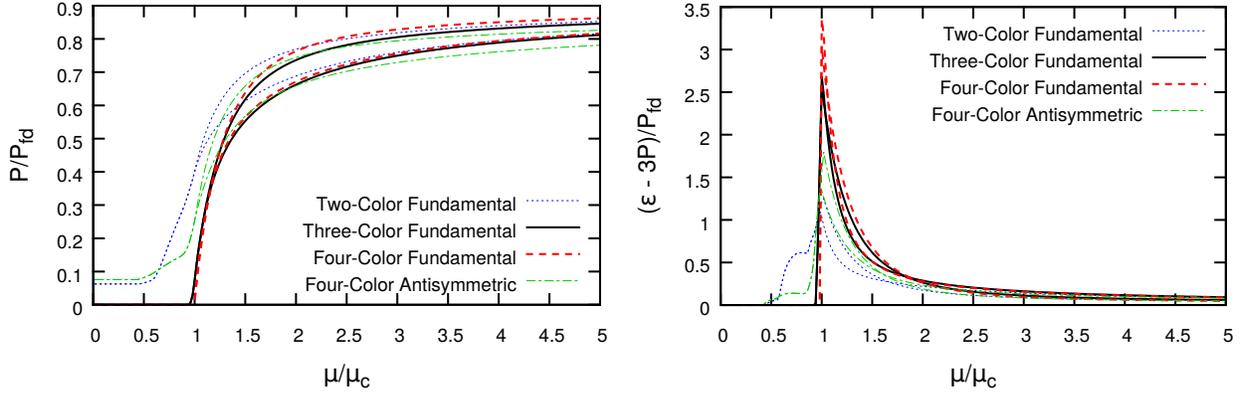

FIGURE 4.3: Normalized pressure (left) and trace anomaly (right) at $T = 0$ for the two-color, three-color, four-color fundamental, and the four-color antisymmetric theories in HRG+pQCD. Note that the $\mu$-axis has been scaled by the critical chemical potential (see main text).

While the results suggests that the $T_c$ values for the different theories are within 20 percent of each other, the extracted $\mu_c$ values span a much broader range.

We wish to remind the reader here that, in the $T = 0$ case, we have not included the CSC phase in the high-$\mu$ EoS, which will introduce a correction to the pressure on the ten-percent level (see the discussion near Eq. (4.16)). We also note that a ten-percent change in the plots of the bulk thermodynamic properties will not affect them in a noticeable way, for the error bands are already at least of this order.

We stress that in the four-color antisymmetric case with $\Lambda_{\overline{MS}}/\sqrt{\sigma} = 0.723$, we were unable to carry out the matching procedure at $\mu = 0$ in the chiral limit. We found that in this case, the HRG pressure rose too sharply and never intersected the pQCD pressure-band. Thus, we have only plotted the $\Lambda_{\overline{MS}}/\sqrt{\sigma} = 0.527$ results for the four-color antisymmetric theory in the figures.

| Group, Representation, $n_f$ | $T_c/\sqrt{\sigma}$ | $\mu_c/\sqrt{\sigma}$ |
|---|---|---|
| SU(2), fundamental, 2 | 0.400 | 3.24 |
| SU(3), fundamental, 3 | 0.47 | 2.382 |
| SU(4), fundamental, 2 | 0.44 | 2.853 |
| SU(4), antisymmetric, 2 ($\Lambda_{\overline{MS}}/\sqrt{\sigma} = 0.527$) | 0.29 | 5.09 |
| SU(4), antisymmetric, 2 ($\Lambda_{\overline{MS}}/\sqrt{\sigma} = 0.723$) | no matching | 5.0 |

TABLE 4.1: The ratios $T_c/\sqrt{\sigma}$ and $\mu_c/\sqrt{\sigma}$ for the theories analyzed in this section. Errors are given by the number of significant figures.

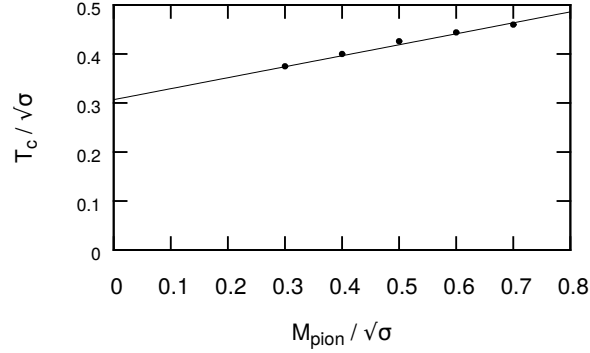

FIGURE 4.4: Deconfinement-transition temperature $T_c$ as a function of pion mass for the four-color, antisymmetric theory in the $\mu = 0$, $\Lambda_{\overline{MS}}/\sqrt{\sigma} = 0.723$ case. The straight line is a fit to the results where matching could be performed, and defines the extrapolation to the chiral limit (see main text).

We feel this is justified for a few reasons. First of all, the values of $\mu_c$ are equal within uncertainties for the two different values of $\Lambda_{\overline{MS}}/\sqrt{\sigma}$. Secondly, in the case where $\Lambda_{\overline{MS}}/\sqrt{\sigma} = 0.723$, we were able to carry out the HRG+pQCD matching procedure when we increased the mass of the lightest bosons (the pion mass). By varying the pion mass, we were able to extrapolate to the chiral limit, obtaining a value of $\mu_c/\sqrt{\sigma} = 0.3$, which agrees with the value found for $\Lambda_{\overline{MS}}/\sqrt{\sigma} = 0.527$ (see Tab. 4.1). In light of this agreement, and in light of how the four-color antisymmetric theory was the only theory where the matching was strained, we conjecture that the true value of $\Lambda_{\overline{MS}}/\sqrt{\sigma}$ in this case is closer to the pure-glue value than it is in the real-world, three-color case. We point out that this prediction could be tested in future lattice-gauge-theory calculations.

Finally, we have calculated the speed of sound $c_s$ at $T = 0$ in all four QCD-like theories using the HRG+pQCD scheme, shown in Fig. 4.5. We note that in some cases, $c_s$ exceeds the speed of light, and thus these particular matching results from HRG+pQCD should be considered unphysical (a standard constraint when using cold-nuclear-matter EoSs). Nevertheless, our results indicate that it is generally possible to obtain physical EoSs wherein $c_s^2 > 1/3$ for all fundamental QCD-like theories. This finding could be of interest because restricting $c_s^2 < 1/3$ has previously been noted to be in tension with astrophysical observations [75]. Again, we point out that this is a property which could be tested in future lattice-gauge-theory calculations.

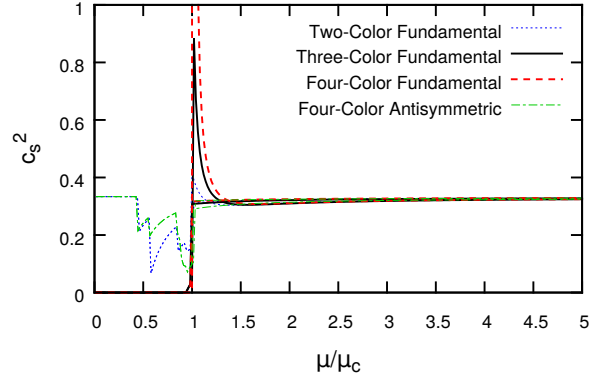

FIGURE 4.5: The speed of sound squared at T = 0 for the two-color, three-color, four-color fundamental, and the four-color antisymmetric theories in HRG+pQCD. Note that the μ-axis has been scaled by the critical chemical potential (see main text).

### 4.1.5    Conclusions: HRG+pQCD matching

We have calculated the EoS at non-zero temperatures and densities in a first-principles approach: by matching physics from the hadron resonance gas at low energies to perturbative QCD at high energies for two-, three-, and four-color 'QCD'. In particular, we have provided predictions for results in future lattice studies at zero temperature and non-zero chemical potential for two-color QCD with two fundamental fermions and four-color QCD with two flavors of fermions in the two-index, antisymmetric representation. While some aspects of this study are systematically improvable (in the ways discussed in the opening paragraphs of this chapter), we expect the current HRG+pQCD results to be sufficiently robust that a direct comparison with future lattice-QCD studies in the two- and four-color cases could validate or rule out the HRG+pQCD method, depending on the quantitative agreement. In the case of agreement, one could thus also reasonably expect HRG+pQCD results to be quantitatively accurate in the physically-relevant, three-color-QCD case.

The results of this systematic study have been made electronically available [76] so that they may be more easily accessible.

## 4.2    Kurkela *et al.* [1] matching: ChEFT to pQCD

The second topic that we address in this chapter is the current state-of-the-art matched EoS of Kurkela *et al.* [1], which demonstrates the more sophisticated matching approach discussed in the introductory paragraphs of this chapter. In the papers in Refs. [1, 11], the authors conducted a careful matching procedure to constrain the QCD EoS between the ChEFT and pQCD limits. The authors only assumed the validity of the perturbative EoSs up to a point, and between these two extremes they approximated the QCD EoS by two or more (though see below) **polytropic EoSs**, i.e., EoSs of the form

$$P(n) = \kappa_i n^{\gamma_i}, \tag{4.29}$$

where $n$ is the density, $\kappa_i$ and $\gamma_i$ are constants, and $i$ labels the different polytropes. The exponent $\gamma_i$ is referred to as the **polytropic index**. The matching to the ChEFT and pQCD EoSs was performed at $n = 1.1 n_s$ and $\mu = 2.6$ GeV, respectively, where the relative uncertainties for each perturbative EoS reach $\pm 24\%$ [77, 10]. Here, $n_s \approx 0.16 / \text{fm}^3$ is the **nuclear saturation density**. Below $n = 1.1 n_s$, the authors used the state-of-the-art ChEFT EoS of Tews *et al.* [77], and above $\mu = 2.6$ GeV, the authors used the state-of-the-art pQCD EoS from Ref. [10] in the compact form presented in Ref. [78].

Between these two controlled regimes, the authors of Ref. [1] used either two or three polytopes of the form (4.29) both with and without latent heat (i.e., a first-order phase transition) at the matching points. The authors eventually concluded that the addition of latent heat was actually *more* restrictive on the matching, and, in addition, a third polytrope only minimally increased the range of allowed EoSs (see Fig. 4.6, taken from their paper). Thus, in the rest of our descriptions here, we will outline the procedure used in Ref. [1] to match *two* intermediate polytropes.

To match the two polytropic EoSs, the authors chose a random intermediate matching value $\mu_c$ and a matching point in the pQCD band (obtained by varying the renormalization scale $\overline{\Lambda}$ about $\sqrt{(2\pi T)^2 + (\mu)^2}$ by a factor of two in both directions, as is customary) and attempted to solve for the polytropic indices $\gamma_i$ that would provide a matched EoS for those random values. If such values existed, the authors then checked to see if the speed of sound in the resulting EoS was subluminal throughout, and if it was not, rejected it. In this way, 3500 EoSs were generated, with $\gamma_1 \in [2.23, 9.2]$ and $\gamma_2 \in [1.0, 1.5]$. Fig. 4.6 shows the resulting band of allowed EoSs in the form of a P vs. $\mu$ plot (note that P has been scaled by the Stefan–Boltzmann pressure $P_{SB}$, and $\mu_B$ is the baryon chemical potential).

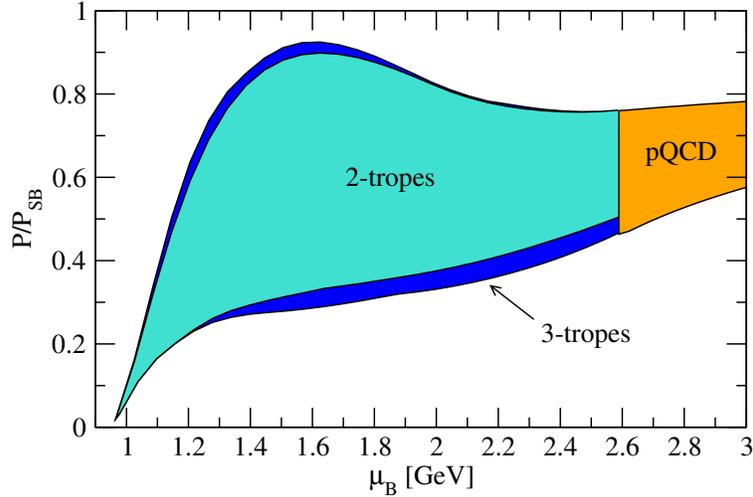

FIGURE 4.6: The allowed band of EoSs determined by Kurkela *et al.* [1], consisting of two- and three-trope intermediate EoSs. From Ref. [1]. The ChEFT EoS is not shown on the plot; it connects to the bands shown in the lower-left corner. In this plot, $\mu_B$ is the baryon chemical potential, and P has been scaled by $P_{SB}$, the Stefan–Boltzmann pressure.

In Ref. [1], the authors also used one final constraint; namely, that matter obeying the constructed EoS should be able to support a non-rotating NS of two solar masses ($2M_\odot$). This is a constraint imposed by observation rather than theory, for a $2M_\odot$ neutron star has been observed [79, 80]. This actually provides a quite stringent constraint on the QCD EoS beyond the matching constraint itself, and it allowed the authors to constrain the QCD EoS to $\pm30\%$ throughout the entire range of $\mu$.

Rather than continue from the theoretical point of view, as the study of NS properties is one of our goals, let us now turn to NSs themselves. We shall return to the observationally-constrained QCD EoS band of Kurkela *et al.* [1] in the following chapter, where we will analyze the full range of observational constrains coming from both static and rotating NSs. This will allow us to more fully appreciate how highly one may constrain the QCD EoS by applying it to NSs.

# CHAPTER 5

# NEUTRON STARS AND APPLICATIONS

Neutron stars (NSs) are one of the most extreme physical systems in the cosmos. Within a sphere of radius ~10 km lies over $1M_\odot$ of matter. In the outer layers of NSs, controlled techniques such as ChEFT [77] or quantum Monte Carlo [81] are applicable and can yield insights into both the static properties of the bulk matter (such as the equation of state or EoS) and some transport properties. Currently, these low-density calculations are valid up to about 1.1 times the nuclear saturation density $n_s \approx 0.16/\text{fm}^3$, corresponding to a baryon chemical potential of about $\mu \approx$ 0.97 GeV [77]. Deep in the core, however, such controlled, direct theoretical calculations are not possible. This is because the densities and chemical potentials at the center of the star, though extreme, are not large enough to fall into the range accessible by pQCD. In the state-of-the-art pQCD calculations at zero temperature in Ref. [10], the errors associated with varying the mass scale reach 30% at around $\mu = 2.6$ GeV. The value of $\mu$ in the cores of NSs lies within a subset of this $0.97 - 2.6$ GeV range.

The problem of the interiors of NSs is thus currently a non-perturbative one. However, as discussed in Chap. 4, one can hope to reach the intermediate values of $\mu$ by matching the low-density EoS from a low-energy effective theory to the pQCD results in a thermodynamically consistent way to investigate the makeup of NSs. As also discussed in that chapter, this has been carried out in the work of Kurkela *et al.* [1] and Fraga *et al.* [11], who, in addition, incorporated the $2M_\odot$ constraint from Refs. [79, 80]. (See also Ref. [82], in which the authors use only ChEFT and the $2M_\odot$ constraint to extend the low-energy EoS.) In these works, the authors used their matched EoSs to analyze non-rotating NSs only. It is known [83, 84] that slowly-rotating NSs can be approximated as non-rotating for frequencies of rotation less than about f $\approx$ 200 Hz. Beyond this, however, one must use numerical codes to analyze the structure of the stars. Such a numerical approach has been recently used by Cipolletta *et al.* [84] and Haensel *et al.* [85] in the context of phenomenological EoSs, and one of the purposes of this chapter is to extend these analyses to include EoSs that are more fully constrained by first-principles physics.

Broadly speaking, the purpose of this chapter is to investigate the structure of NSs at frequencies from zero all the way up to the mass-shedding limit using the constraints on the QCD EoS determined in Refs. [1, 11]. We are particularly interested in constraining NS properties that are relevant observationally. In Sec. 5.1, we quickly review the general procedure for constraining global NS structure within the framework of general relativity (GR). Then, in Sec. 5.2, we consider applications of the QCD EoS band of Kurkela *et al.* [1]. In this section, we first discuss the work done in Ref. [1] to incorporate the $2M_\odot$ constraint into the authors' matching procedure and highlight their conclusions about observable NS properties. We follow this discussion with original work that extends these results to rotating NSs. We investigate the maximum allowed NS masses, as well as the allowed regions for mass–radius curves, mass–frequency curves, and radius–frequency curves for a typical $1.4M_\odot$ star. In addition, we investigate the allowed values of the moment of inertia of the double pulsar PSR J0737-3039A [86, 87] and study how this is correlated with the radius. In this way, we aim to provide a strong direct link between astronomical observations and the allowed QCD EoSs coming from current state-of-the-art pQCD and ChEFT calculations.

This chapter draws heavily from work that will soon be published in Ref. [88].

# 5.1  The QCD EoS and the structure of NSs: overview

To determine the global structure of a NS in GR, one often starts with symmetry assumptions. The two simplest assumptions are that the star is non-rotating, or rotating uniformly. Let us begin with the former case.

A non-rotating, spherically-symmetric object in GR can be described by the metric [89]

$$ds^2 = e^{2\alpha(r)}\,dt^2 - e^{2\beta(r)}\,dr^2 - r^2\big(d\theta^2 + \sin^2\theta\,d\phi^2\big). \tag{5.1}$$

Here, $\alpha$ and $\beta$ are arbitrary functions of $r$. To compute the structure of a NS, one must solve Einstein's Equations,

$$R_{\mu\nu} - \frac{1}{2}g_{\mu\nu}R = 8\pi G\,T_{\mu\nu}, \tag{5.2}$$

inside the star, where we assume that the matter in the star is a perfect fluid with

$$T_{\mu\nu} = (\varepsilon + P)u_\mu u_\nu - Pg_{\mu\nu}. \tag{5.3}$$

In these equations, G is Newton's gravitational constant, $u_\mu$ is the four-velocity of a fluid element, and, as before, $\varepsilon$ is the energy density and P is the pressure. Using algebra, the equations (5.1)-(5.3) can be reduced to the following two equations, called the **Tolman–Oppenheimer–Volkoff** or **TOV equations** [89]:

$$\frac{dP}{dr} = -\frac{G(\varepsilon + P)\big[m(r) + 4\pi r^3 P\big]}{r\big[r - 2Gm(r)\big]}, \tag{5.4}$$

$$\frac{dm}{dr} = 4\pi r^2 \varepsilon. \tag{5.5}$$

Note that these are simply the equations describing hydrostatic equilibrium for a static, spherically-symmetric fluid. To solve for the structure of a non-rotating NS, one must prescribe an EoS (i.e., $P(\varepsilon)$). With this specified, one may construct a star by specifying a central energy density $\varepsilon^*$ and then solving the TOV equations outward from the center until the radius R such that $\varepsilon(R) = 0$; this is the stellar surface.

For a uniformly-rotating star, the procedure is in principle the same, though the equations are more complicated. The metric of a stationary, axisymmetric space-time (i.e., one describing a uniformly-rotating, axisymmetric star with angular velocity $\Omega$ as judged by an observer at infinity) is given by [90, 91, 92]

$$ds^2 = e^{2\nu}\,dt^2 - e^{2\alpha}\left(dr^2 + r^2\,d\theta^2\right) - e^{2\beta}r^2\sin^2\theta\left(d\phi - \omega\,dt\right)^2, \tag{5.6}$$

where $\nu$, $\omega$, $\alpha$, and $\beta$ are functions of $r$ and $\theta$. In this case, one must again solve Einstein's Equations (5.2) with the assumption (5.3) for $T_{\mu\nu}$. This leads to the following equation for hydrostationary equilibrium (as opposed to *hydrostatic* equilibrium when there was no motion at all) [93, 92]:

$$\frac{1}{\varepsilon + P}\nabla P + \nabla\nu - \frac{1}{2}\nabla\ln(1 - w^2) = 0, \tag{5.7}$$

where here, $w$ is the proper velocity of a fluid element with respect to a local zero-angular-momentum observer:

$$w = r\sin\theta e^{\beta - \nu}(\Omega - \omega). \tag{5.8}$$

Note that in this rotational case, the hydrostationary-equilibrium equation is not decoupled from the Einstein equations.

The coupled system of (5.8) and Einstein's Equations (5.2) can be solved numerically with the use of the publicly-available RNS code of Ref. [94]. This code takes as input an EoS in the form $P(\varepsilon)$ and two parameters: a central energy density $\varepsilon^*$ and the ratio of the polar coordinate radius to the equatorial coordinate radius $r^*$. Other inputs can be used as well (see Sec. 5.2.2, where we use this code in combination with the QCD EoS of Kurkela *et al.* [1]), but internally each NS that is constructed is specified by the parameters $\varepsilon^*$ and $r^*$. From this input, the code can calculate various global properties of the star, including the total (or gravitational) mass $M$, the circumferential equatorial radius $R_e$, the frequency of rotation $f$, and the moment of inertia $I$.

## 5.2   Global properties of NSs with QCD EoSs

In this section, we discuss applications of the QCD EoS band of Refs. [1, 11] to NS. These applications take two complementary forms. First, the QCD EoS band can be used to constrain observational properties of NSs. Second, observations can be used to further constrain the QCD EoS band. Both of these applications will be discussed below. Following all of the discussions, we review the main conclusions of both types of applications in Sec. 5.2.3, including ones that are most relevant to astrophysical observation.

### 5.2.1  Non-rotating case

Applications of the QCD EoS band derived in Refs. [1, 11] were already discussed in that reference itself. In addition, both types of applications were considered. In the aforementioned references, the authors constrained their QCD EoS band beyond what was described in Sec. 4.2 above by requiring that the QCD EoS be able to support a non-rotating, $2M_\odot$ star. This is required to agree with observations of a $2M_\odot$ star with $f \ll 200$ Hz [80]. (Note that the other $2M_\odot$ NS that has been detected has $f = 317.45$ Hz [79], which is too large to be considered non-rotating.)

Using this constraint, the authors were able to significantly reduce the uncertainties on their resulting QCD EoS band (see Fig. 5.1). Throughout the entire intermediate (matching) region, the QCD EoS was constrained to within ±30% using the observational constraint. In addition to constraining the QCD EoS band, the authors also determined mass–radius curves for the constructed

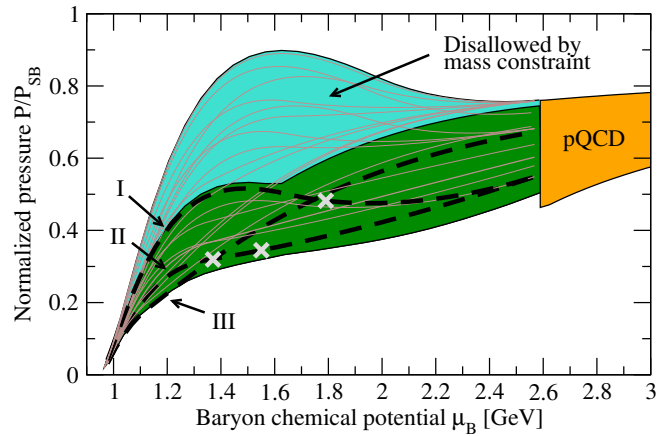

FIGURE 5.1: The allowed band of EoSs determined by Kurkela *et al.* [1], including the $2M_\odot$ mass constraint. From Ref. [1]. The lines indicate individual constructed EoSs, and the bold, dashed lines are tabulated in Ref. [1]. The crosses denote the largest value of $\mu$ reached within a non-rotating NS constructed from each of the bold, dashed EoSs. The ChEFT EoS is not shown on the plot; it connects to the bands shown in the lower-left corner. Note that P has been scaled by $P_{SB}$, the Stefan–Boltzmann pressure.

EoSs. We do not include their mass–radius plots here, for in the next section we reproduce the non-rotating region calculated in Ref. [1] as the horizontally-striped area in Fig. 5.2 below. However, we do note that the authors concluded that any non-rotating NS constructed from their QCD EoS band must satisfy $M < 2.75 M_\odot$ (or $M < 2.5 M_\odot$ for bitropic EoSs), with the radius of a typical $1.4 M_\odot$ star falling between 11 and 14.5 km.

Let us now turn to our generalizations of these results to include rotating NSs.

### 5.2.2    General rotating case

In this section, we discuss generalizations of the work of Ref. [1] to include rotating NSs. We begin by briefly describing how the RNS code mentioned above in Sec. 5.1 was used to construct mass–radius curves, mass–frequency curves, radius–frequency curves for a typical $1.4 M_\odot$ star, and moment-of-inertia–radius plots of the double pulsar PSR J0737-3039A. Following this, we present our results and all of our plots in detail.

#### Methodology

To conduct our analysis of rotating NSs, we used the publicly available RNS code [94]. In addition to constructing a single star specified by $\varepsilon^*$ and $r^*$ (see the discussion in Sec. 5.1 above), the RNS code can construct sequences of stars as well as accept other stellar properties as input to construct internal sequences and find stars satisfying those inputs. It can also calculate the **mass-shedding frequency** for a given central energy density $\varepsilon_0$, which is the fastest rotation rate possible before the star begins to throw off mass from its equator. This provides an upper bound on the rotation rate for the central energy density $\varepsilon_0$. Rotating stars have both a larger maximum mass and a larger maximum equatorial radius, and so the mass-shedding limit can be used to investigate larger, more massive stars than were possible in the non-rotating limit.

The approach used in this investigation was to take the EoSs used in Refs. [1, 11] in the form $P(\varepsilon)$ and feed them into the RNS code to calculate various properties of physical interest. A comment is in order here. Since in Refs. [1, 11], the authors concluded that adding latent heat was actually *more* restrictive on the matching, and, in addition, they found that a third polytrope only minimally increased the range of allowed EoSs, we have also only used the bitropic EoSs without latent heat in this section.

To construct our data, we first ran the RNS code on the static and mass-shedding sequences. From this, we could construct the mass–radius curves and one boundary of the allowed mass–frequency region for NSs. The rest of our numerical data involved either fixed-frequency runs, fixed-mass runs (or both), or coding a binary search to fill in the gaps where the code was unable to generate the star. This was necessary in the cases of very small frequencies, as internally the code always uses $r^*$ as a parameter instead of f. (This behavior was also noted in Ref. [84].) The fixed-frequency runs were used to determine the other boundary of the allowed mass–frequency region, and the fixed-mass runs were used to determine the radius–frequency relations for a typical, $1.4 M_\odot$ NS. Finally, the runs at fixed mass and frequency were used for investigating PSR J0737-3039A.

### Results: Rotating case

We present first our results for mass vs. equatorial radius curves in Fig. 5.2. The non-rotating region is the same as in Ref. [1], and has a maximum mass of about $2.5 M_\odot$. As seen in the figure, rotating NSs have a larger radius and a larger maximum mass than non-rotating ones. This can be thought of as a consequence of centrifugal force: the stars with large central energy densities that are unstable past the maximum-mass point for non-rotating stars are stabilized (and their central energy densities are lowered) by the outward centrifugal force in the rotating case. The larger radius is a consequence of the eccentricity of the star caused by the centrifugal force as well. We see that the maximum-mass star now has a mass of about $3.25 M_\odot$, and the largest stellar radius is about 21km.

As one might expect, the boundaries of the non-rotating region and the mass-shedding regions in Fig. 5.2 are formed from the same EoSs; e.g., the EoS that contains the highest-mass stars in the non-rotating case also contains the highest-mass stars in the mass-shedding case. This means that any further observational constraints that restrict the left, horizontally-striped region in Fig. 5.2 will also restrict the right, vertically-striped region in the same way.

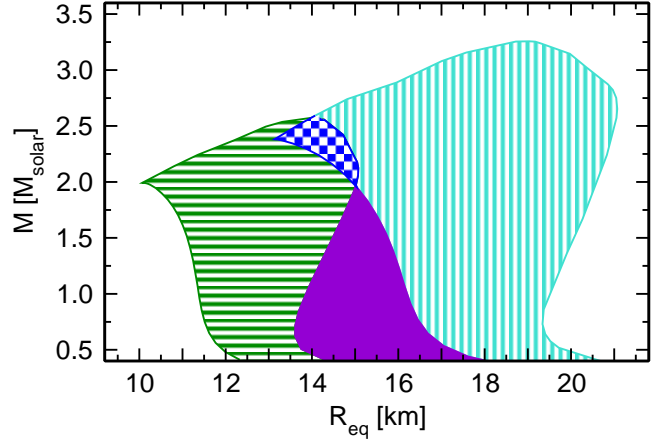

FIGURE 5.2: Mass vs. equatorial radius regions for non-rotating stars (horizontal stripes) and mass-shedding stars (vertical stripes). The upper, checkered region is an overlap between the non-rotating and mass-shedding regions. The lower, solid region is only accessible to non-mass-shedding rotating NSs.

In Fig. 5.3, we show the allowed regions for NSs in the mass–frequency plane. The inner, solid region is allowed for every EoS, and the outer, checkered band shows where the possible boundaries are for each EoS. The right boundary of the checkered region is constrained by the mass-shedding stars: beyond a certain limiting frequency at a given mass, stars become unstable. The upper boundary of the checkered region consists of the curves $M_{max}(f)$, the maximum NS mass as a function of frequency. We also include a dashed line in Fig. 5.3, which is the boundary of the mass–frequency region for a sample EoS. This is to illustrate the shape of the boundary for each EoS. Every EoS is shaped similarly: the top boundary rises towards the sloped, upper-right edge of the checkered region, comes to a point, and then curves back down. Note that this implies that the outermost boundary of the checkered region is not formed from a single EoS; in fact, even the upper edge and lower-right edge of the checkered region are formed by different EoSs.

We also show in Fig. 5.3 data points for NSs with frequencies above 100Hz, taken from Ref. [85]. A star located in the checkered band would eliminate some of the EoSs (namely, the ones whose curves in the checkered region are closer to the inner, solid region than the data point of the star). We see that there is only one star that is pushing into the checkered band: this is B1516+02B. If the mass of this star were further constrained, it could potentially eliminate a sizeable number of additional EoSs. Note, however, that $f = 125.83$Hz for B1516+02B, so this is still within the regime where approximating the star as non-rotating is valid. Thus, this constraint is not fundamentally one of rotation.

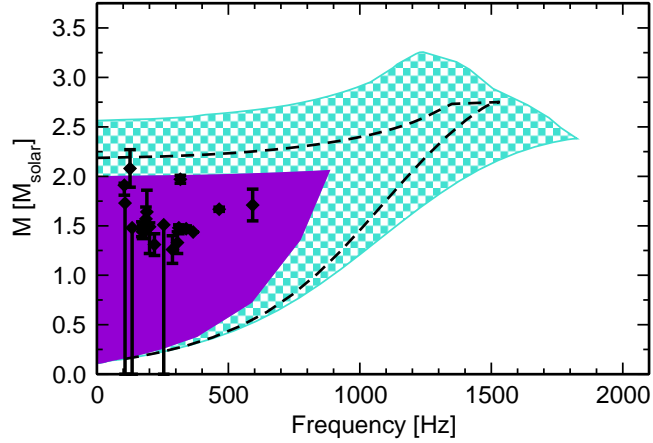

FIGURE 5.3: The allowed mass–frequency region for all of the possible QCD EoSs. The inner, solid region is allowed for every EoS, and the outer, checkered band shows where the possible boundaries are for each EoS. The dashed line is the outer boundary of the mass–frequency region for a sample EoS. Data points for NSs with f > 100Hz, taken from a table in Ref. [85], are also plotted.

From Fig. 5.3, however, we see that for high-f stars, there *is* a constraint coming from rotation. The clearest example of this is the upper-right corner of the inner, solid region with coordinates $(M, f) = (2.06 M_\odot, 883\text{Hz})$. This frequency, $f = 883\text{Hz}$, signifies the highest frequency that all of the EoSs can support. Thus, if a NS is ever found with $f > 883\text{Hz}$, this would eliminate some of the possible EoSs of Refs. [1, 11]. We note, however, that this is the highest frequency that would eliminate some EoSs: lower-frequency NSs could also rule out some EoSs if their masses could be measured and were sufficiently low. For example, PSR J1748-2446ad, currently the fastest rotating NS known ($f = 716\text{Hz}$) [95], would eliminate some EoSs if its mass is less than about $1 M_\odot$.

For a $1.4 M_\odot$ NS, the largest frequency that all EoSs can support is lower, $f = 780\text{Hz}$, as show in Fig. 5.4. In this figure, we have plotted the equatorial radius as a function of frequency $R_e(f)$ for a typical $1.4 M_\odot$ NS for each EoS. This plot serves as a prediction for observational astronomers. Furthermore, when consistent, reliable data of NS radii are available, a plot of this type could be overlaid with observational data to further constrain the QCD EoS (similar to Fig. 5.3 above). One other comment we wish to make here is that this radius–frequency band agrees with the result of the minimum-$\chi^2$, hybrid EoS of Kurkela *et al.* in Ref. [96]. That result lies directly in the center of our band in Fig. 5.4. We do note, however, that their *mass*–frequency boundary only partially agrees with our band: The boundary of the mass–frequency region coming from the mass-shedding curve in Ref. [96] lies in the center of our checkered band coming from our mass-shedding curves, but their upper boundary cuts into our solid band. This is because the minimum-$\chi^2$, hybrid EoS obtained in Ref. [96] does not permit a $2 M_\odot$ NS.

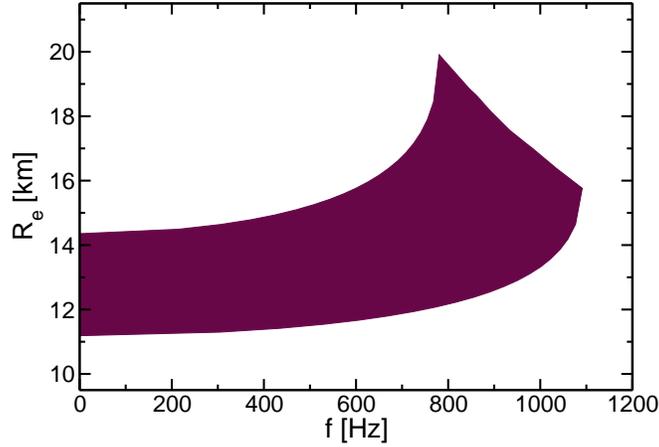

FIGURE 5.4: The region of allowed circumferential equatorial radius vs. frequency curves for a $1.4 M_\odot$ star.

The final plot that we have generated from the EoSs is shown in Fig. 5.5. In this figure, we show the allowed region for the moment of inertia and equatorial radius of PSR J0737-3039A. The moment of inertia of this star may be measured in a few years [86, 87], and so it is natural to investigate what the QCD EoSs predicts its value should be. We find that $I \in [1.2, 1.8] \times 10^{45} \mathrm{g\,cm^2}$. Work of this type has been performed previously assuming phenomenological EoSs, e.g. in Refs. [97, 87, 98]; and, more recently, Raithel *et al.* [99] have performed an analysis in which an EoS is only assumed up to $n_s$, and the remaining mass is shifted around to minimize and maximize $I$ for the star. This allows the authors to plot the largest allowed region in the $R_e$–$I$ plane constrained by controlled first-principles low-energy physics. Our allowed region in Fig. 5.5 does fall within the larger-$R_e$, larger-$I$ (i.e., upper-right) portion of the region calculated in the aforementioned work, and it also falls roughly in the center of the forty EoS data points presented in an earlier figure in that work.

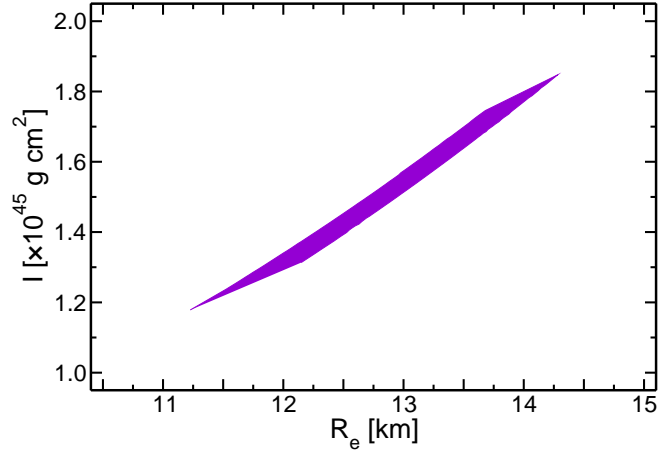

FIGURE 5.5: The allowed region of moment of inertia vs. circumferential equatorial radius for PSR J0737-3039A.

We also find that all of the "hard" and "soft" EoSs from Refs. [1, 11] fall on the two boundaries of our allowed region: the "hard" EoSs form the right boundary and the "soft" ones form the left boundary. In other words, the "hard" and "soft" EoSs each lie on their own fixed curve. This is not surprising, since the largest contribution to I comes from the matter at the largest radii (in the low-density crust region), and there, all the "hard" or "soft" EoSs agree by construction. Note, however, that since these EoSs form the vertical boundaries of the region, even a relatively imprecise measurement of the moment of inertia of PSR J0737-3039A (e.g., one with a precision of 10%) will *significantly* constrain which EoSs are consistent with the measurement. Since the allowed region spans $0.6 \times 10^{45}$ g cm$^2$ in I, a 10% measurement will only be consistent with about $0.15/0.6 = 25\%$ of the EoSs.

This percentage is not a physical meaningful result, but we translate it into a statement about the QCD EoS band in Fig. 5.6. In this figure, we display the QCD EoS band of Kurkela *et al*. [1], along with the subset of it that is consistent with $I = 1.5 \times 10^{45}$ g cm$^2$ to a precision of 10%, as an example. We see that such a measurement would shrink the percent errors of the band by up to 50% in some places, especially in the lowest-density regime. Again, this makes sense because it is the low-density material farthest from the rotation axis that contributes most to I. This reduction in the QCD EoS band would then, by extension, significantly constrain *all* of the NS properties mentioned in this section. This makes a measurement of the moment of inertia of the double pulsar PSR J0737-3039A of extreme interest. Such a measurement would also constrain the radius of the pulsar to within about $\pm 0.5$ km.

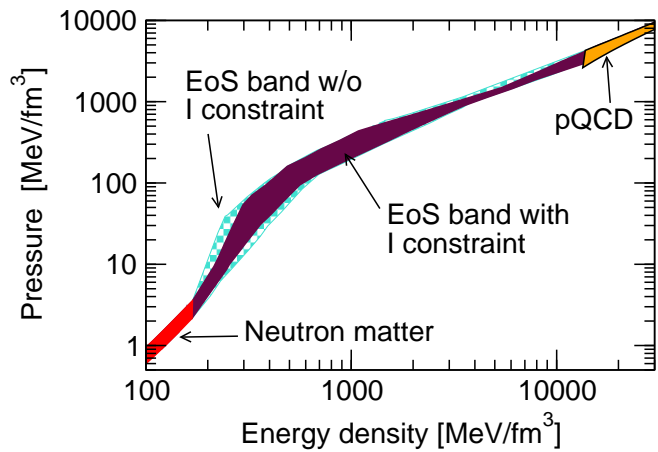

FIGURE 5.6: A plot illustrating how much the QCD EoS band of Ref. [1] would be restricted by a hypothetical measurement of $I = 1.5 \times 10^{45}$ g cm$^2$ with a precision of 10% for PSR J0737-3039A.

### 5.2.3    Conclusions: Applications to NS

In this section, we have investigated the effects of rotation on global properties of NSs constructed from the EoSs of Refs. [1, 11]. We have found the maximum allowed NS mass to be about $3.25 M_\odot$, and the maximum allowed NS radius to be about 21 km. From investigations of mass–frequency relations, we have have identified B1516+02B as a NS of particular interest: constraining its mass more precisely could potentially eliminate many allowed QCD EoSs. From mass–frequency relations, we also have identified $f = 883$ Hz as the maximum allowed NS rotation frequency consistent with every EoS. In the case of a canonical $1.4 M_\odot$ NS, we have found that $f = 780$ Hz is the maximum allowed rotation frequency consistent with every EoS. We have also determined the allowed $R_e$ vs. $f$ region for a $1.4 M_\odot$ NS, which may serve has a prediction for astronomers, and may also be overlaid with future precise radius measurements to further constrain the QCD EoS. Finally, we have calculated the moment of inertia and radius of PSR J0737-3039A for each EoS and found it to be consistent with the minimally constrained results of Ref. [99]. We have found that $I \in [1.2, 1.8] \times 10^{45}$ g cm$^2$ for the allowed QCD EoSs. Most excitingly, we have concluded that even a measurement of the moment of inertia of this star with a precision of 10% would reduce the percent errors on the band of allowed QCD EoSs that are consistent with observations to 50% of its current size at low densities. We thus conclude that a measurement of the moment of inertia of PSR J0737-3039A would be of extreme interest.

# CHAPTER 6

# HIGHER-ORDER TERMS IN THE PQCD PRESSURE AT ZERO TEMPERATURE

The matching procedure outlined in Chap. 4 and used in Chap. 5 to investigate NSs can be improved by theoretical advancements as well as observations. In this chapter, we will detail some improvements to pQCD at $T = 0$ that have already been achieved by the author and collaborators since the work of Kurkela *et al.* [1]. In particular, we derive in this chapter the $\mathcal{O}(g^6 \ln^2 g)$ piece of the pQCD pressure for $n_f$ massless quarks: a contribution that comes entirely from the plasmon term already discussed in Chap. 3. In doing this calculation, we will also extract a piece of the full $\mathcal{O}(g^6 \ln g)$ term and even parts of the $\mathcal{O}(g^6)$ term.

## 6.1 Higher orders for a single massless fermion

We start with a single massless fermion, where things are slightly simpler. To improve on the $T = 0$, $\mathcal{O}(g^4 \ln g)$ result of Sec. 3.1.8, we use the formulae (and notation) listed in that section to isolate the terms higher than $\mathcal{O}(g^4)$:

$$\frac{d_A}{(2\pi)^3} \int_0^\infty dK^2 K^2 \int_0^{\pi/2} d\Phi \sin^2(\Phi) \left\{ \ln\left[1 - \frac{F_{\text{mat}}(K, \Phi)}{K^2}\right] + \frac{F_{\text{mat}}(K, \Phi)}{K^2} + \frac{F_{\text{mat}}^2(K, \Phi)}{2K^4} \right.$$
$$\left. - \ln\left[1 - \frac{F_{\text{mat}}(K = 0, \Phi)}{K^2}\right] - \frac{F_{\text{mat}}(K = 0, \Phi)}{K^2} - \frac{F_{\text{mat}}^2(K = 0, \Phi)}{2K^4} \right\}. \quad (6.1)$$

As before, we will leave out the 2G terms except in the final equations in our derivations. We investigate the leading infrared behavior of this integrand. To do this, we require the expansion of the functions $F_{mat}$ and $G_{mat}$ about $K = 0$ to order $K^2$, which are found, in the case of a single massless quark with chemical potential $\mu$, to be

$$F_{mat}(K, \Phi) = F_0(\Phi) + \left(K^2 \ln \frac{K^2}{4\mu^2}\right) \widetilde{F}_1(\Phi) + \left(K^2\right) F_1(\Phi), \tag{6.2}$$

$$G_{mat}(K, \Phi) = G_0(\Phi) + \left(K^2 \ln \frac{K^2}{4\mu^2}\right) \widetilde{G}_1(\Phi) + \left(K^2\right) G_1(\Phi), \tag{6.3}$$

where the functions $F_0(\Phi), \widetilde{F}_1(\Phi), F_1(\Phi), G_0(\Phi), \widetilde{G}_1(\Phi)$, and $G_1(\Phi)$ are given by

$$F_0(\Phi) = \frac{g^2\mu^2}{2\pi^2} \csc^2\Phi \left(\Phi \cot\Phi - 1\right), \tag{6.4}$$

$$G_0(\Phi) = \frac{g^2\mu^2}{8\pi^2} \cot\Phi \csc^2\Phi \left(\sin 2\Phi - 2\Phi\right), \tag{6.5}$$

$$\widetilde{F}_1(\Phi) = -\frac{g^2}{24\pi^2}, \tag{6.6}$$

$$\widetilde{G}_1(\Phi) = -\frac{g^2}{24\pi^2}, \tag{6.7}$$

$$F_1(\Phi) = \frac{g^2}{72\pi^2} \left[5 + 3\csc^2\Phi - 3\Phi\cot\Phi(2 + \csc^2\Phi)\right], \tag{6.8}$$

$$G_1(\Phi) = \frac{g^2}{576\pi^2} \csc^3\Phi \left(27\sin\Phi - 13\sin 3\Phi + 12\Phi\cos 3\Phi\right). \tag{6.9}$$

Note that $F_0(\Phi)$ is what we were formally calling $F_{mat}(K = 0, \Phi)$, and similarly for $G_0(\Phi)$. Using the expansions (6.2) and (6.3), we find that the leading order infrared divergent piece of the integral (6.1) is

$$\frac{d_A}{(2\pi)^3} \int_0^{\pi/2} d\Phi \sin^2(\Phi) \int_0^\infty dK^2 \left[-\frac{F_0^2(\Phi)\widetilde{F}_1(\Phi)}{K^2} \ln\left(\frac{K^2}{4\mu^2}\right) - \frac{F_0^2(\Phi)F_1(\Phi)}{K^2}\right]. \tag{6.10}$$

In fact, we find that there is an infinite sequence of terms similar to this that can be resummed:

$$\left[-\widetilde{F}_1(\Phi) \ln\left(\frac{K^2}{4\mu^2}\right) - F_1(\Phi)\right] \sum_{n=3}^\infty F_0(\Phi) \left(\frac{F_0(\Phi)}{K^2}\right)^{n-2} = \frac{F_0(\Phi)\widetilde{F}_1(\Phi)}{F_0(\Phi) - K^2} \ln\left(\frac{K^2}{4\mu^2}\right) + \frac{F_0(\Phi)F_1(\Phi)}{F_0(\Phi) - K^2}. \tag{6.11}$$

Note that $F_0$ and $G_0$ are negative on $(0, \pi/2)$, and so the denominators here are never zero. Despite this, the integral of these terms over $K^2$ diverges. To regulate it, we subtract terms of the similar form

$$\frac{F_0(\Phi)\widetilde{F}_1(\Phi)}{-\widetilde{\chi}_1^2 - K^2} \ln\left(\frac{K^2}{4\mu^2}\right) + \frac{F_0(\Phi)F_1(\Phi)}{-\chi_1^2 - K^2}. \tag{6.12}$$

Here, $\widetilde{\chi}_1$ and $\chi_1$ are two further fictitious mass scales that must also drop out of the calculation of physical observables. The regulated integral over $K^2$ evaluates to

$$\int_0^\infty dK^2 \left\{ \frac{F_0(\Phi)\widetilde{F}_1(\Phi)}{F_0(\Phi) - K^2} \ln\left(\frac{K^2}{4\mu^2}\right) + \frac{F_0(\Phi)F_1(\Phi)}{F_0(\Phi) - K^2} - \frac{F_0(\Phi)\widetilde{F}_1(\Phi)}{-\widetilde{\chi}_1^2 - K^2} \ln\left(\frac{K^2}{4\mu^2}\right) - \frac{F_0(\Phi)F_1(\Phi)}{-\chi_1^2 - K^2} \right\}$$

$$= \frac{1}{2} F_0(\Phi)^2 \widetilde{F}_1(\Phi) \left[ \ln^2\left(\frac{-F_0(\Phi)}{4\mu^2}\right) - \ln^2\left(\frac{4\mu^2}{\widetilde{\chi}_1^2}\right) \right] + F_0^2(\Phi)F_1(\Phi) \ln\left(\frac{-F_0(\Phi)}{\chi_1^2}\right). \quad (6.13)$$

Let us be clear what we have done. Just as in Sec. 3.1.8 above, we isolated a piece of the full integral (6.1), namely the terms in Eq. (6.11), and regulated the result to obtain a finite answer. All that is left is to perform the integral over $\Phi$; this leads to the complete contributions to $\Omega_{\text{plas}}$ at orders $g^6 \ln^2 g$,

$$\frac{\Omega_{\text{plas}}^{(1)}}{V} = (g^6 \ln^2 g) \frac{d_A}{(2\pi)^3} \int_0^{\pi/2} d\Phi \sin^2(\Phi) \left[ \frac{2F_0(\Phi)^2 \widetilde{F}_1(\Phi)}{g^6} + 2 \cdot \frac{2G_0(\Phi)^2 \widetilde{G}_1(\Phi)}{g^6} \right], \quad (6.14)$$

and $g^6 \ln g$,

$$\frac{\Omega_{\text{plas}}^{(2)}}{V} = (g^6 \ln g) \frac{d_A}{(2\pi)^3} \int_0^{\pi/2} d\Phi \sin^2(\Phi) \left\{ \frac{2F_0^2(\Phi)F_1(\Phi)}{g^6} + \frac{2F_0^2(\Phi)\widetilde{F}_1(\Phi)}{g^6} \ln\left(\frac{-F_0(\Phi)}{4\mu^2 g^2}\right) \right.$$

$$\left. + 2 \cdot \left[ \frac{2G_0^2(\Phi)G_1(\Phi)}{g^6} + \frac{2G_0^2(\Phi)\widetilde{G}_1(\Phi)}{g^6} \ln\left(\frac{-G_0(\Phi)}{4\mu^2 g^2}\right) \right] \right\}, \quad (6.15)$$

with an additional piece contributing at order $g^6$:

$$\frac{\Omega_{\text{plas}}^{(3)}}{V} = (g^6) \frac{d_A}{(2\pi)^3} \int_0^{\pi/2} d\Phi \sin^2(\Phi) \left\{ \frac{1}{2} \frac{F_0^2(\Phi)\widetilde{F}_1(\Phi)}{g^6} \left[ \ln^2\left(\frac{-F_0(\Phi)}{4\mu^2 g^2}\right) - \ln^2\left(\frac{4\mu^2}{\widetilde{\chi}_1^2}\right) \right] \right.$$

$$+ \frac{F_0^2(\Phi)F_1(\Phi)}{g^6} \ln\left(\frac{-F_0(\Phi)}{\chi_1^2 g^2}\right) + \frac{G_0^2(\Phi)\widetilde{G}_1(\Phi)}{g^6} \left[ \ln^2\left(\frac{-G_0(\Phi)}{4\mu^2 g^2}\right) - \ln^2\left(\frac{4\mu^2}{\widetilde{\chi}_1^2}\right) \right]$$

$$\left. + \frac{2G_0^2(\Phi)G_1(\Phi)}{g^6} \ln\left(\frac{-G_0(\Phi)}{\chi_1^2 g^2}\right) \right\}. \quad (6.16)$$

The final contribution at $\mathcal{O}(g^6)$ is given by the remaining four terms of $\mathcal{O}(g^6)$ contained in the original integral (6.1) that were not included in the sum (6.11), *plus* the regulating terms that we subtracted to obtain the finite integral (6.13) These are:

$$
\begin{aligned}
\frac{\Omega_{\text{plas}}^{(4)}}{V} =& (g^6) \frac{d_A}{(2\pi)^3} \int_0^{\pi/2} d\Phi \sin^2(\Phi) \int_0^\infty dK^2 \Bigg\{ \frac{F_0^3(\Phi)}{3K^4 g^6} - \frac{F_{\text{mat}}^3(K,\Phi)}{3K^4 g^6} + 2 \cdot \left( \frac{G_0^3(\Phi)}{3K^4 g^6} - \frac{G_{\text{mat}}^3(K,\Phi)}{3K^4 g^6} \right) \\
&+ \frac{F_0(\Phi)\widetilde{F}_1^2(\Phi)}{g^6} \left( \frac{1}{-\widetilde{\chi}_1^2 - K^2} + \frac{1}{K^2} \right) \ln\left( \frac{K^2}{4\mu^2} \right) + \frac{F_0(\Phi)F_1^2(\Phi)}{g^6} \left( \frac{1}{-\chi_1^2 - K^2} + \frac{1}{K^2} \right) \\
&+ 2 \cdot \left[ \frac{G_0(\Phi)\widetilde{G}_1^2(\Phi)}{g^6} \left( \frac{1}{-\widetilde{\chi}_1^2 - K^2} + \frac{1}{K^2} \right) \ln\left( \frac{K^2}{4\mu^2} \right) + \frac{G_0(\Phi)G_1^2(\Phi)}{g^6} \left( \frac{1}{-\chi_1^2 - K^2} + \frac{1}{K^2} \right) \right] \Bigg\}.
\end{aligned}
$$
(6.17)

Thus, to $\mathcal{O}(g^6)$

$$
\Omega_{\text{plas}} = \Omega_{\text{plas}}^{(1)} + \Omega_{\text{plas}}^{(2)} + \Omega_{\text{plas}}^{(3)} + \Omega_{\text{plas}}^{(4)}.
$$
(6.18)

For the single, massless quark flavor described above, some of these integrals can be done analytically, while others can be partially done:

$$
\frac{\Omega_{\text{plas}}^{(1)}}{V} = -\frac{d_A \mu^4}{12 (2\pi)^8} g^6 \ln^2 g,
$$
(6.19)

$$
\frac{\Omega_{\text{plas}}^{(2)}}{V} = \frac{d_A \mu^4}{24 (2\pi)^8} \left( \frac{35}{6} - \frac{7\pi^2}{24} + \frac{16 \ln^2 2}{3} + \frac{4 \ln(4\pi^3)}{3} - \delta \right) g^6 \ln g.
$$
(6.20)

Here $\delta$ is defined exactly as in [39], namely,

$$
\begin{aligned}
\delta \equiv& \frac{16}{\pi} \int_0^{\pi/2} dx \sin^2 x \Bigg\{ \left( \frac{1 - x \cot x}{\sin^2 x} \right)^2 \ln \frac{1 - x \cot x}{\sin^2 x} + \frac{1}{2} \left( 1 - \frac{1 - x \cot x}{\sin^2 x} \right)^2 \ln \left[ 1 - \frac{1 - x \cot x}{\sin^2 x} \right] \Bigg\} \\
\approx& -0.856383209326942806848310232915940358847279097111357608993090 8673.
\end{aligned}
$$
(6.21)

Some of the remaining $\Phi$ integrals in $\Omega_{\text{plas}}^{(3)} + \Omega_{\text{plas}}^{(4)}$ can be done analytically too, but since we have only managed to do the integral over $K^2$ in Eq. (6.17) numerically due to the structure of $F_{\text{mat}}$ and $G_{\text{mat}}$, we do not reproduce them here. We do, however, include the numerical integrals for a single massless quark flavor:

$$
\frac{\Omega_{\text{plas}}^{(1)}}{V} = -3.430676811637164 \times 10^{-8} d_A \mu^4 g^6 \ln^2 g,
$$
(6.22)

$$
\frac{\Omega_{\text{plas}}^{(2)}}{V} = 2.195772106112194 \times 10^{-7} d_A \mu^4 g^6 \ln g,
$$
(6.23)

$$
\frac{\Omega_{\text{plas}}^{(3)} + \Omega_{\text{plas}}^{(4)}}{V} = -3.088466558527530 \times 10^{-7} d_A \mu^4 g^6.
$$
(6.24)

## 6.2  Higher-order terms for multiple massless fermions

We now tackle multiple massless fermions. For $n_f$ massless quark flavors, the expansions (6.2)-(6.3) become

$$F_{mat}(K, \Phi) = F_0(\Phi) + \left( \sum_f K^2 \ln \frac{K^2}{4\mu_f^2} \right) \widetilde{F}_1(\Phi) + (K^2)\, F_1(\Phi), \tag{6.25}$$

$$G_{mat}(K, \Phi) = G_0(\Phi) + \left( \sum_f K^2 \ln \frac{K^2}{4\mu_f^2} \right) \widetilde{G}_1(\Phi) + (K^2)\, G_1(\Phi), \tag{6.26}$$

with

$$F_0(\Phi) = \frac{g^2 \mu^2}{2\pi^2} \csc^2 \Phi\, (\Phi \cot \Phi - 1), \tag{6.27}$$

$$G_0(\Phi) = \frac{g^2 \mu^2}{8\pi^2} \cot \Phi \csc^2 \Phi\, (\sin 2\Phi - 2\Phi), \tag{6.28}$$

$$\widetilde{F}_1(\Phi) = -\frac{g^2}{24\pi^2}, \tag{6.29}$$

$$\widetilde{G}_1(\Phi) = -\frac{g^2}{24\pi^2}, \tag{6.30}$$

$$F_1(\Phi) = \frac{n_f g^2}{72\pi^2} \left[ 5 + 3 \csc^2 \Phi - 3\Phi \cot \Phi (2 + \csc^2 \Phi) \right], \tag{6.31}$$

$$G_1(\Phi) = \frac{n_f g^2}{576\pi^2} \csc^3 \Phi\, (27 \sin \Phi - 13 \sin 3\Phi + 12\Phi \cos 3\Phi), \tag{6.32}$$

where we have defined $\mu^2 = \sum_f \mu_f^2$. This means that the integrals for $\Omega_{plas}^{(1)}$-$\Omega_{plas}^{(4)}$ become

$$\frac{\Omega_{plas}^{(1)}}{V} = (g^6 \ln^2 g) \frac{d_A}{(2\pi)^3} \int_0^{\pi/2} d\Phi \sin^2(\Phi) \left[ \frac{2F_0(\Phi)^2 \widetilde{F}_1(\Phi)}{g^6} + 2 \cdot \frac{2G_0(\Phi)^2 \widetilde{G}_1(\Phi)}{g^6} \right], \tag{6.33}$$

$$\frac{\Omega_{plas}^{(2)}}{V} = (g^6 \ln g) \frac{d_A}{(2\pi)^3} \int_0^{\pi/2} d\Phi \sin^2(\Phi) \left\{ \frac{2F_0^2(\Phi) F_1(\Phi)}{g^6} + \frac{2F_0^2(\Phi) \widetilde{F}_1(\Phi)}{g^6} \sum_f \ln\left( \frac{-F_0(\Phi)}{4\mu_f^2 g^2} \right) \right.$$
$$\left. + 2 \left[ \frac{2G_0^2(\Phi) G_1(\Phi)}{g^6} + \frac{2G_0^2(\Phi) \widetilde{G}_1(\Phi)}{g^6} \sum_f \ln\left( \frac{-G_0(\Phi)}{4\mu_f^2 g^2} \right) \right] \right\}, \tag{6.34}$$

$$\frac{\Omega_{plas}^{(3)}}{V} = (g^6) \frac{d_A}{(2\pi)^3} \int_0^{\pi/2} d\Phi \sin^2(\Phi) \left\{ \frac{1}{2} \frac{F_0^2(\Phi) \widetilde{F}_1(\Phi)}{g^6} \sum_f \left[ \ln^2\left( \frac{-F_0(\Phi)}{4\mu_f^2 g^2} \right) - \ln^2\left( \frac{4\mu_f^2}{\widetilde{\chi}_1^2} \right) \right] \right.$$
$$+ \frac{F_0^2(\Phi) F_1(\Phi)}{g^6} \ln\left( \frac{-F_0(\Phi)}{\chi_1^2 g^2} \right) + \frac{G_0^2(\Phi) \widetilde{G}_1(\Phi)}{g^6} \sum_f \left[ \ln^2\left( \frac{-G_0(\Phi)}{4\mu_f^2 g^2} \right) - \ln^2\left( \frac{4\mu_f^2}{\widetilde{\chi}_1^2} \right) \right]$$
$$\left. + \frac{2G_0^2(\Phi) G_1(\Phi)}{g^6} \ln\left( \frac{-G_0(\Phi)}{\chi_1^2 g^2} \right) \right\}, \tag{6.35}$$

$$\frac{\Omega_{\text{plas}}^{(4)}}{V} = (g^6) \frac{d_A}{(2\pi)^3} \int_0^{\pi/2} d\Phi \sin^2(\Phi) \int_0^\infty dK^2 \left\{ \frac{F_0^3(\Phi)}{3K^4 g^6} - \frac{F_{\text{mat}}^3(K,\Phi)}{3K^4 g^6} + 2 \cdot \left( \frac{G_0^3(\Phi)}{3K^4 g^6} - \frac{G_{\text{mat}}^3(K,\Phi)}{3K^4 g^6} \right) \right.$$

$$+ \frac{F_0(\Phi)\widetilde{F}_1^2(\Phi)}{g^6} \left( \frac{1}{-\widetilde{\chi}_1^2 - K^2} + \frac{1}{K^2} \right) \sum_f \ln\left( \frac{K^2}{4\mu_f^2} \right) + \frac{F_0(\Phi)F_1^2(\Phi)}{g^6} \left( \frac{1}{-\chi_1^2 - K^2} + \frac{1}{K^2} \right)$$

$$\left. + 2 \left[ \frac{G_0(\Phi)\widetilde{G}_1^2(\Phi)}{g^6} \left( \frac{1}{-\widetilde{\chi}_1^2 - K^2} + \frac{1}{K^2} \right) \sum_f \ln\left( \frac{K^2}{4\mu_f^2} \right) + \frac{G_0(\Phi)G_1^2(\Phi)}{g^6} \left( \frac{1}{-\chi_1^2 - K^2} + \frac{1}{K^2} \right) \right] \right\}.$$

$$(6.36)$$

$\Omega_{\text{plas}}^{(1)}$ is identical to the single-flavor case, with $\mu^2$ replaced by $\boldsymbol{\mu}^2$, while $\Omega_{\text{plas}}^{(2)}$ may be made very similar to its form in the single-flavor case by using the identity $\mu_f^2 = (\mu_f^2/\boldsymbol{\mu}^2)\cdot\boldsymbol{\mu}^2$ in the logarithms and expanding. Finally, for simplicity we may choose $\widetilde{\chi}^2 = \chi^2 = 4\mu^2$, since these mass scales may be chosen arbitrary. These lead to

$$\frac{\Omega_{\text{plas}}^{(1)}}{V} = -\frac{d_A \left(\mu^2\right)^2}{12 \left(2\pi\right)^8} g^6 \ln^2 g,$$

$$(6.37)$$

$$\frac{\Omega_{\text{plas}}^{(2)}}{V} = \frac{d_A \left(\mu^2\right)^2}{24 \left(2\pi\right)^8} n_f \left[ \left( \frac{35}{6} - \frac{7\pi^2}{24} + \frac{16}{3}\ln^2 2 + \frac{4}{3}\ln(4\pi^3) - \delta \right) - \frac{2}{n_f} \sum_f \ln\left( \frac{\mu^2}{\mu_f^2} \right) \right] g^6 \ln g,$$

$$(6.38)$$

$$\frac{\Omega_{\text{plas}}^{(3)} + \Omega_{\text{plas}}^{(4)}}{V} = \frac{d_A \left(\mu^2\right)^2}{12 \left(2\pi\right)^8} n_f \left[ \frac{7\ln\pi}{48}(\pi^2 - 20) - \frac{\ln 2}{9}(19 + \ln 4) + \frac{2\ln^2 2}{3}(1 - \ln 128 - 4\ln\pi) \right.$$

$$- \frac{\ln\pi}{3}\ln(16\pi^3) + \frac{\ln(8\pi^2)}{4}\delta + \frac{\zeta}{3} - \frac{\eta}{8} + \frac{\ln 2}{8}\theta$$

$$+ \left( \frac{4}{3}\ln^2 2 + \frac{1}{3}\ln(4\pi^3) - \frac{\delta}{4} \right) \frac{1}{n_f} \sum_f \ln\left( \frac{\mu^2}{\mu_f^2} \right) \right] g^6$$

$$+ \frac{d_A}{(2\pi)^3} \int_0^{\pi/2} d\Phi \sin^2(\Phi) \int_0^\infty dK^2 \left\{ \frac{F_0^3(\Phi)}{3K^4 g^6} - \frac{F_{\text{mat}}^3(K,\Phi)}{3K^4 g^6} + 2 \cdot \left( \frac{G_0^3(\Phi)}{3K^4 g^6} - \frac{G_{\text{mat}}^3(K,\Phi)}{3K^4 g^6} \right) \right.$$

$$+ \left( \frac{1}{-4\mu^2 - K^2} + \frac{1}{K^2} \right) \sum_f \left[ \frac{F_0^2(\Phi)F_1(\Phi)}{g^6} + \frac{F_0^2(\Phi)\widetilde{F}_1(\Phi)}{g^6} \ln\left( \frac{K^2}{4\mu_f} \right) \right]$$

$$\left. + 2 \cdot \left( \frac{1}{-4\mu^2 - K^2} + \frac{1}{K^2} \right) \sum_f \left[ \frac{G_0^2(\Phi)G_1(\Phi)}{g^6} + \frac{G_0^2(\Phi)\widetilde{G}_1(\Phi)}{g^6} \ln\left( \frac{K^2}{4\mu_f} \right) \right] \right\} g^6,$$

$$(6.39)$$

where the final contribution to the $\mathcal{O}(g^6)$ term has not yet been analytically evaluated. Here, $\delta$ is as defined above, and $\zeta$, $\eta$, and $\theta$ are defined as the following integrals:

$$\zeta \equiv \frac{4}{\pi} \int_0^{\pi/2} dx \sin^2 x \left\{ \left[ 5 + 3 \csc^2 x - 3x \cot x (2 + \csc^2 x) \right] \left( \frac{1 - x \cot x}{\sin^2 x} \right)^2 \ln \frac{1 - x \cot x}{\sin^2 x} \right.$$

$$\left. + \frac{1}{16} \frac{12x \cos 3x + 27 \sin x - 13 \sin 3x}{\sin^3 x} \left( 1 - \frac{1 - x \cot x}{\sin^2 x} \right)^2 \ln \left[ 1 - \frac{1 - x \cot x}{\sin^2 x} \right] \right\}$$

$$\approx -0.663092750844561011343238360662050961220020228051329183397267483, \tag{6.40}$$

$$\eta \equiv \frac{16}{\pi} \int_0^{\pi/2} dx \sin^2 x \left\{ \left( \frac{1 - x \cot x}{\sin^2 x} \right)^2 \ln^2 \frac{1 - x \cot x}{\sin^2 x} + \frac{1}{2} \left( 1 - \frac{1 - x \cot x}{\sin^2 x} \right)^2 \ln^2 \left[ 1 - \frac{1 - x \cot x}{\sin^2 x} \right] \right\}$$

$$\approx 0.548647839714571966619636574085354909461606151212207302064972094 0, \tag{6.41}$$

$$\theta \equiv \frac{16}{\pi} \int_0^{\pi/2} dx \sin^2 x \left\{ \left( 1 - \frac{1 - x \cot x}{\sin^2 x} \right)^2 \ln \left[ 1 - \frac{1 - x \cot x}{\sin^2 x} \right] \right\}$$

$$\approx -0.511365586749237545388551517966599806276922762525302264053070365 7, \tag{6.42}$$

which are all quite similar to the defining integral for $\delta$.

## 6.3 Contribution of the two-loop self-energy

We also note here that there will be a contribution to the ring sum at $\mathcal{O}(g^6 \ln g)$ coming from an $\mathcal{O}(g^4)$ contribution to $\Pi^{\mu\nu}(K = 0, \Phi)$. This can be seen directly from Eq. (3.91). Suppose that $F_{\text{mat}}(K = 0, \Phi)$ is of the form

$$F_{\text{mat}}(K = 0, \Phi) = F_2(\Phi) g^2 + F_4(\Phi) g^4 \tag{6.43}$$

with $G_{\text{mat}}(K = 0, \Phi)$ of an analogous form. Then the $F_{\text{mat}}^2(K = 0, \Phi)$ term in (3.91) will contain a term of $\mathcal{O}(g^6)$, namely

$$2 F_2(\Phi) F_4(\Phi) g^6, \tag{6.44}$$

which contributes a term

$$\frac{\Omega_{\text{plas}}^{\text{2-loop}}}{V} = \frac{2 d_A g^6 \ln g}{(2\pi)^3} \int_0^{\pi/2} d\Phi \sin^2(\Phi) \left[ F_2(\Phi) F_4(\Phi) + 2 G_2(\Phi) G_4(\Phi) \right] \tag{6.45}$$

to the pressure. Any other contributions to $\Omega_{plas}$ coming from the $\mathcal{O}(g^4)$ piece of $\Pi^{\mu\nu}$ will be of at least $\mathcal{O}(g^8 \ln^2 g)$ (coming from cross terms in the $\mathcal{O}(g^6 \ln^2 g)$ piece derived above). This contribution to the $\mathcal{O}(g^6 \ln g)$ piece of the zero-temperature pQCD pressure is currently being worked on by the author and collaborators. This contribution, in fact, is the *only other contribution* to the full $\mathcal{O}(g^6 \ln g)$ result. (This follows from the fact that all the other diagrams that contribute at $\mathcal{O}(g^6)$ are infrared-safe, and it is only through infrared divergences of individual diagrams that the non-analytic logarithmic terms can arise.) The full $\mathcal{O}(g^6)$ result, however, has many more diagrams that contribute, which makes that term *much* more formidable to calculate.

Note, however, that an $\mathcal{O}(g^6 \ln^2 g)$ result cannot be reproduced by higher-order terms in the gluon self-energy, since the lowest-order contribution from the two-loop self-energy enters at $\mathcal{O}(g^6 \ln g)$. This means that the $\mathcal{O}(g^6 \ln^2 g)$ contribution given in this chapter in Eq. (6.37) is the *full* contribution at that order, and by itself constitutes a quantitative improvement to the pQCD pressure at $T = 0$.

# CHAPTER 7

# CONCLUSIONS

In this thesis, we have advocated for thermodynamic matching as a way to constrain the zero-temperature QCD EoS in the intermediate, non-perturbative regime, corresponding roughly to $\mu \in (0.97\,\text{GeV}, 2.6\,\text{GeV})$. Such a matching procedure can be carried out with various levels of sophistication, as has been illustrated in Chap. 4. The simplicity of the approach is noteworthy, for one may quantitatively constrain the intermediate EoS with little or no knowledge of the microphysics in the regime of interest. This is of course not to say that we are not interested in the microphysics, but a quantitative constraint of any non-perturbative physical property is a significant step.

We began this thesis with three goals in mind, beyond simply identifying thermodynamic matching as a legitimate approach to physically interesting problems: First, we desired a method to check if such a simplistic approach is valid; second, we wished to extend applications of the state-of-the-art QCD EoS of Refs. [1, 11] to rotating NSs; and third, we wished to make quantitative improvements to the zero-temperature pQCD pressure. All three of these goals have been accomplished.

On the first topic, we have shown in Sec. 4.1 that EoS matching in certain SU(N) ("QCD-like") theories can be verified or refuted by lattice simulations. In particular, the theories $(N, n_f) = (2, 2)$ with quarks in the fundamental representation and $(N, n_f) = (4, 2)$ with quarks in the two-index, antisymmetric representation can both be simulated on the lattice without a sign problem. We have produced matching results within the HRG+pQCD framework for the pressure and trace anomaly along the T- and $\mu$-axes in both of these theories, as well as in the theories $(N, n_f) = (4, 2)$ and $(3, 3)$ with quarks in the fundamental representation. The last of these theories is an approximation to real-world QCD. While some aspects of our HRG+pQCD study are systematically improvable, we expect the results in Sec. 4.1 to be sufficiently robust that a direct comparison with future lattice-QCD studies in the aforementioned theories could validate or rule out the HRG+pQCD method.

On the second topic, we have extended the results of Refs. [1, 11] to rotating NSs. We identified B1516+02B as a NS that could place more stringent constraints on the QCD EoS if its mass could be determined more precisely. We also identified $f = 882$ Hz as the maximum allowed rotation rate for a NS (or $f = 780$ Hz if restricted to a $1.4 M_\odot$ star). In addition, we derived the allowed equatorial radius vs. frequency band of a $1.4 M_\odot$ star, which serves as a prediction for astronomers, and can be used in the future to further constrain the QCD EoS. Lastly, we identified the binary pulsar PSR J0737-3039A as a physical object of extreme interest: a measurement of the moment of inertia of this pulsar, even to low precision, would *significantly* constrain the QCD EoS band of Refs. [1, 11].

On the third and final topic, we have calculated the *full* $\mathcal{O}(g^6 \ln^2 g)$ piece of the pQCD pressure at $T = 0$, along with a significant portion of the $\mathcal{O}(g^6 \ln g)$ piece and even a portion of the $\mathcal{O}(g^6)$ piece. These improvements can be incorporated into future matching work to improve the entire zero-temperature QCD EoS. In addition, these improvements are of interest in their own right, for they constitute fundamental improvements to our knowledge of pQCD at $T = 0$.

QCD matching is therefore an effective, verifiable, and systematically improvable method to explore non-perturbative regimes of the QCD EoS. It provides a window into regions of the QCD phase diagram where, as yet, no microphysical descriptions exist. This makes it a powerful, controlled tool for probing areas of the universe that are currently inaccessible to direct, microscopic calculations.

# APPENDIX  A

# PARTICLE TABLES

| Mesons | | |
|---|---|---|
| Mass/$\sqrt{\sigma}$ | Spin | Isospin |
| 0.00 | 0 | 1 |
| 1.43 | 0 | 0 |
| 1.60 | 1 | 1 |
| 1.86 | 1 | 0 |
| 2.79 | 1 | 0 |
| 3.02 | 2 | 0 |
| 3.06 | 1 | 0 |
| 3.08 | 0 | 0 |
| 3.10 | 1 | 1 |
| 3.14 | 2 | 1 |
| 3.25 | 1 | 1 |
| 3.26 | 0 | 0 |
| 3.38 | 1 | 0 |
| 3.50 | 0 | 1 |
| 3.50 | 0 | 1 |

| Mesons (continued) | | |
|---|---|---|
| Mass/$\sqrt{\sigma}$ | Spin | Isospin |
| 3.65 | 1 | 1 |
| 3.92 | 2 | 0 |
| 3.93 | 1 | 0 |
| 3.98 | 2 | 1 |
| 3.98 | 3 | 0 |
| 4.02 | 3 | 1 |
| 4.05 | 1 | 1 |
| 4.25 | 0 | 1 |
| 4.25 | 0 | 0 |
| 4.35 | 1 | 1 |
| 4.35 | 1 | 0 |
| 4.86 | 4 | 1 |
| 4.88 | 4 | 0 |
| 5.00 | 1 | 1 |
| 5.00 | 1 | 0 |

| Baryons | | |
|---|---|---|
| Mass/$\sqrt{\sigma}$ | Spin | Isospin |
| 0.00 | 0 | 0 |
| 1.43 | 0 | 0 |
| 1.86 | 1 | 1 |
| 2.79 | 1 | 1 |
| 3.26 | 0 | 0 |
| 3.06 | 1 | 0 |
| 3.02 | 2 | 0 |
| 3.92 | 2 | 0 |
| 3.93 | 1 | 1 |
| 3.98 | 3 | 1 |
| 4.88 | 4 | 0 |
| 3.08 | 0 | 0 |
| 3.38 | 1 | 1 |
| 5.00 | 1 | 1 |
| 4.25 | 0 | 0 |
| 4.35 | 1 | 0 |

TABLE A.1: The included particle spectrum in the two-color fundamental theory.

| Mesons | | |
|---|---|---|
| Mass/$\sqrt{\sigma}$ | Spin | Isospin |
| 0.00 | 0 | 1 |
| 1.43 | 0 | 0 |
| 1.83 | 1 | 1 |
| 1.86 | 1 | 0 |
| 2.33 | 0 | 1 |
| 2.79 | 1 | 0 |
| 2.94 | 1 | 1 |
| 3.00 | 1 | 1 |
| 3.02 | 2 | 0 |
| 3.06 | 1 | 0 |
| 3.08 | 0 | 0 |
| 3.10 | 0 | 1 |
| 3.14 | 2 | 1 |
| 3.26 | 0 | 0 |
| 3.38 | 1 | 0 |
| 3.45 | 1 | 1 |
| 3.45 | 0 | 1 |
| 3.90 | 1 | 1 |
| 3.92 | 2 | 0 |
| 3.93 | 1 | 0 |
| 3.98 | 2 | 1 |

| Mesons (continued) | | |
|---|---|---|
| Mass/$\sqrt{\sigma}$ | Spin | Isospin |
| 3.98 | 3 | 0 |
| 4.02 | 3 | 1 |
| 4.05 | 1 | 1 |
| 4.25 | 0 | 0 |
| 4.35 | 1 | 0 |
| 4.86 | 4 | 1 |
| 4.88 | 4 | 0 |
| 5.00 | 1 | 1 |
| 5.00 | 1 | 0 |

| Baryons | | |
|---|---|---|
| Mass/$\sqrt{\sigma}$ | Spin | Isospin |
| 2.84 | 0 | 1 |
| 3.05 | 1 | 3 |
| 3.47 | 2 | 5 |
| 2.84 | 0 | 5 |
| 3.05 | 1 | 5 |
| 3.47 | 2 | 5 |
| 3.05 | 1 | 3 |
| 3.47 | 2 | 3 |
| 4.10 | 3 | 3 |

TABLE A.2: The included particle spectrum in the four-color fundamental theory.

| Mesons | | |
|---|---|---|
| Mass/$\sqrt{\sigma}$ | Spin | Isospin |
| 0.00 | 0 | 1 |
| 1.43 | 0 | 0 |
| 1.83 | 1 | 1 |
| 1.86 | 1 | 0 |
| 2.33 | 0 | 1 |
| 2.79 | 1 | 0 |
| 2.94 | 1 | 1 |
| 3.00 | 1 | 1 |
| 3.02 | 2 | 0 |
| 3.06 | 1 | 0 |
| 3.08 | 0 | 0 |
| 3.10 | 0 | 1 |
| 3.14 | 2 | 1 |
| 3.26 | 0 | 0 |
| 3.38 | 1 | 0 |
| 3.45 | 1 | 1 |
| 3.45 | 0 | 1 |
| 3.90 | 1 | 1 |
| 3.92 | 2 | 0 |
| 3.93 | 1 | 0 |
| 3.98 | 2 | 1 |
| 3.98 | 3 | 0 |
| 4.02 | 3 | 1 |
| 4.05 | 1 | 1 |
| 4.25 | 0 | 0 |
| 4.35 | 1 | 0 |
| 4.86 | 4 | 1 |
| 4.88 | 4 | 0 |

| Mesons (continued) | | |
|---|---|---|
| Mass/$\sqrt{\sigma}$ | Spin | Isospin |
| 5.00 | 1 | 1 |
| 5.00 | 1 | 0 |

| Diquarks | | |
|---|---|---|
| Mass/$\sqrt{\sigma}$ | Spin | Isospin |
| 0.00 | 0 | 1 |
| 1.43 | 0 | 1 |
| 1.86 | 1 | 0 |
| 2.79 | 1 | 0 |
| 3.02 | 2 | 1 |
| 3.06 | 1 | 1 |
| 3.08 | 0 | 1 |
| 3.26 | 0 | 1 |
| 3.38 | 1 | 0 |
| 3.92 | 2 | 1 |
| 3.93 | 1 | 0 |
| 3.98 | 3 | 0 |
| 4.25 | 0 | 1 |
| 4.35 | 1 | 1 |
| 4.88 | 4 | 1 |
| 5.00 | 1 | 0 |

| Baryons | | |
|---|---|---|
| Mass/$\sqrt{\sigma}$ | Spin | Isospin |
| 0.71 | 0 | 0 |
| 1.55 | 1 | 1 |
| 3.24 | 2 | 2 |
| 5.77 | 3 | 3 |

TABLE A.3: The mesons, diquarks, and baryons in the four-color antisymmetric theory. (See Table A.4 for remaining particles in this theory.)

| Four-quark objects | | | (continued) | | | (continued) | | | (continued) | | |
| --- | --- | --- | --- | --- | --- | --- | --- | --- | --- | --- | --- |
| $m/\sqrt{\sigma}$ | $g_S$ | $g_I$ | $m/\sqrt{\sigma}$ | $g_S$ | $g_I$ | $m/\sqrt{\sigma}$ | $g_S$ | $g_I$ | $m/\sqrt{\sigma}$ | $g_S$ | $g_I$ |
| 0.00 | 1 | 1 | 5.36 | 3 | 3 | 6.86 | 9 | 9 | 7.90 | 66 | 66 |
| 1.43 | 1 | 1 | 5.40 | 7 | 7 | 6.94 | 25 | 25 | 7.94 | 27 | 27 |
| 1.86 | 3 | 3 | 5.57 | 9 | 9 | 6.95 | 15 | 15 | 7.95 | 49 | 49 |
| 2.79 | 3 | 3 | 5.68 | 1 | 1 | 6.98 | 15 | 15 | 7.96 | 9 | 9 |
| 2.86 | 1 | 1 | 5.78 | 18 | 18 | 6.99 | 9 | 9 | 8.02 | 15 | 15 |
| 3.02 | 5 | 5 | 5.79 | 9 | 9 | 7.00 | 40 | 40 | 8.06 | 9 | 9 |
| 3.06 | 3 | 3 | 5.81 | 15 | 15 | 7.01 | 3 | 3 | 8.08 | 3 | 3 |
| 3.08 | 1 | 1 | 5.84 | 21 | 21 | 7.04 | 24 | 24 | 8.14 | 9 | 9 |
| 3.26 | 1 | 1 | 5.85 | 9 | 9 | 7.06 | 7 | 7 | 8.17 | 5 | 5 |
| 3.29 | 3 | 3 | 5.87 | 3 | 3 | 7.14 | 9 | 9 | 8.18 | 3 | 3 |
| 3.38 | 3 | 3 | 6.05 | 28 | 28 | 7.18 | 5 | 5 | 8.23 | 7 | 7 |
| 3.72 | 9 | 9 | 6.08 | 15 | 15 | 7.19 | 3 | 3 | 8.26 | 30 | 30 |
| 3.92 | 5 | 5 | 6.11 | 5 | 5 | 7.24 | 7 | 7 | 8.27 | 15 | 15 |
| 3.93 | 3 | 3 | 6.11 | 3 | 3 | 7.27 | 5 | 5 | 8.28 | 9 | 9 |
| 3.98 | 7 | 7 | 6.12 | 9 | 9 | 7.30 | 15 | 15 | 8.33 | 21 | 21 |
| 4.21 | 3 | 3 | 6.14 | 3 | 3 | 7.31 | 12 | 12 | 8.38 | 9 | 9 |
| 4.25 | 1 | 1 | 6.17 | 10 | 10 | 7.33 | 1 | 1 | 8.50 | 1 | 1 |
| 4.35 | 3 | 3 | 6.21 | 9 | 9 | 7.36 | 21 | 21 | 8.60 | 3 | 3 |
| 4.45 | 5 | 5 | 6.29 | 5 | 5 | 7.37 | 15 | 15 | 8.70 | 9 | 9 |
| 4.49 | 3 | 3 | 6.31 | 9 | 9 | 7.41 | 9 | 9 | 8.80 | 45 | 45 |
| 4.51 | 1 | 1 | 6.32 | 3 | 3 | 7.43 | 3 | 3 | 8.81 | 27 | 27 |
| 4.65 | 9 | 9 | 6.35 | 1 | 1 | 7.51 | 1 | 1 | 8.86 | 63 | 63 |
| 4.69 | 1 | 1 | 6.40 | 15 | 15 | 7.61 | 3 | 3 | 8.92 | 15 | 15 |
| 4.81 | 3 | 3 | 6.43 | 3 | 3 | 7.63 | 3 | 3 | 8.93 | 9 | 9 |
| 4.88 | 9 | 9 | 6.44 | 9 | 9 | 7.67 | 27 | 27 | 8.98 | 21 | 21 |
| 4.89 | 15 | 15 | 6.46 | 3 | 3 | 7.73 | 9 | 9 | 9.13 | 9 | 9 |
| 4.92 | 9 | 9 | 6.52 | 1 | 1 | 7.79 | 9 | 9 | 9.23 | 27 | 27 |
| 4.95 | 3 | 3 | 6.64 | 3 | 3 | 7.83 | 25 | 25 | 9.25 | 3 | 3 |
| 5.00 | 3 | 3 | 6.70 | 15 | 15 | 7.85 | 15 | 15 | 9.35 | 9 | 9 |
| 5.12 | 3 | 3 | 6.71 | 9 | 9 | 7.86 | 9 | 9 | 9.76 | 81 | 81 |
| 5.24 | 9 | 9 | 6.74 | 27 | 27 | 7.89 | 35 | 35 | 9.88 | 27 | 27 |
| 5.35 | 5 | 5 | 6.76 | 30 | 30 | | | | | | |

TABLE A.4: The included tetraquarks, di-mesons, and diquark-mesons in the four-color antisymmetric theory. (There is one of each of these particle types for each line in this table.) Here, $m$, $g_S$, and $g_I$ are the mass, total spin, and isospin degeneracies, respectively. As noted above in Section 4.1.2, we need not determine how all of the four-quark-object degrees of freedom break up into spin and isospin multiplets because of the mass degeneracy.